\documentclass[a4paper,fleqn]{cas-dc}

\usepackage{lineno}
\usepackage[labelfont=normalfont]{caption}
\usepackage[authoryear]{natbib}
\usepackage{setspace}
\usepackage{siunitx}
\usepackage[caption=false]{subfig}
\usepackage{tikz}

\newcommand{\cmmnt}[1]{}

\newcommand\bmm[1]{\boldsymbol{#1}}
\hypersetup{
    linktoc     = all,
    colorlinks  = true,
    allcolors = [rgb]{0,0,0.75}, %%BLUE
}
\hypersetup{final}

\usetikzlibrary{arrows,chains,matrix,positioning,scopes}

\catcode`@=11
\def\caseswithdelim#1#2{\left#1\,\vcenter{\normalbaselines\m@th
        \ialign{\strut$##\hfil$&\quad##\hfil\crcr#2\crcr}}\right.}% you might like it without the \strut
\catcode`@=12

\def\breakloop{\fi\iffalse}

\newcommand\drm[1]{\mathrm{d} #1}

\newcommand{\file}[1]{\texttt{#1}}

\newcommand{\fsize}{}

\newcommand{\figtype}{}

\newcommand{\ftime}{}

\newcommand{\forloop}[5][1]%
{%
    \setcounter{#2}{#3}%
    \ifthenelse{#4}%
    {%
        #5%
        \addtocounter{#2}{#1}%
        \forloop[#1]{#2}{\value{#2}}{#4}{#5}%
    }%
    {%
    }%
}%

\newcommand*\overbar[1]{%
    \vbox{%
        \hrule height 0.9pt%
        \kern0.35ex%
        \hbox{%
            \kern-0.1em%
            \ifmmode#1\else\ensuremath{#1}\fi%
            \kern0.0em%
        }% end of hbox
    }% end of vbox
}

\ExplSyntaxOn
\keys_set:nn {stm/mktitle} {nologo}
\ExplSyntaxOff

\renewcommand{\figtype}{pdf}
\begin{document}
\let\WriteBookmarks\relax
\def\floatpagepagefraction{1}
\def\textpagefraction{.001}
\shorttitle{The influence of physiological flow development on popular wall shear stress metrics in an idealized curved artery}
\shortauthors{C. Cox and M.W. Plesniak}

\title[mode = title]{The influence of physiological flow development on popular wall shear stress metrics in an idealized curved artery}
%\tnotemark[1,2]
%\tnotetext[2]{}

\author[1]{Christopher Cox}[orcid=0000-0001-6552-4890]
\cormark[1]
\ead{ccox@gwmail.gwu.edu}
%\ead[]{ccox@gwmail.gwu.edu}

\author[2]{Michael W. Plesniak}[orcid=0000-0001-6575-5074]
%\cormark[2]
\ead{plesniak@gwu.edu}
%\ead[]{plesniak@gwu.edu}

\address[1]{Lawrence Livermore National Laboratory,
            Physical and Life Sciences,
            Livermore, CA 94550, USA }
\address[2]{The George Washington University,
            Department of Mechanical and Aerospace Engineering,
            Washington, DC 20052, USA }

\cortext[cor1]{Corresponding author}
%\cortext[cor2]{Principal corresponding author}
%\fntext[fn1]{footnote text.}
%\fntext[fn2]{footnote text.}

\begin{abstract}
    %\doublespacing
    \singlespacing
    We numerically investigate the influence of flow development on secondary flow patterns and subsequent wall shear stress distributions in a curved artery model, and we compute vascular metrics commonly used to assess variations in blood flow characteristics as it applies to arterial disease. We model a human artery with a simple, rigid $180^\circ$ curved tube with circular cross-section and constant curvature, neglecting effects of taper, torsion and elasticity. High-fidelity numerical results are computed from an in-house discontinuous spectral element flow solver. The flow rate used in this study is physiological. We perform this study using a Newtonian blood-analog fluid subjected to a pulsatile flow with two inflow conditions. The first flow condition is fully developed while the second condition is undeveloped (i.e. uniform). We observe and discuss differences in secondary flow patterns that emerge over the rapid acceleration and deceleration phases of the physiological waveform, and we directly connect the variation in intensity of these secondary flow patterns along the curvature to differences in the wall shear stress metrics for each entrance condition. Results indicate that decreased axial velocities under an undeveloped condition produce less intense secondary flow that, in turn, reduces both the oscillatory and multidirectional nature of the wall shear stress vector, and we link this effect to abnormalities in computed stress metrics. These results suggest potentially lower prevalence of disease in curvatures where entrance flow is rather undeveloped---a physiologically relevant result to further understand the influence of blood flow development on disease.
\end{abstract}


\maketitle
%\doublespacing
\singlespacing

%%%%%%%%%%%%%%%%%%%%%%%%%%%%%%%%%%%%%%%%%%%%%%%%%%
\section{Introduction}
\label{s:introduction}

%%%%%%%%%%%%%%%%%%%%%%%%%%%%%%%%%%%%%%%%%%%%%%%%%%
%\include{F-curved-pipe-geometry}
%\input{F-curved-pipe-geometry.tex}
%
\begin{figure*}
   \renewcommand{\fsize}{0.24\textwidth}
   \centering\setcounter{subfigure}{0}
   \subfloat[]{
       \includegraphics[height=\fsize,keepaspectratio]
       {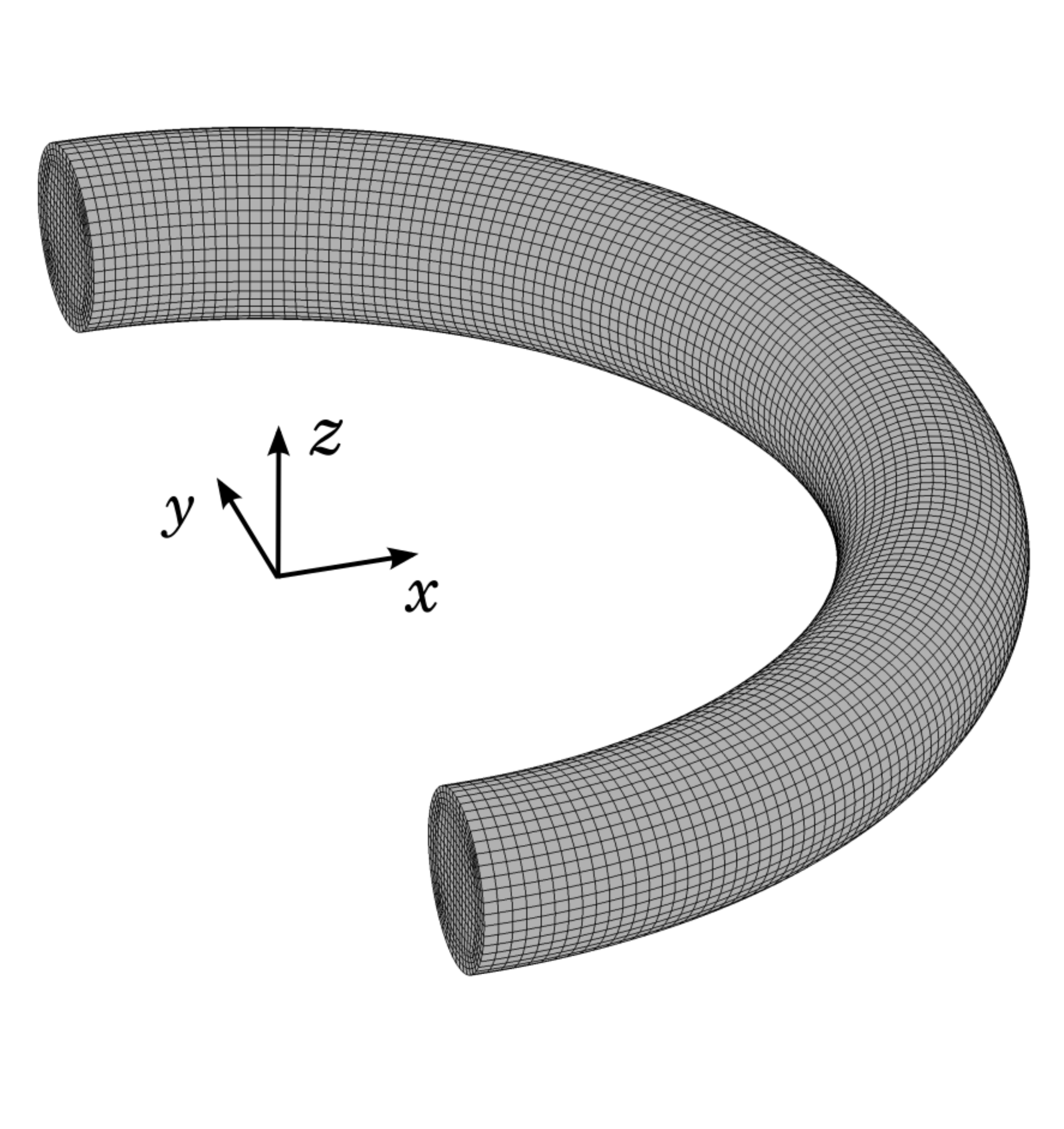}
       \label{f:pipe_mesh}
   }
   \subfloat[]{
       \includegraphics[height=\fsize,keepaspectratio]
       {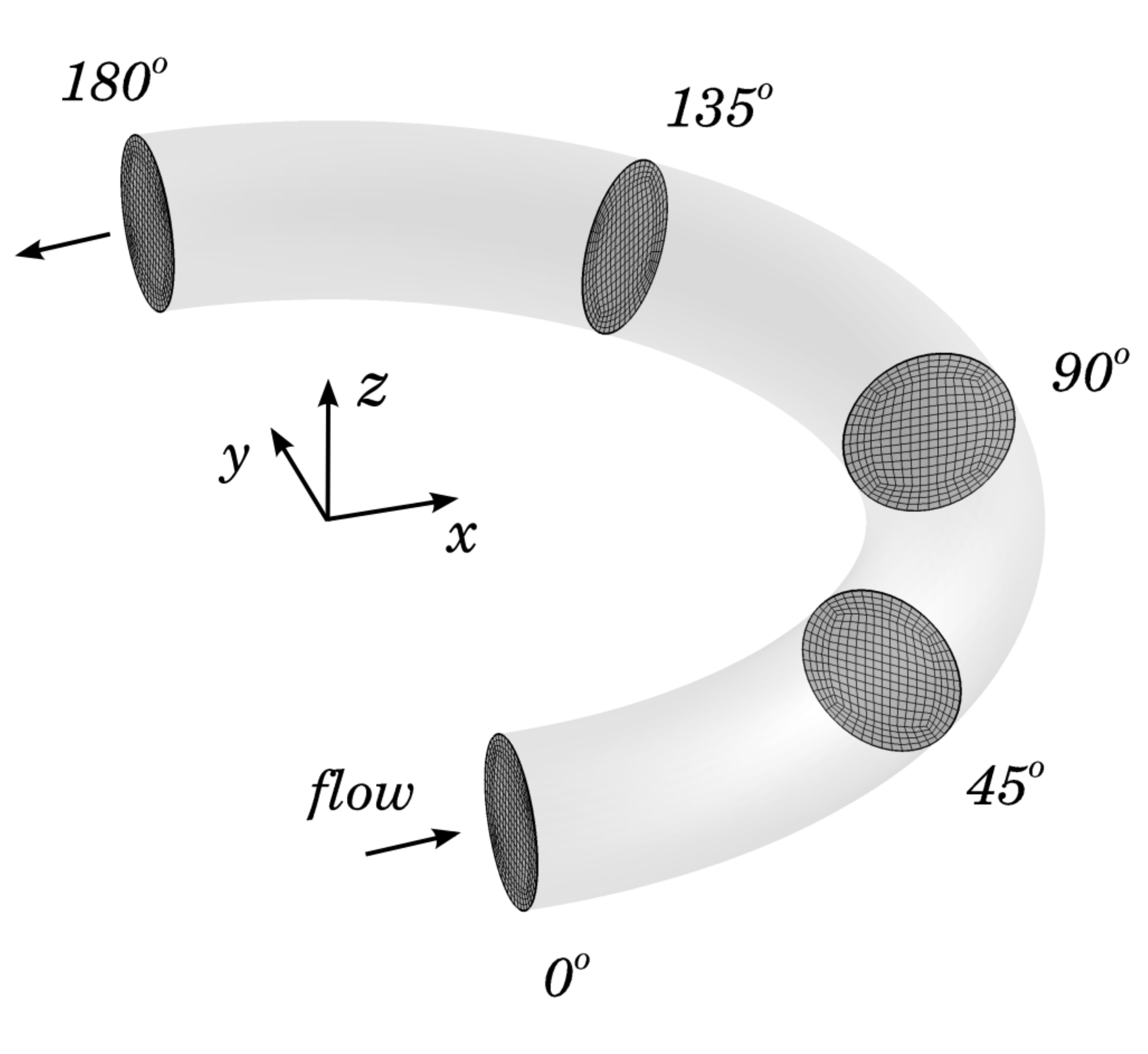}
       \label{f:pipe_cross_section}
   }
   \subfloat[]{
       \includegraphics[height=\fsize,keepaspectratio]
       {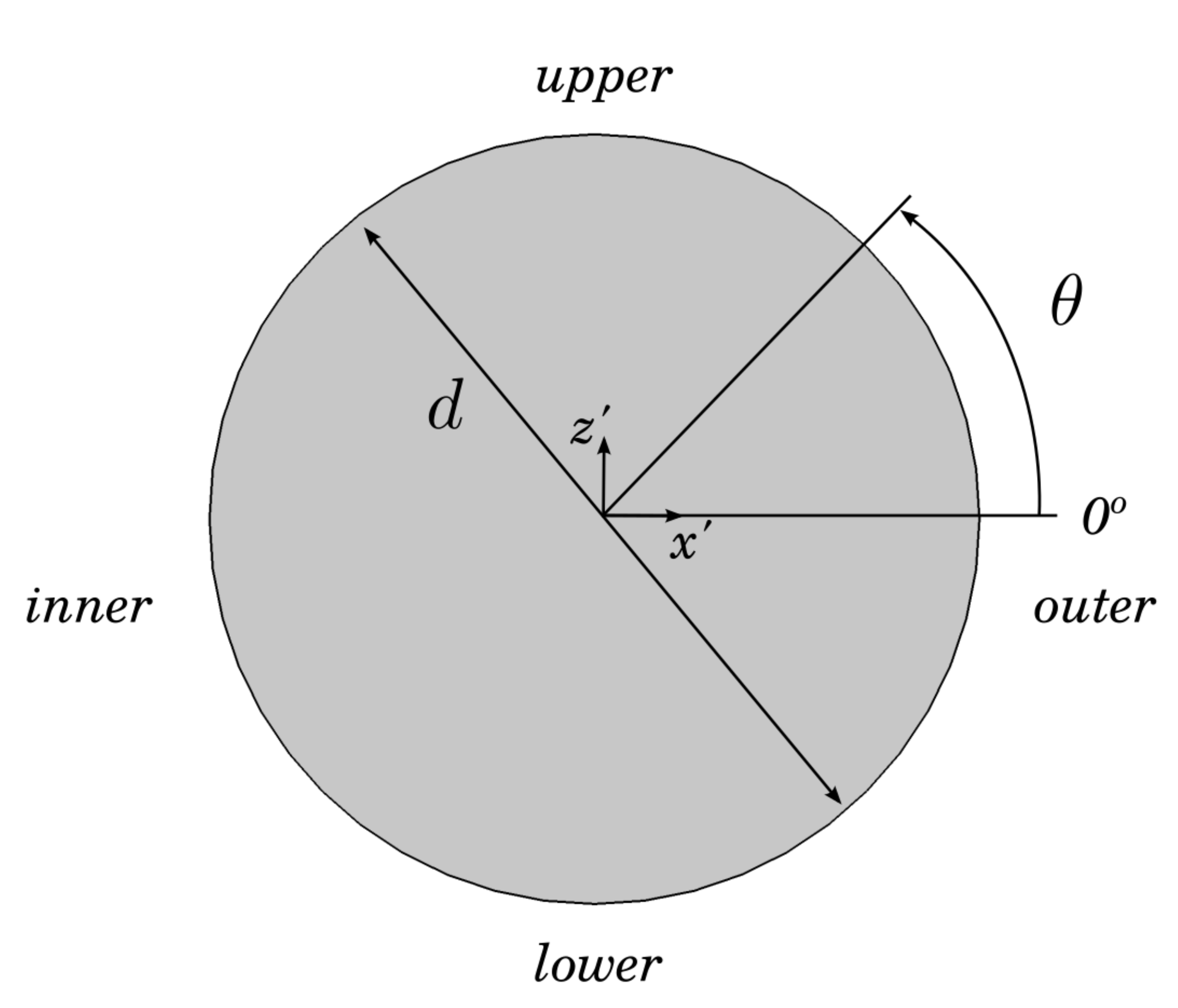}
       \label{f:theta_planar_view}
   }
   \subfloat[]{
       \includegraphics[height=\fsize,keepaspectratio]
       {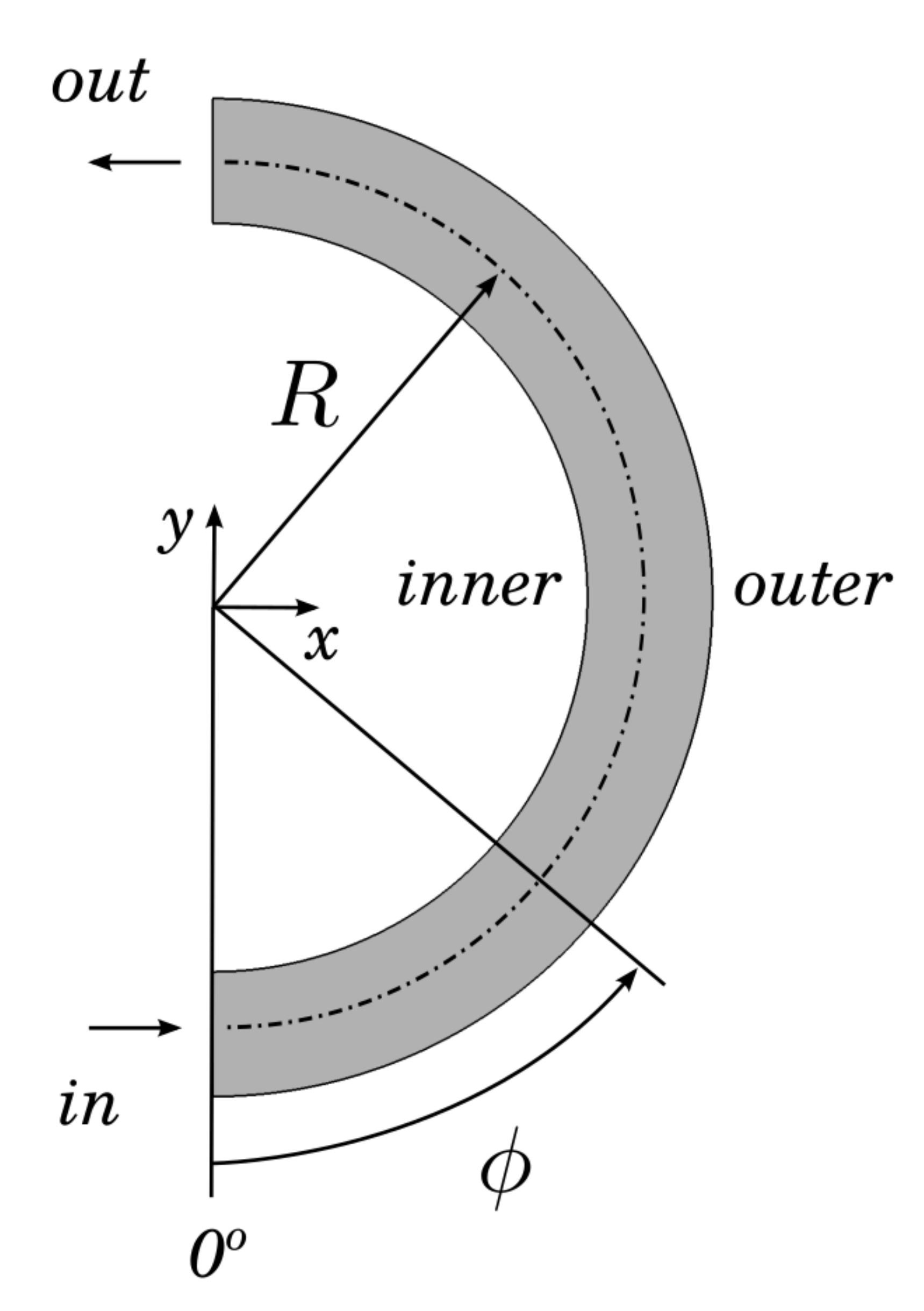}
       \label{f:phi_top_view}
   }
   
   \subfloat[]{
       \includegraphics[width=0.6\textwidth,keepaspectratio]
       {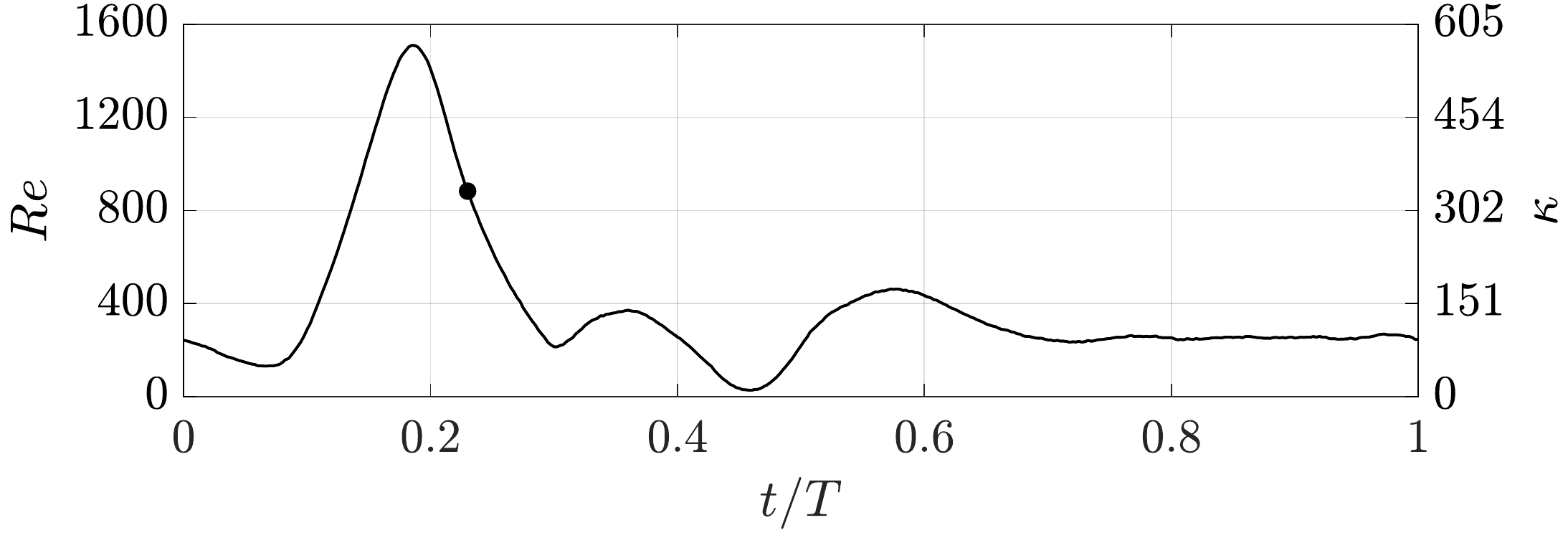}
       \label{f:waveform_Re}
   }
   
   \caption{Curved pipe geometry and orientation. (a) curvature mesh, (b) curved pipe with flow direction and various cross-sections, (c) $x'z'$ cross-sectional plane viewed from upstream, (d) $xy$ plane of symmetry, (e) physiological waveform in terms of Reynolds $Re$ and Dean $\kappa$ numbers as a function of nondimensional pulse period $t^\star=t/T$, displaying rapid acceleration $0.10 < t^\star < 0.19$ and deceleration $0.19 < t^\star < 0.30$. Mid-deceleration occurs at approximately $t^\star=0.23$ (black dot).}
   \label{f:curved_pipe_mesh_geometry_N4}
\end{figure*}
%%%%%%%%%%%%%%%%%%%%%%%%%%%%%%%%%%%%%%%%%%%%%%%%%%

Cardiovascular flows are pulsatile, incompressible flows that exist in complex geometries with compliant walls. Together, these factors are conducive to complex flow physics that affect wall shear stress---a physiologically and pathologically relevant force at play within the cardiovascular system that acts on the endothelium. This force is important because atherosclerotic regions are strongly correlated with curvature and branching in the human vasculature, where there exists oscillatory shear stress with a low time-averaged value and spatial/temporal shear gradients (\cite{davies:2008,glagov-zarins-giddens-ku:1988}).

Multidirectionality of the flow may also play an important role in the prevalence of atherosclerotic disease (\cite{mantha:2006,chakraborty:2012,peiffer-sherwin-weinberg:2013}). Relevant haemodynamic metrics used to assess the local variation in blood flow characteristics as it relates to atherosclerotic lesions are low {\it time-averaged wall shear stress} (TAWSS), {\it oscillatory shear index} (\cite{he-ku:1996}) (OSI) and {\it relative residence time} (\cite{himburg:2004}) (RRT); although, evidence for the low temporal mean and oscillatory shear stress concept may be less robust than previously assumed (\cite{peiffer-sherwin-weinberg:2013b}). In light of this evidence, another metric---{\it transverse wall shear stress} (TransWSS)---was designed to account for multidirectionality of the wall shear stress vector (\cite{peiffer-sherwin-weinberg:2013}), and recent research has applied this complimentary metric (\cite{morbiducci:2015, andersson:2019, hoogendoorn:2020, krishna:2020, dessalles:2021, meschi:2021}).

In this study, we perform numerical simulations of a Newtonian blood-analog fluid using a physiological pulsatile waveform and two inflow conditions to a $180^\circ$ curved tube, neglecting effects of taper, torsion, and elasticity. We adopt this idealized approach without the added complexity of variable geometry by modeling a human artery with a rigid curved tube with circular cross-section and constant curvature. The first pulsatile entrance flow condition is one that is fully developed while the second condition is undeveloped (i.e. uniform). Previous research (\cite{cox-plesniak:2021}) demonstrated that decreased axial velocities under an undeveloped condition produce smaller secondary flows that ultimately inhibit growth of any interior flow vortices and production of abnormal wall shear stresses. Following up this research, the current work investigates how the inflow condition ultimately affects relevant wall shear stress metrics through varying intensities of secondary flow due to pulsatility of the waveform.

%%%%%%%%%%%%%%%%%%%%%%%%%%%%%%%%%%%%%%%%%%%%%%%%%%
%\include{F-waveform-velocity-profile}
%\input{F-waveform-velocity-profile.tex}
%%%%%%%%%%%%%%%%%%%%%%%%%%%%%%%%%%%%%%%%%%%%%%%%%%

% \begin{figure*}
%    \centering
%    \includegraphics[width=0.48\textwidth,keepaspectratio]
%    {./figs/waveform000023_crop.\figtype}
%    
%    \caption{Physiological waveform in terms of Reynolds $Re$ and Dean $\kappa$ numbers as a function of nondimensional pulse period $t^\star=t/T$, displaying rapid acceleration $0.10 < t^\star < 0.19$ and deceleration $0.19 < t^\star < 0.30$. Mid-deceleration occurs at approximately $t^\star=0.23$ (black dot).}
%    \label{f:waveform_Re}
% \end{figure*}

%%%%%%%%%%%%%%%%%%%%%%%%%%%%%%%%%%%%%%%%%%%%%%%%%%
\section{Methods}
\label{s:methods}

\subsection{Geometry}
\label{s:cfd_geometry}

A model of the curved tube that we use for our numerical simulations and presentation of results is depicted in Fig.~\ref{f:curved_pipe_mesh_geometry_N4}. The origin of the Cartesian coordinate system $(x,y,z)=(0,0,0)$ is defined in Fig.~\ref{f:curved_pipe_mesh_geometry_N4}\subref{f:pipe_mesh} along with the various cross-sections of interest in Fig.~\ref{f:curved_pipe_mesh_geometry_N4}\subref{f:pipe_cross_section}. The diameter $d=2r=0.0127$~m and the radius of curvature $R$, flow direction, and location of the inner, outer, upper and lower walls are shown in Fig.~\ref{f:curved_pipe_mesh_geometry_N4}\subref{f:theta_planar_view}\subref{f:phi_top_view}. For these simulations, the curvature ratio $\delta = r/R = 1/7$ matches that used in previous work reported in the literature (\cite{soh-berger:1984,vanwyk:2015,najjari-plesniak:2016,cox-plesniak:2021}). This small curvature ratio is less susceptible to flow separation for this range of Reynolds numbers and facilitates meaningful interpretation of the entrance effect on secondary flow and subsequent wall shear stress metrics. Also, see Fig.~\ref{f:curved_pipe_mesh_geometry_N4}\subref{f:theta_planar_view}\subref{f:phi_top_view} for definitions of the poloidal angle $\theta$ and toroidal angle $\phi$ that are used to present results.

%%%%%%%%%%%%%%%%%%%%%%%%%%%%%%%%%%%%%%%%%%%%%%%%%%
%\include{F-us-profile}
%\input{F-us-profile.tex}
%
\begin{figure*}
   \renewcommand{\fsize}{0.27\textwidth}
   \centering\setcounter{subfigure}{0}
   \subfloat[]{
       \includegraphics[width=\fsize,height=\fsize,keepaspectratio]
       {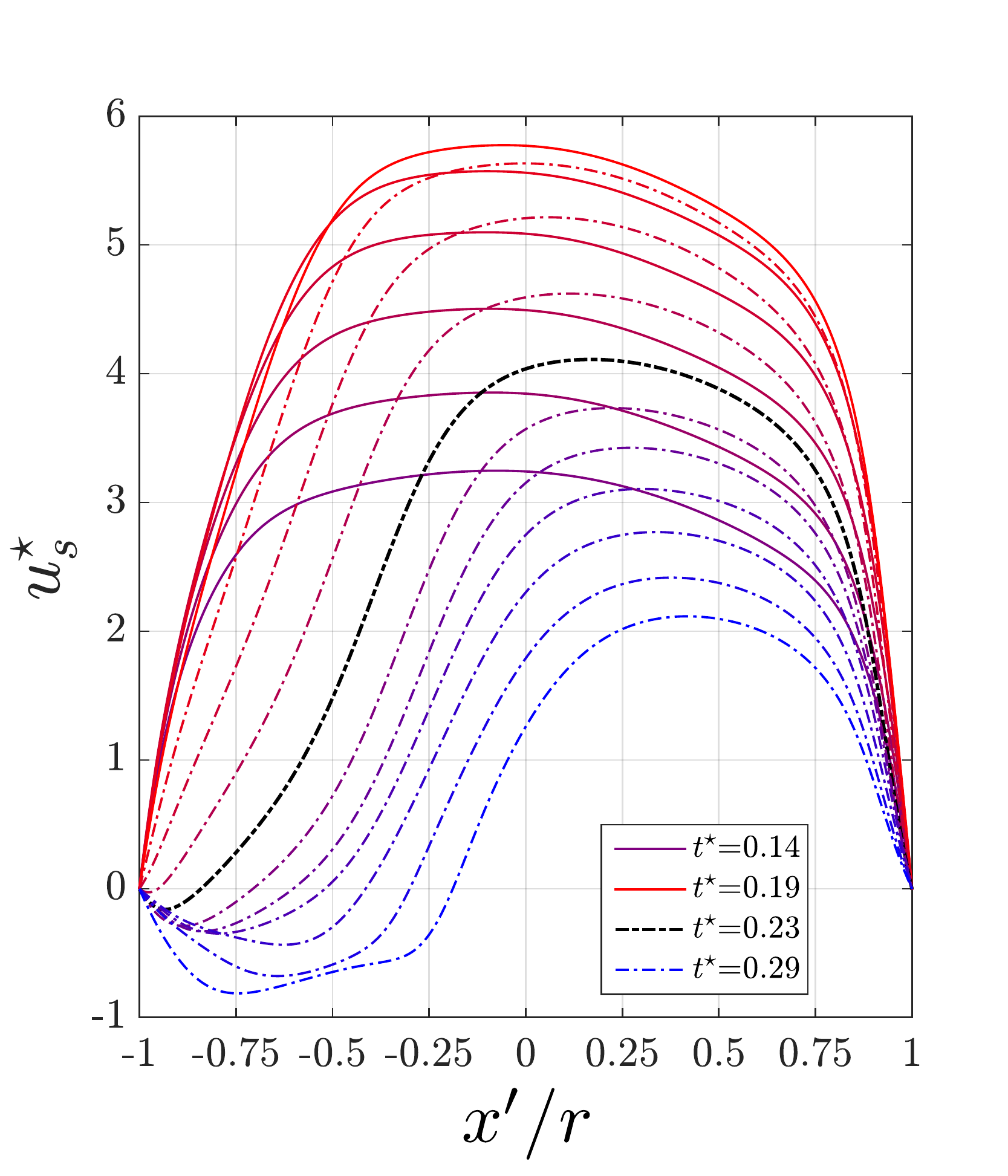}
       \label{f:us_profile_22_womersley_z0}
   }
   \subfloat[]{
       \includegraphics[width=\fsize,height=\fsize,keepaspectratio]
       {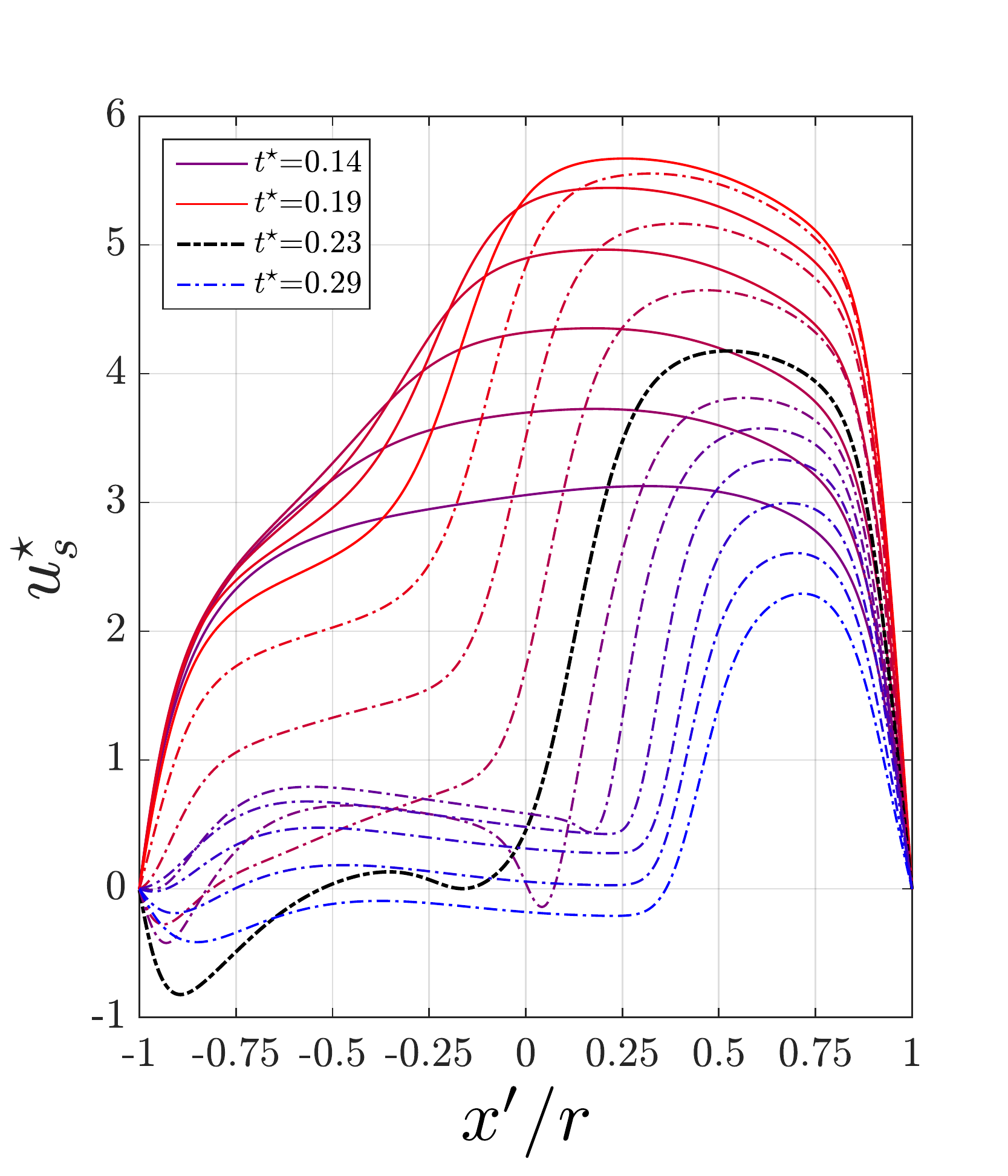}
       \label{f:us_profile_45_womersley_z0}
   }
   \subfloat[]{
       \includegraphics[width=\fsize,height=\fsize,keepaspectratio]
       {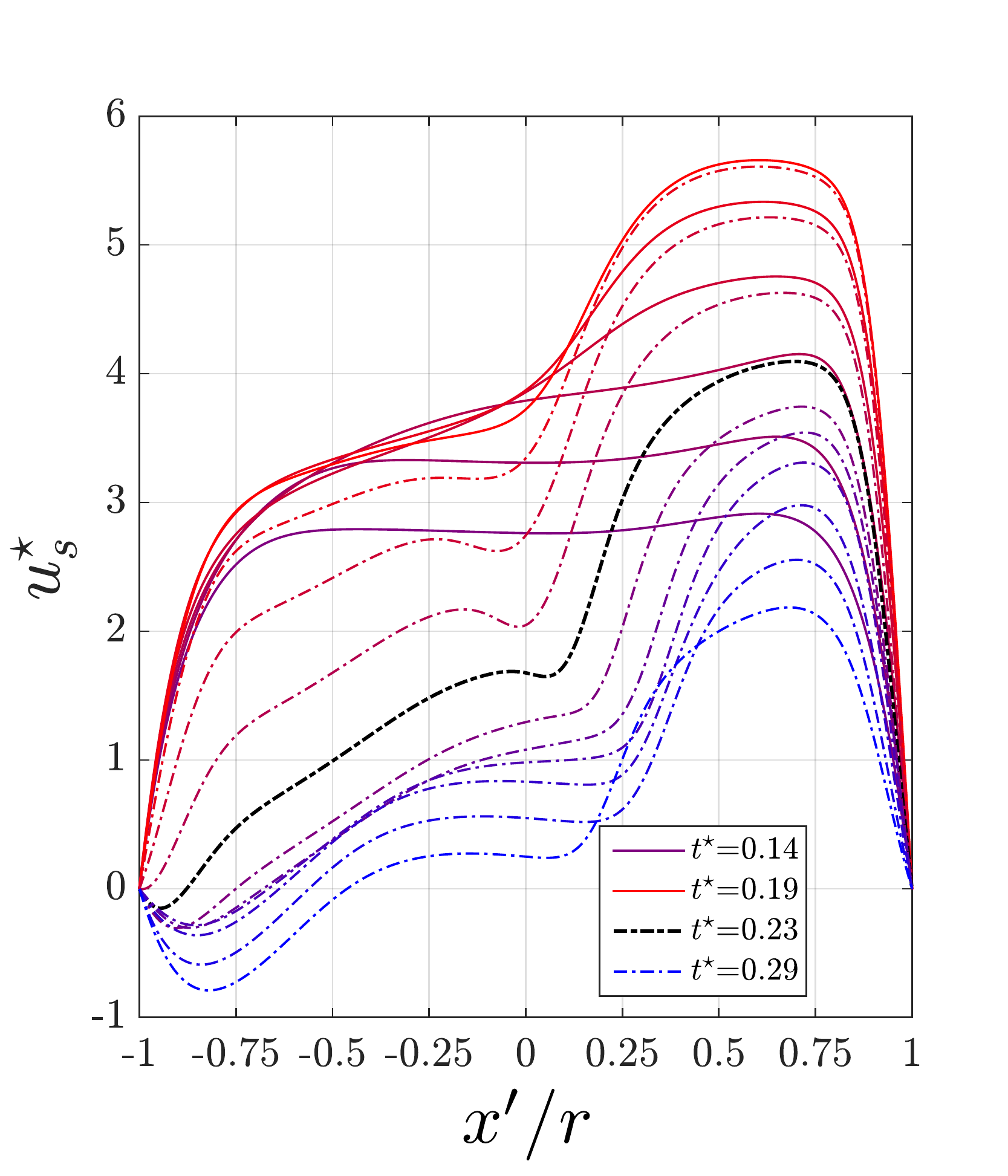}
       \label{f:us_profile_90_womersley_z0}
   }
   \subfloat[]{
       \includegraphics[width=\fsize,height=\fsize,keepaspectratio]
       {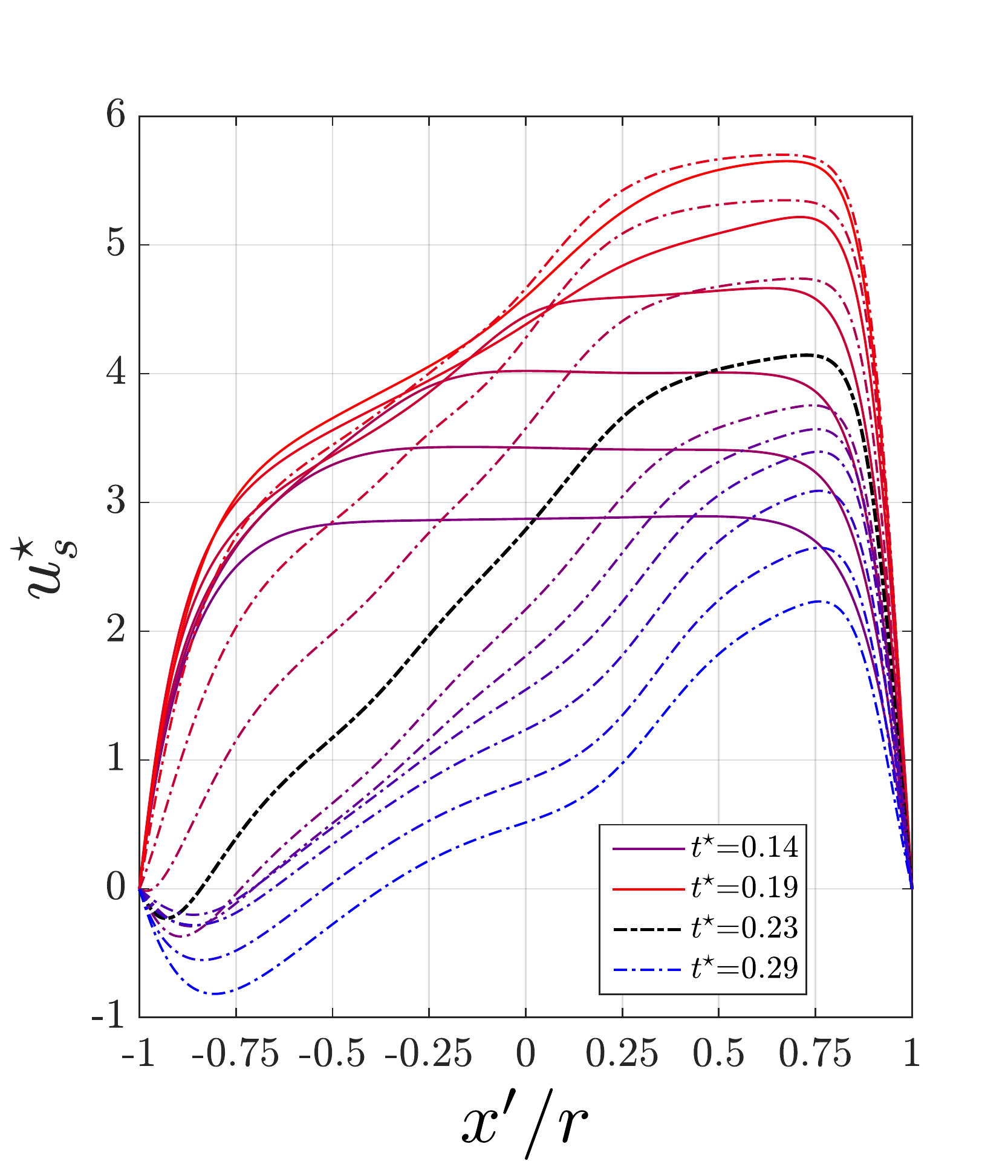}
       \label{f:us_profile_135_womersley_z0}
   }
   \\[-0.16in]
   \subfloat[]{
       \includegraphics[width=\fsize,height=\fsize,keepaspectratio]
       {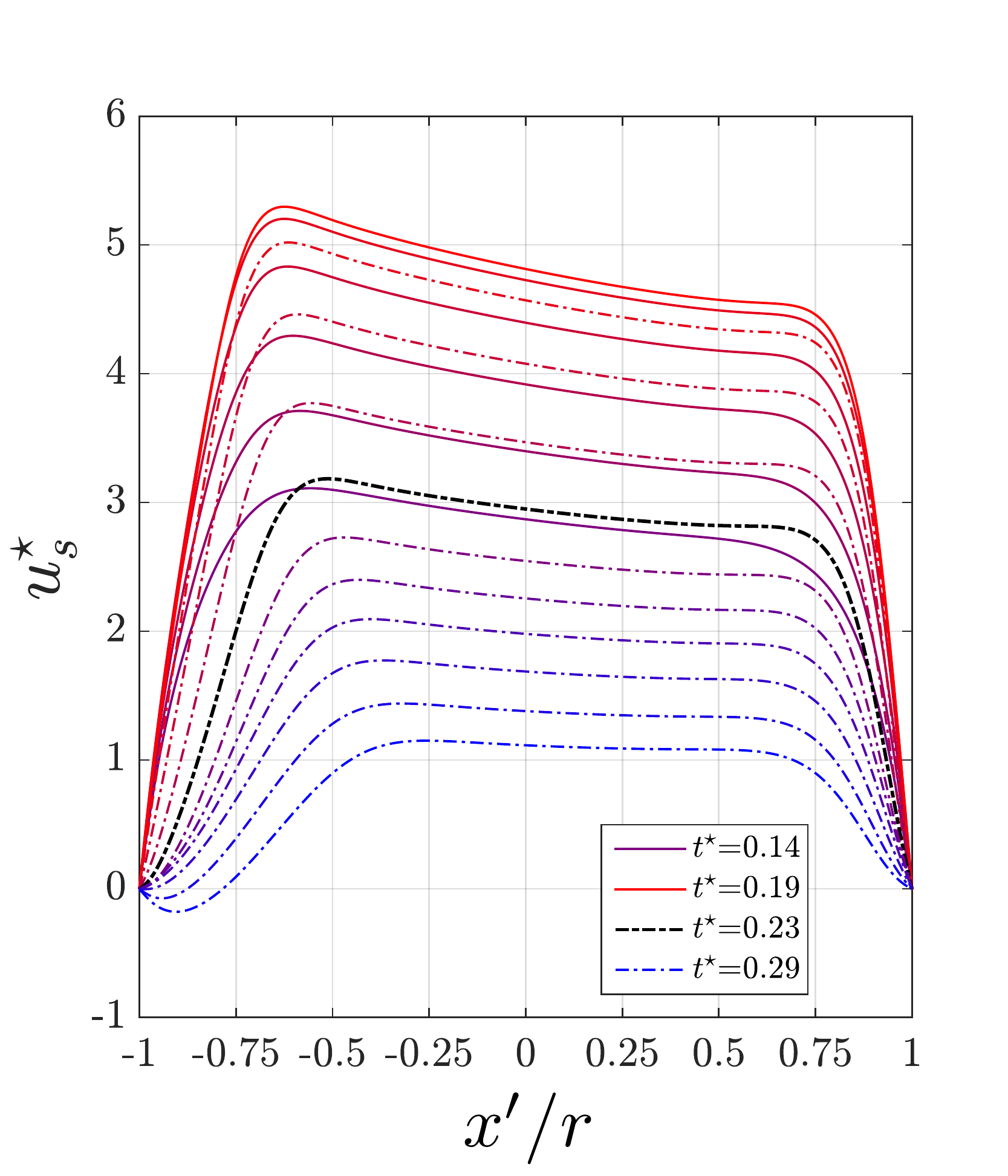}
       \label{f:us_profile_22_plug_z0}
   }
   \subfloat[]{
       \includegraphics[width=\fsize,height=\fsize,keepaspectratio]
       {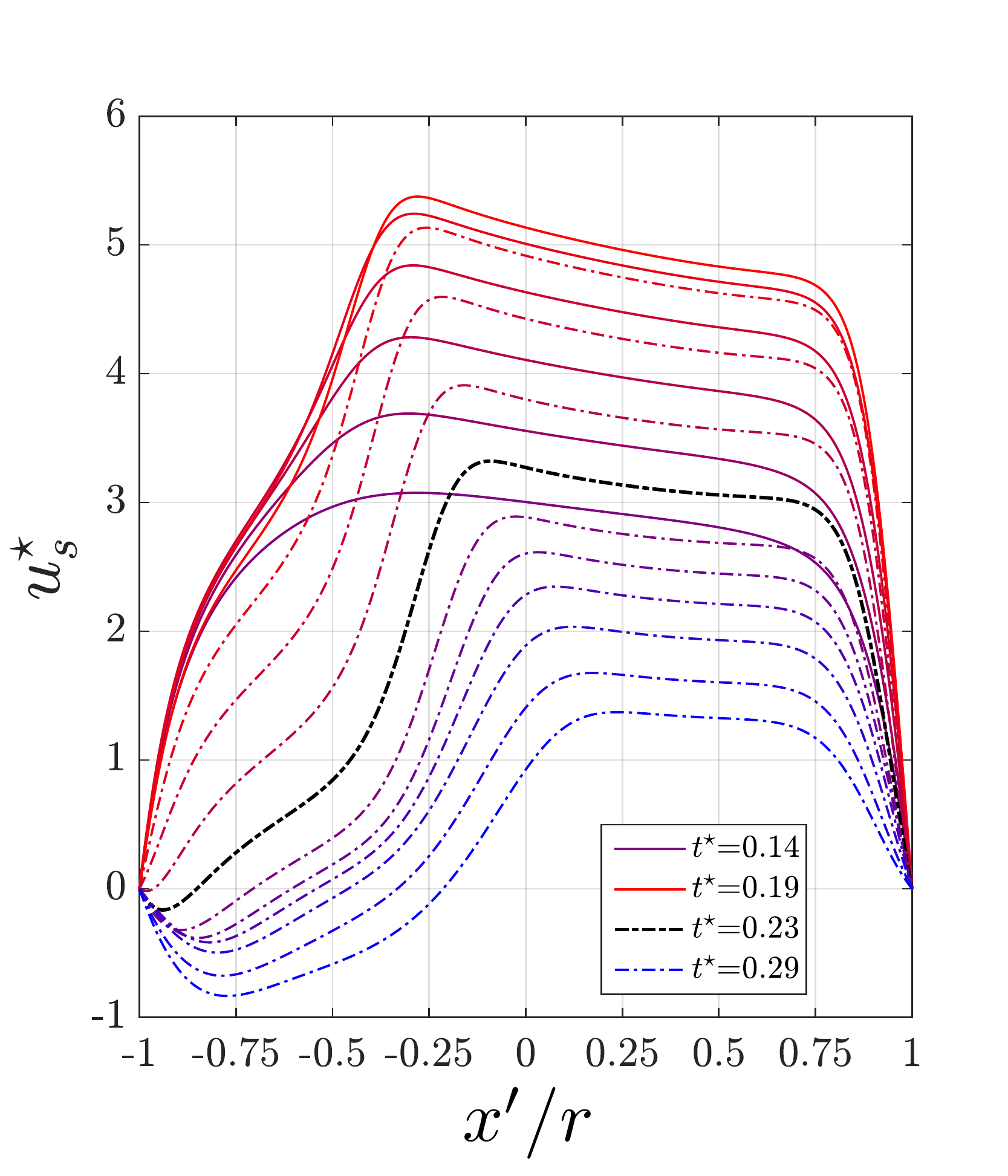}
       \label{f:us_profile_45_plug_z0}
   }
   \subfloat[]{
       \includegraphics[width=\fsize,height=\fsize,keepaspectratio]
       {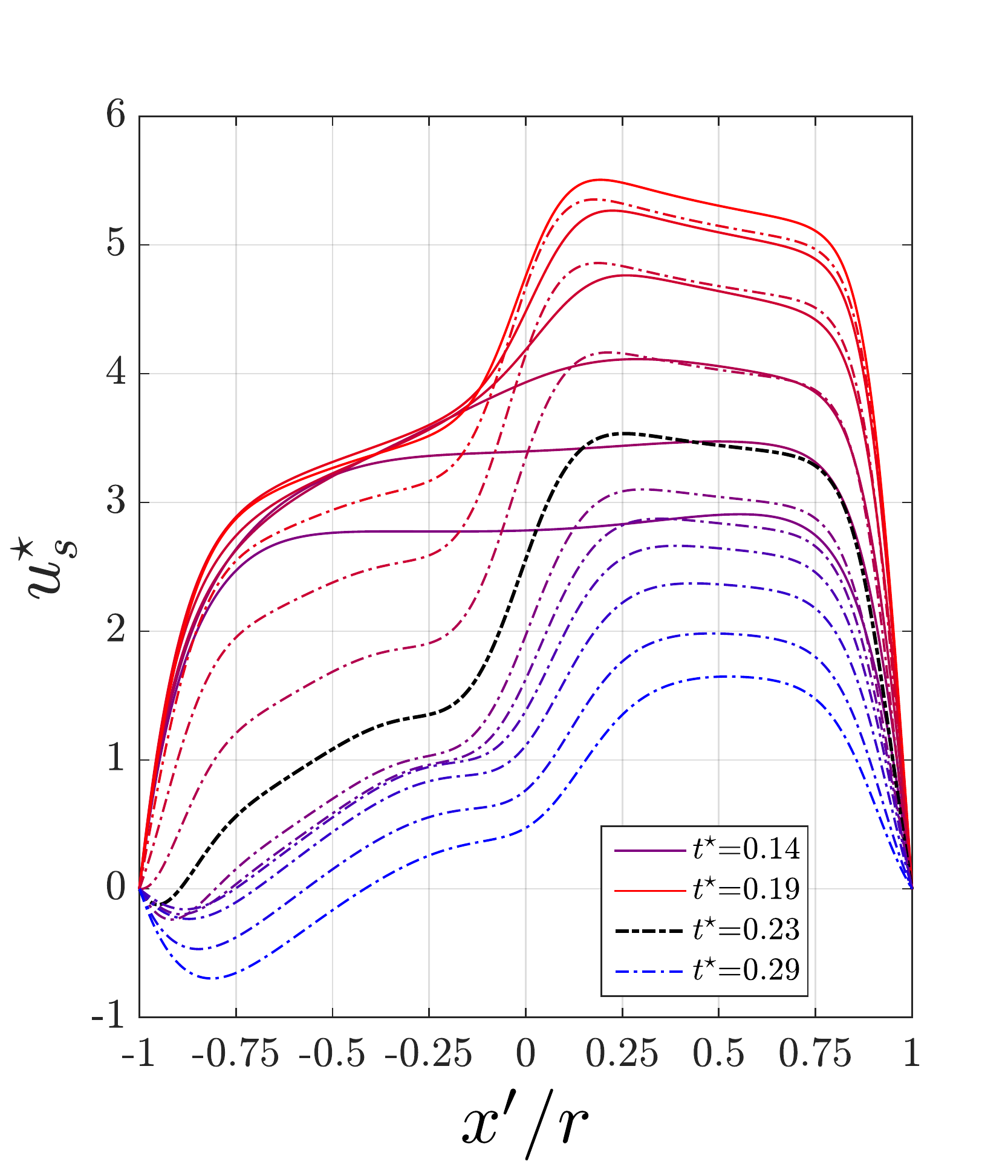}
       \label{f:us_profile_90_plug_z0}
   }
   \subfloat[]{
       \includegraphics[width=\fsize,height=\fsize,keepaspectratio]
       {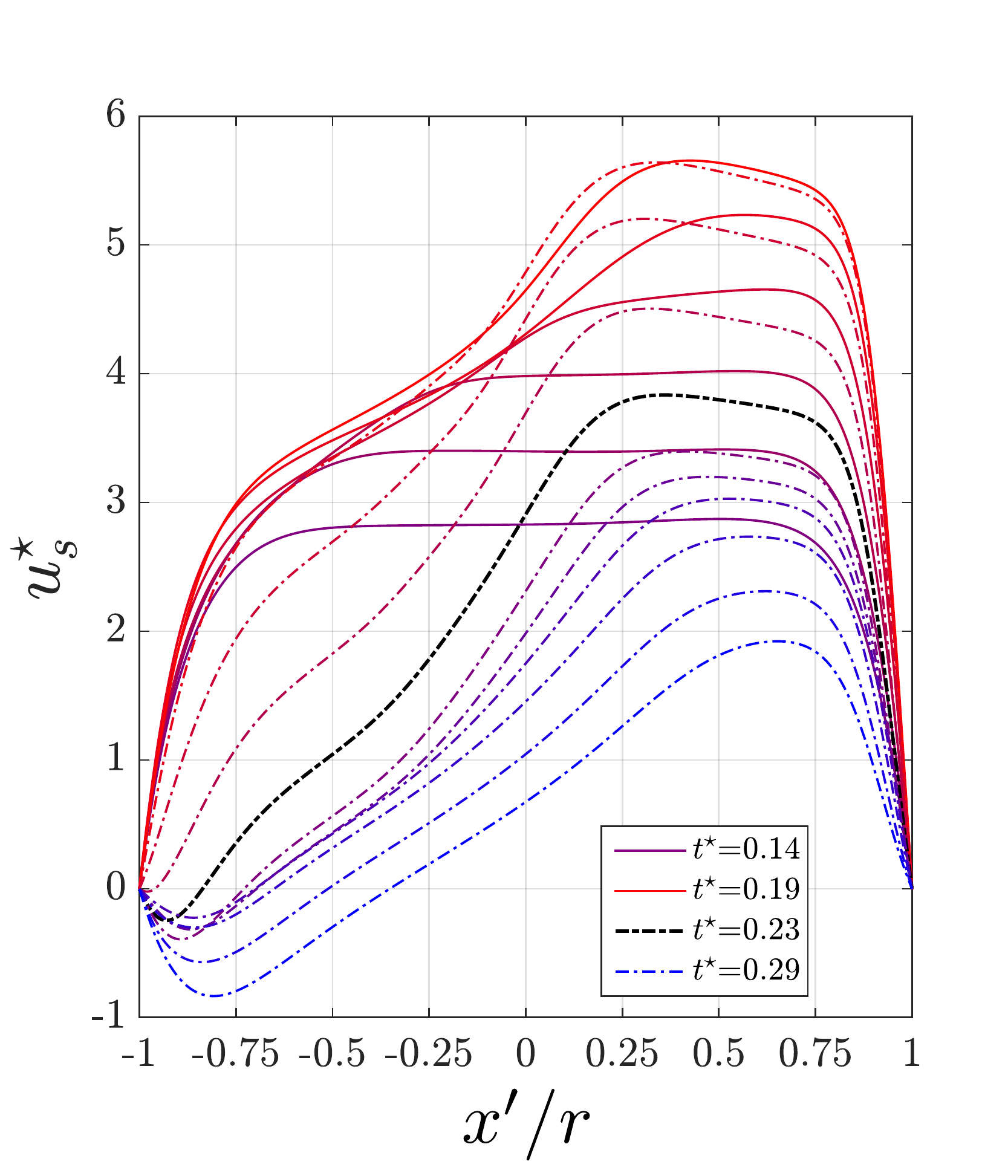}
       \label{f:us_profile_135_plug_z0}
   }
   
   \caption{Profiles of streamwise velocity, $u^\star_s$, under WEC (\textit{top}) and UEC (\textit{bottom}) at $z=0$ plane of symmetry for (a) $\phi=22^\circ$, (b) $\phi=45^\circ$, (c) $\phi=90^\circ$, (d) $\phi=135^\circ$ and $0.14 \leq t^\star \leq 0.29$ showing flow acceleration (solid) and deceleration (dash-dot). Phase $t^\star=0.23$ (black dash-dot) depicts outer wall skewness of the peak velocity that is greater under a fully developed condition and leads to higher intensity secondary flow (see Figs.~\ref{f:var_wec_t23} and \ref{f:var_uec_t23}).
   }
   \label{f:us_profile_z0}
\end{figure*}
%%%%%%%%%%%%%%%%%%%%%%%%%%%%%%%%%%%%%%%%%%%%%%%%%%

\subsection{Physiological flow rate}

The inflow waveform used in these simulations and provided in Fig.~\ref{f:curved_pipe_mesh_geometry_N4}\subref{f:waveform_Re} is that reported by \cite{holdsworth:1999} measured in the carotid artery of healthy human subjects and is prototypical of pulsatile waveforms. The angular frequency $\omega=2 \pi f$ is given by the period of the waveform $T=4$~s, and the kinematic viscosity of the fluid is $\nu=3.5 \times 10^{-6}$~$\mathrm{m^2~s^{-1}}$. We note that systemic compliance is inherent in the waveform and, therefore, taken into account. In the following sections, we use the nondimensional value $t^\star=t/T$ to denote instances of time. The waveform exhibits characteristic flow rate acceleration and deceleration and has been used in previous studies (\cite{vanwyk:2015,plesniak-bulusu:2016,cox-najjari-plesniak:2019,cox-plesniak:2021}), having been scaled to account for tube diameter while maintaining Womersley number and maximum Reynolds numbers found in the carotid artery.

\subsubsection{Dimensionless quantities}

For the given tube diameter, the Womersley number is $\alpha=4.22$ and the peak-to-mean flow rate ratio is approximately 4. The Dean number corresponding to the minimum, mean and maximum flow rate is $\kappa = Re \sqrt{\delta} \in\{10, 143, 567\}$, where Reynolds number is defined in this context based on bulk velocity $\overbar{u}$ and tube diameter. The reduced velocity $u_{red} = \overbar{u}_{mean} T / d = 33.2$, where $\overbar{u}_{mean}$ is the mean velocity over the pulse period $T$, indicates that the distance traveled by the mean flow in one period is nearly three times the length of the entire curve. This signifies that flow structures generated within each pulse period do not interfere with each other and facilitates our investigation of secondary flow patterns and wall shear stress distributions in isolation without flow disturbances from previous waveform cycles.

\subsection{Computational fluid dynamics}
\label{s:numerical_scheme}

We numerically solve the unsteady three-dimensional incompressible Navier-Stokes equations for a Newtonian fluid using an in-house discontinuous spectral element flow solver. Details on the current flow solver are provided in \cite{cox-aiaa:2016,cox-liang-plesniak:2016} and \cite{cox:2017}. The flow solver has been extensively validated using experimental particle image velocimetry (PIV) data (\cite{cox-najjari-plesniak:2019}) under the current waveform and curved geometry by comparing velocity fields, vorticity fields and vortex trajectories. A grid spacing and polynomial convergence study was performed to achieve sufficient spatial resolution to converge the $L^2$-norm of the error in the wall shear stress to within $0.78\%$ and the velocity magnitude to within $0.25\%$. We use the free 3\nobreakdash-D finite element mesh generator GMSH (\cite{geuzaine-remacle:2009}) to create all computational domains.

%%%%%%%%%%%%%%%%%%%%%%%%%%%%%%%%%%%%%%%%%%%%%%%%%%
%\include{F-t23-wec}
%\include{F-t23-uec}
%\input{F-t23-wec.tex}
%\input{F-t23-uec.tex}
%
\begin{figure*}
   \renewcommand{\fsize}{0.162\textwidth}
   \centering\setcounter{subfigure}{0}
   \renewcommand{\ftime}{230}
   \vspace{-0.2in}
   \subfloat[]{
       \includegraphics[height=\fsize,keepaspectratio]
       {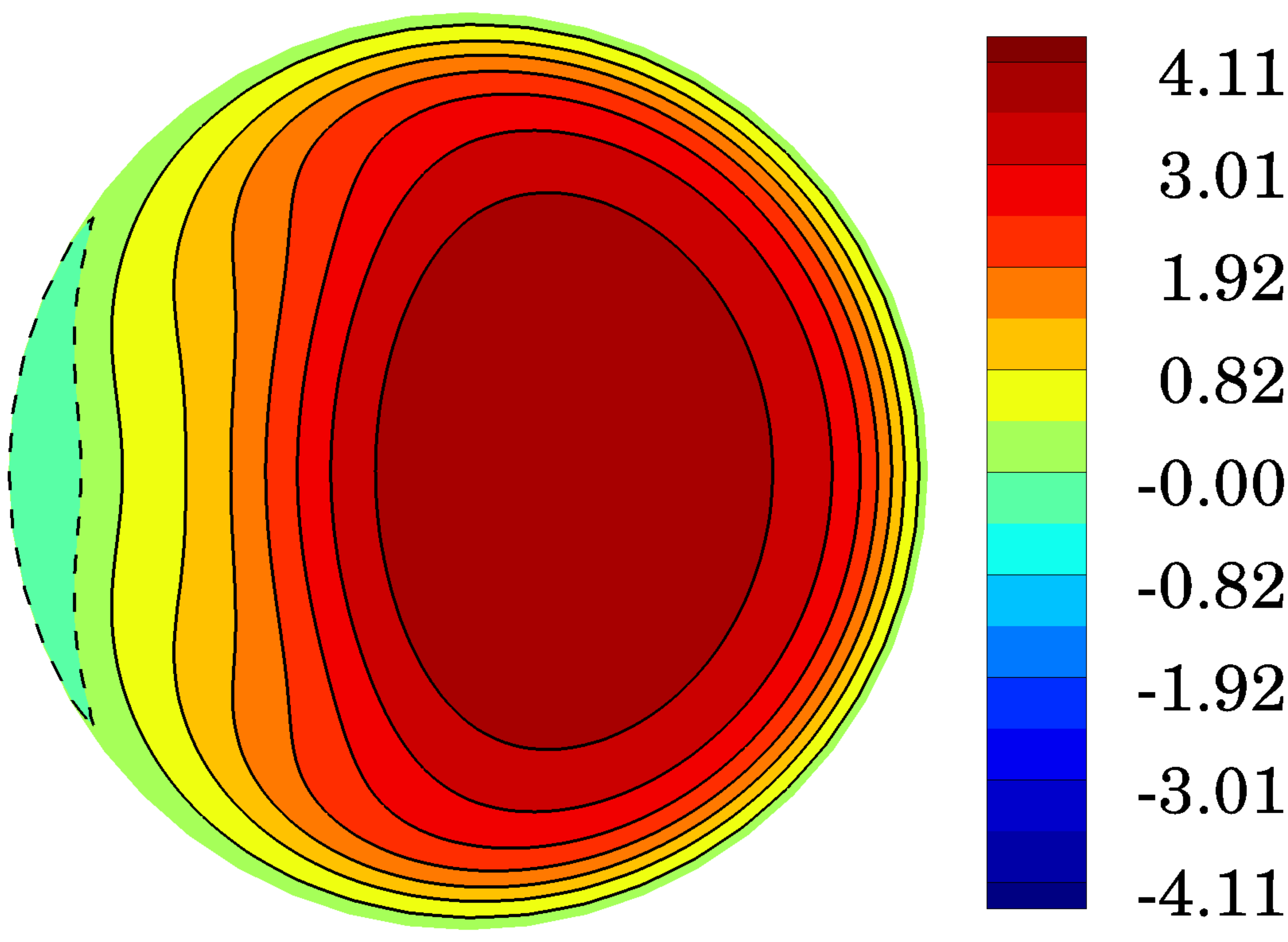}
       \label{f:us_wec_t23_22}}
   \hspace{0.025in}
   \subfloat[]{
       \includegraphics[height=\fsize,keepaspectratio]
       {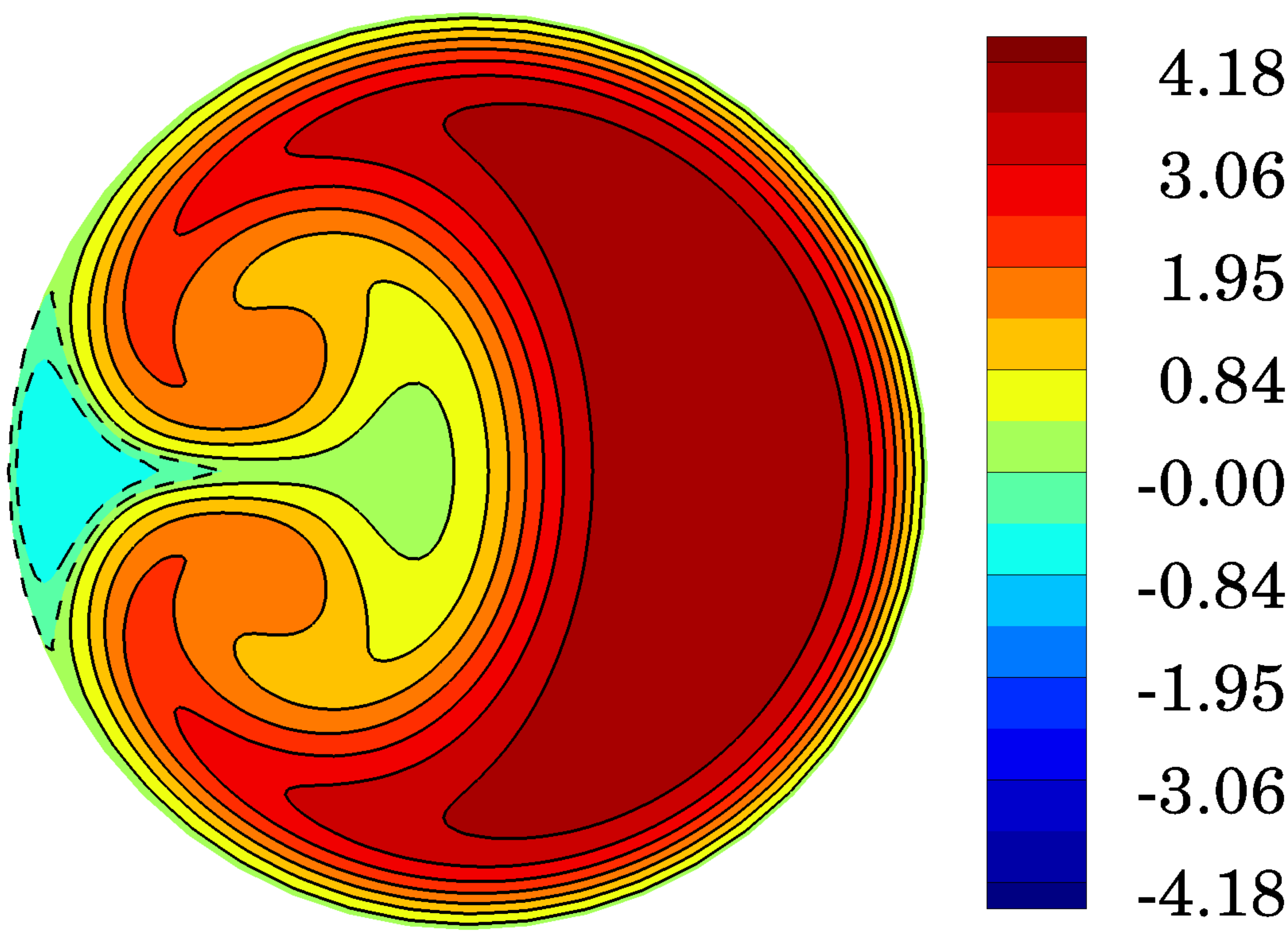}
       \label{f:us_wec_t23_45}}
   \hspace{0.025in}
   \subfloat[]{
       \includegraphics[height=\fsize,keepaspectratio]
       {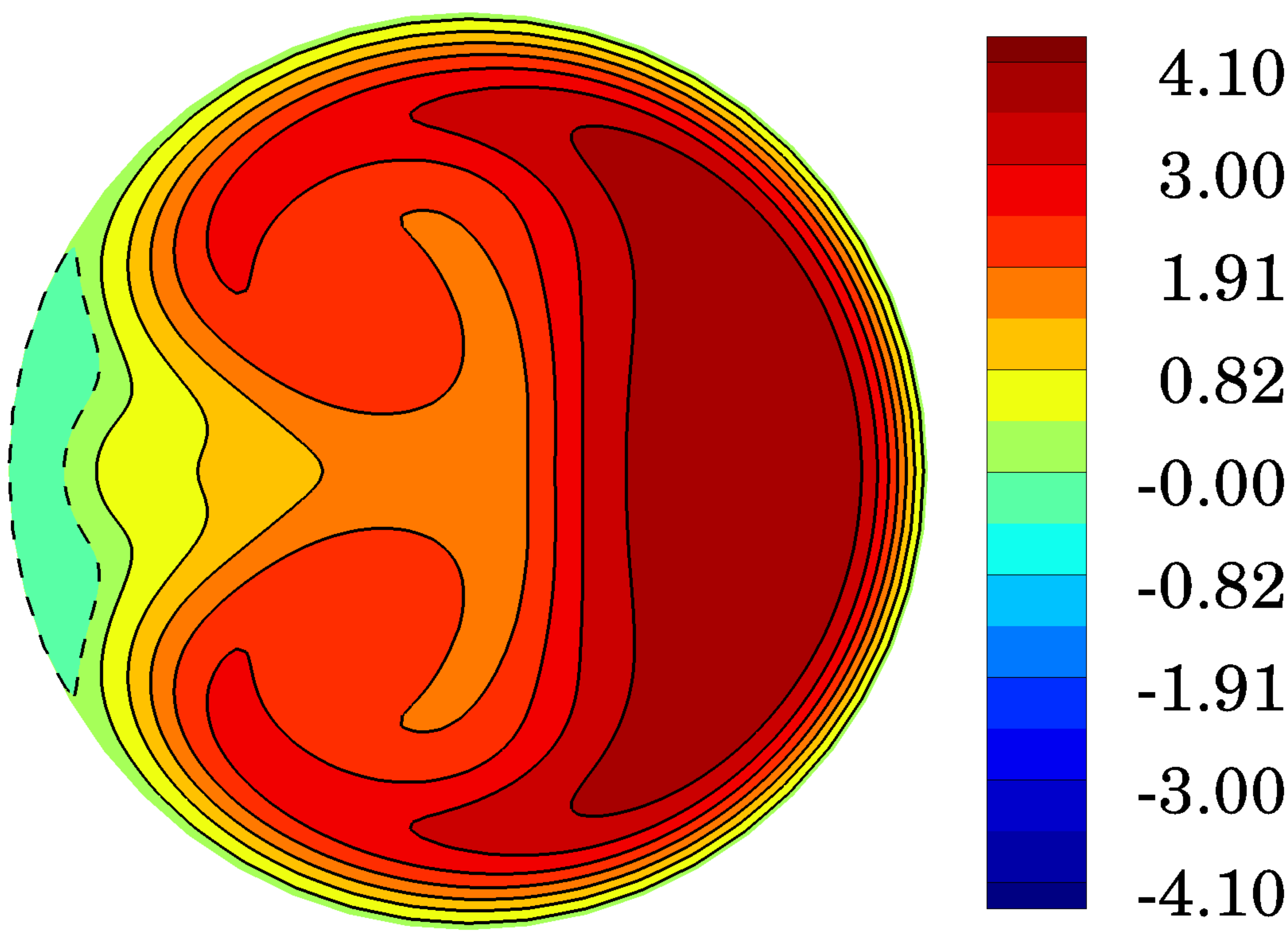}
       \label{f:us_wec_t23_90}}
   \hspace{0.025in}
   \subfloat[]{
       \includegraphics[height=\fsize,keepaspectratio]
       {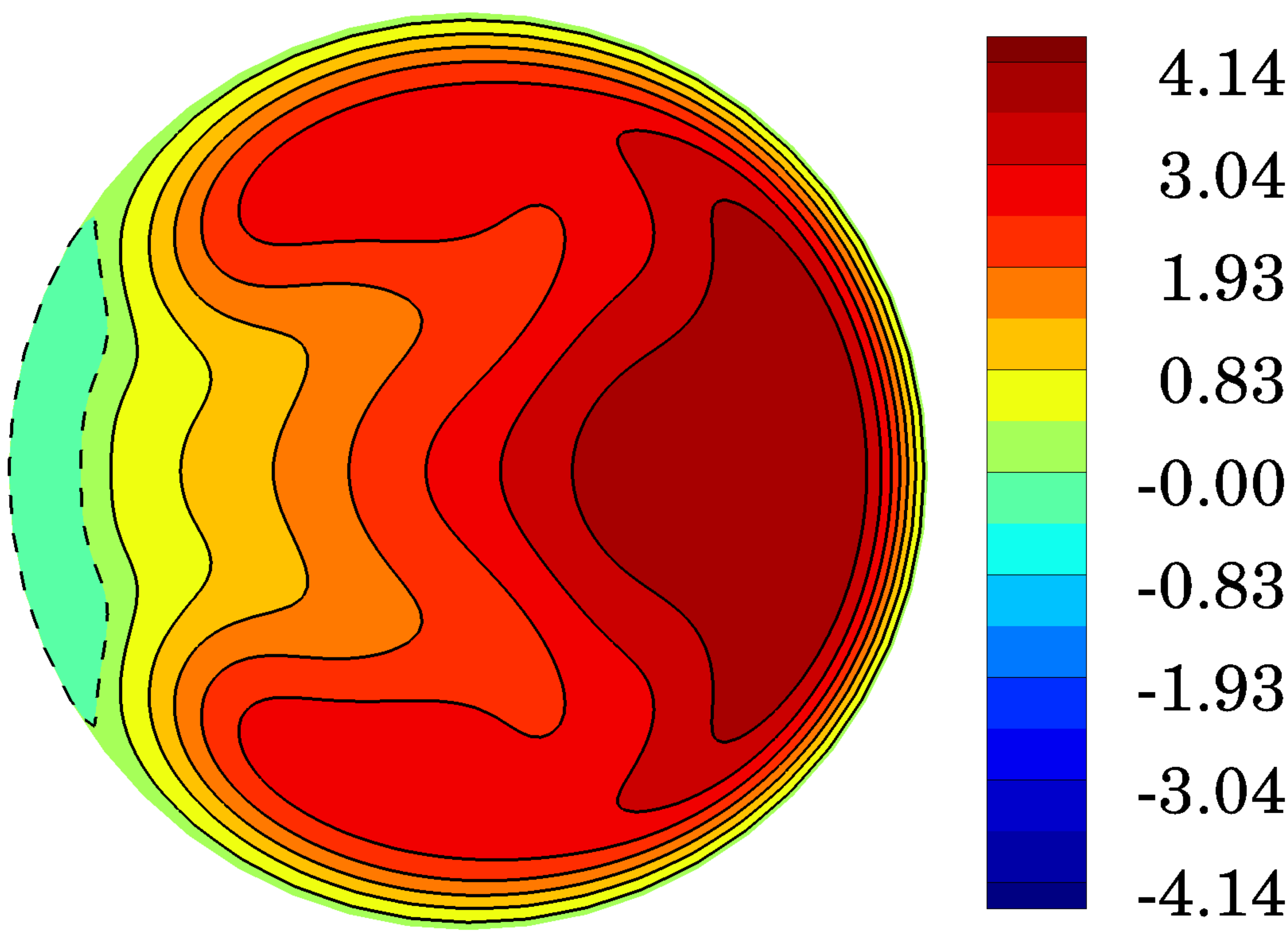}
       \label{f:us_wec_t23_135}}
   \\[-0.1in]
   \subfloat[]{
       \includegraphics[height=\fsize,keepaspectratio]
       {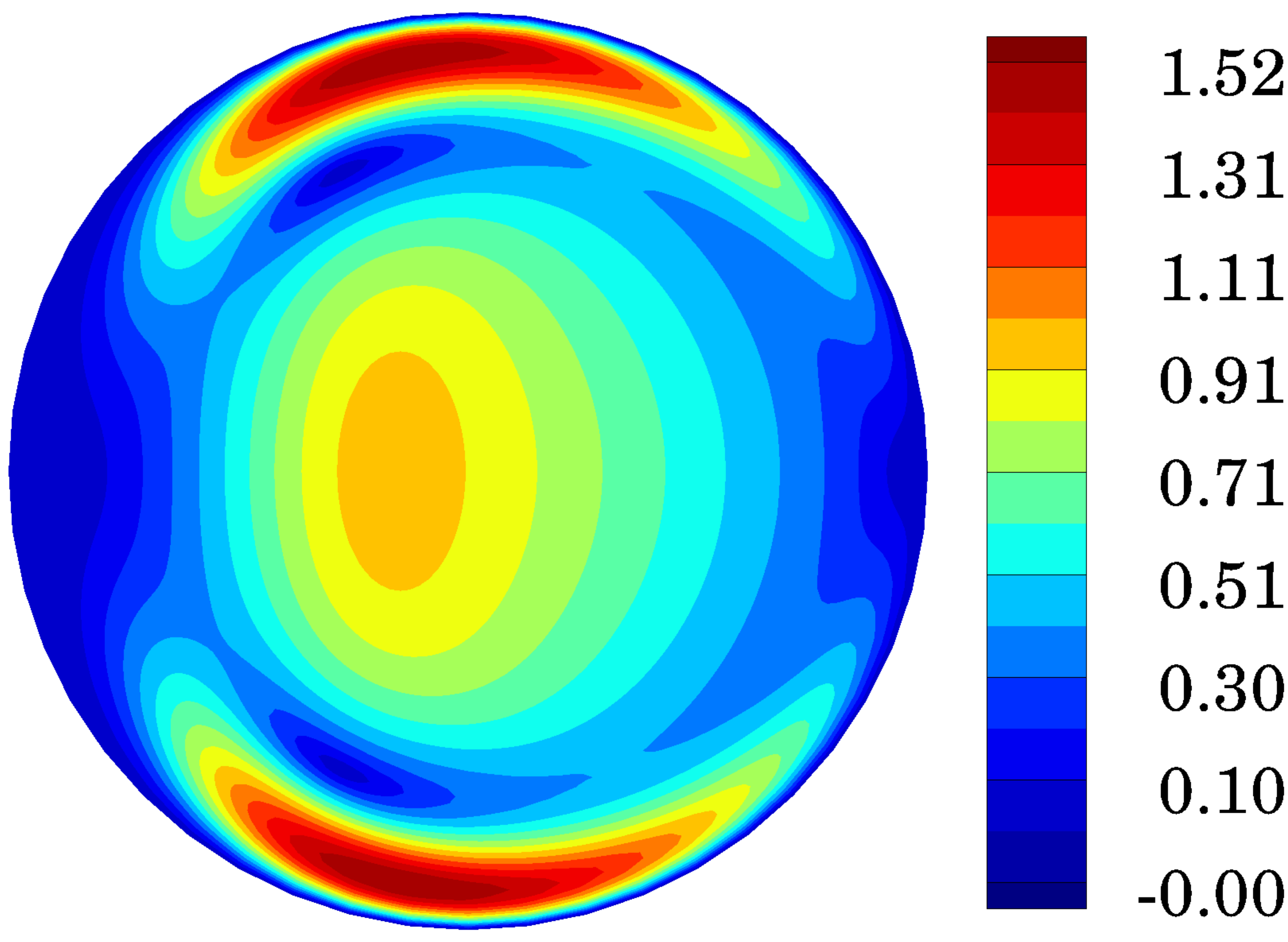}
       \label{f:uplanar_wec_t23_22}}
   \hspace{0.025in}
   \subfloat[]{
       \includegraphics[height=\fsize,keepaspectratio]
       {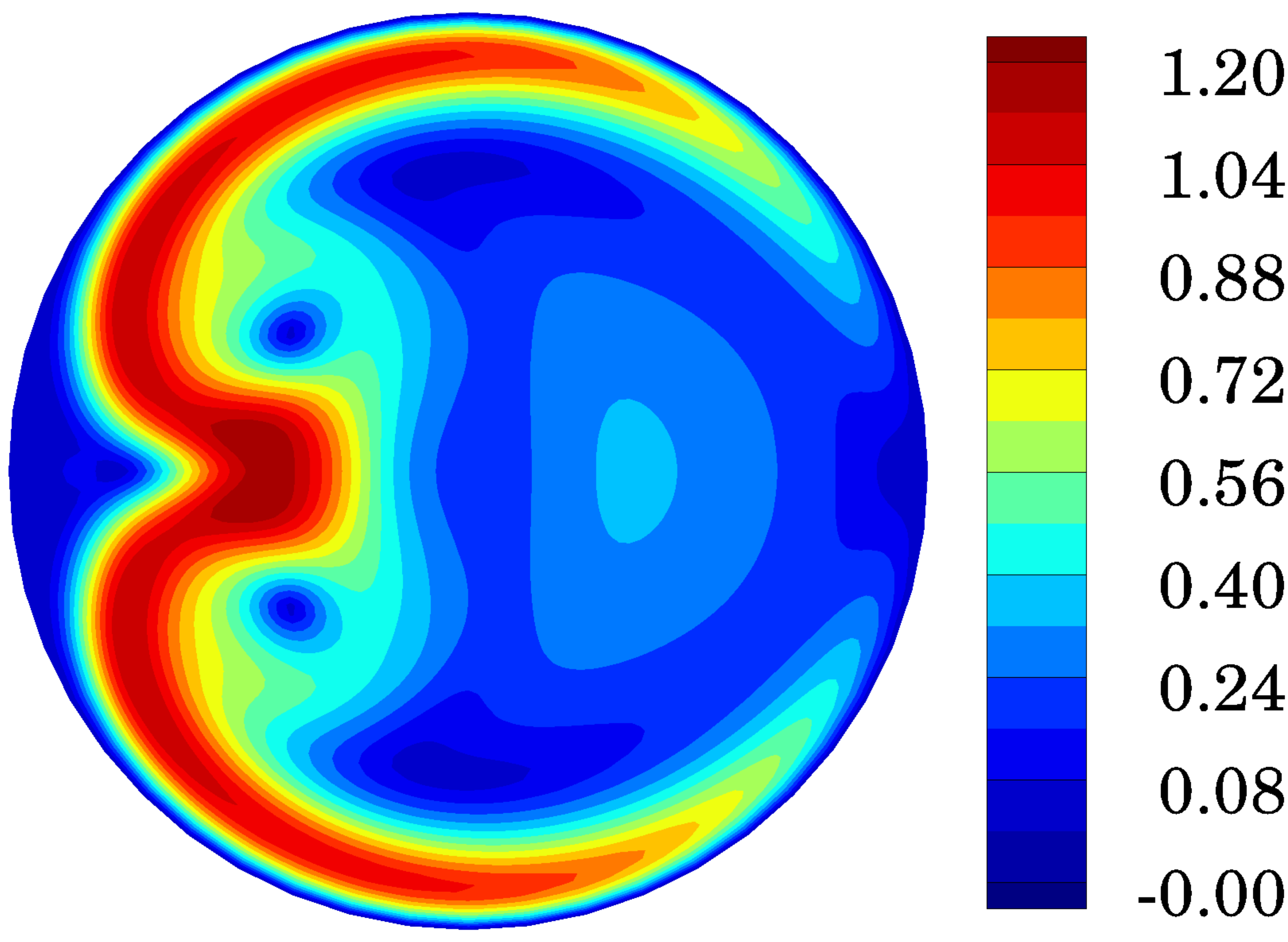}
       \label{f:uplanar_wec_t23_45}}
   \hspace{0.025in}
   \subfloat[]{
       \includegraphics[height=\fsize,keepaspectratio]
       {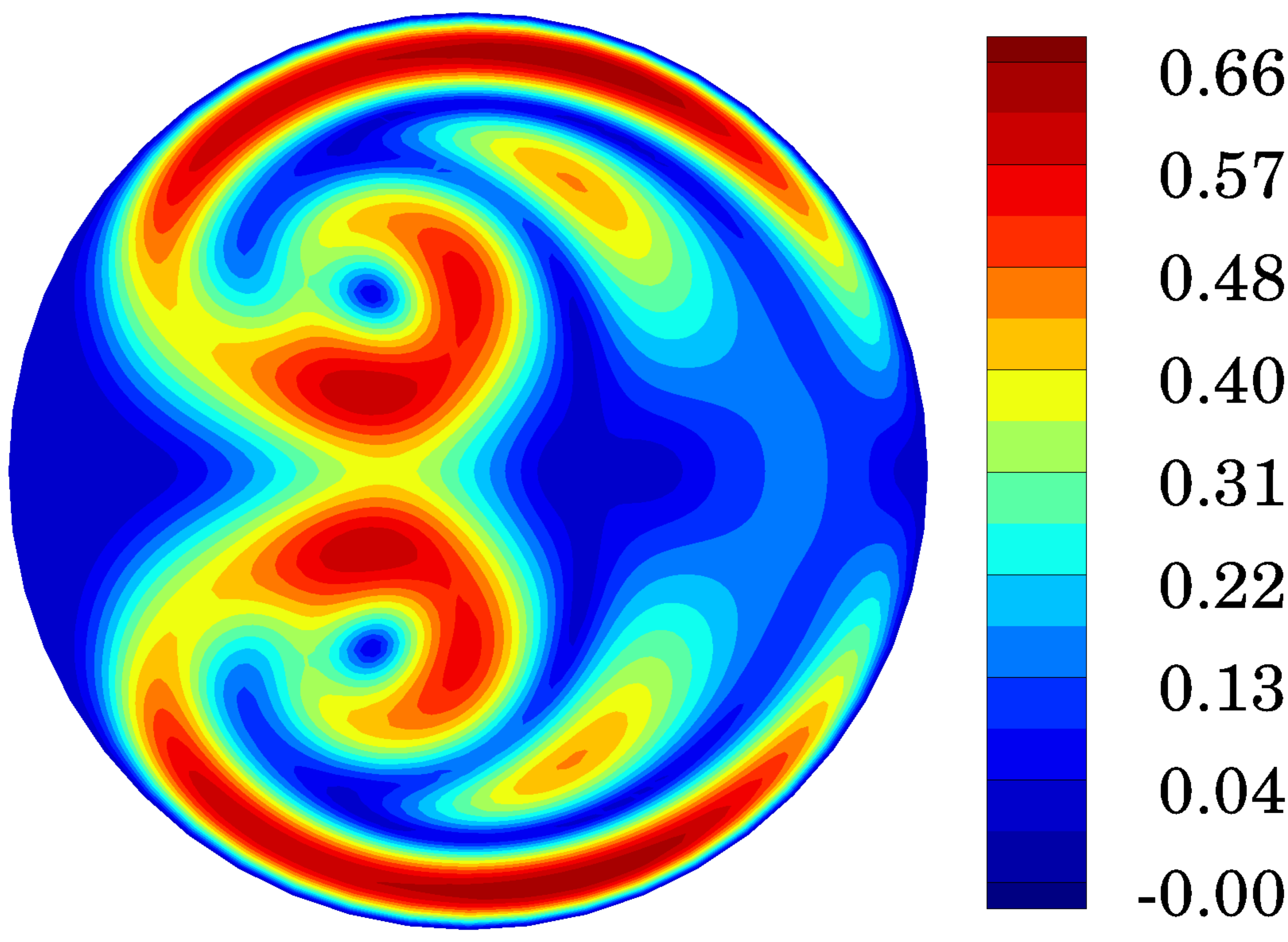}
       \label{f:uplanar_wec_t23_90}}
   \hspace{0.025in}
   \subfloat[]{
       \includegraphics[height=\fsize,keepaspectratio]
       {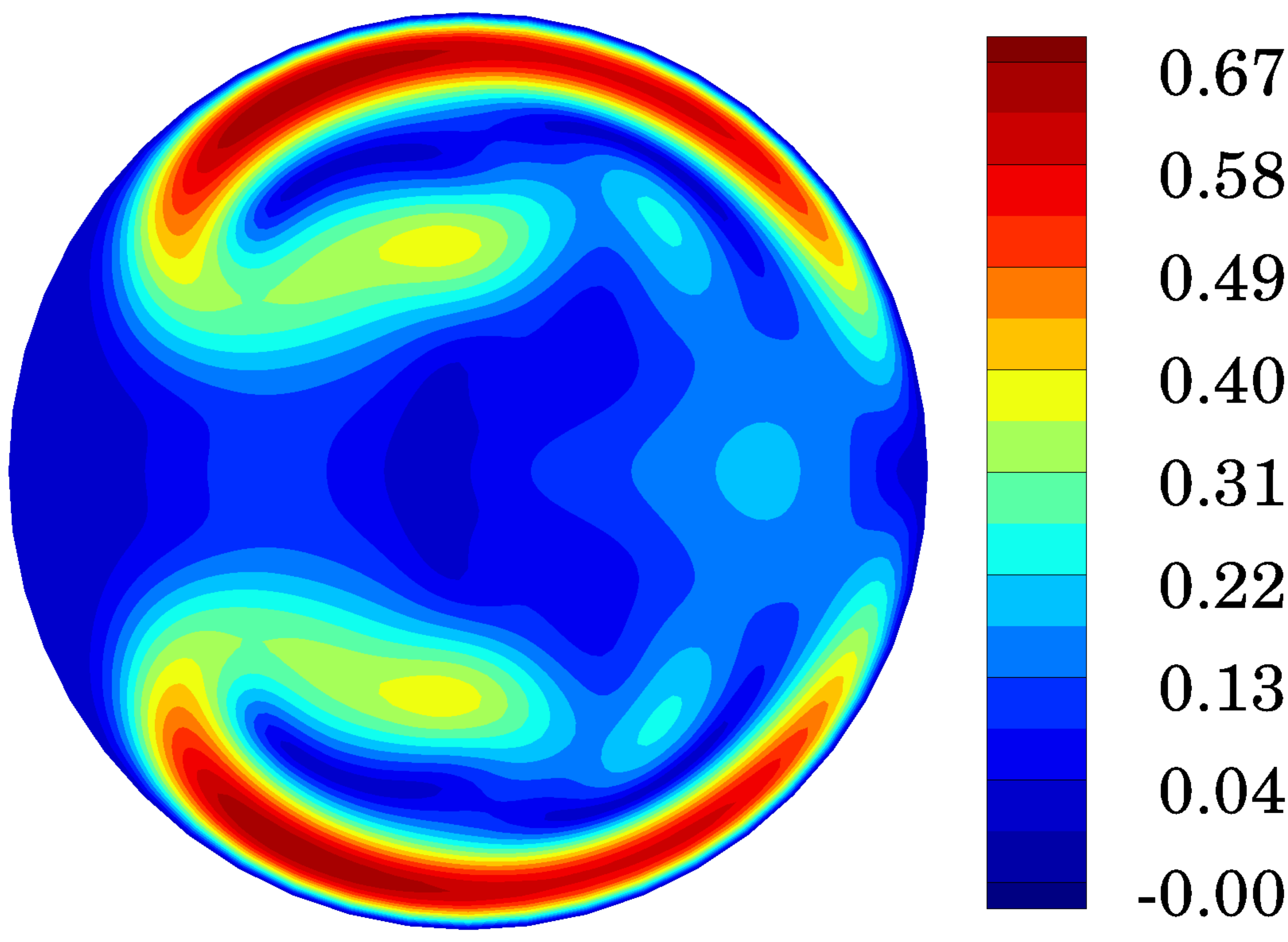}
       \label{f:uplanar_wec_t23_135}}
   \\[-0.1in]
   \renewcommand{\fsize}{0.19\textwidth}
   \hspace{-0.5in}
   \subfloat[]{
       \includegraphics[height=\fsize,keepaspectratio]
       {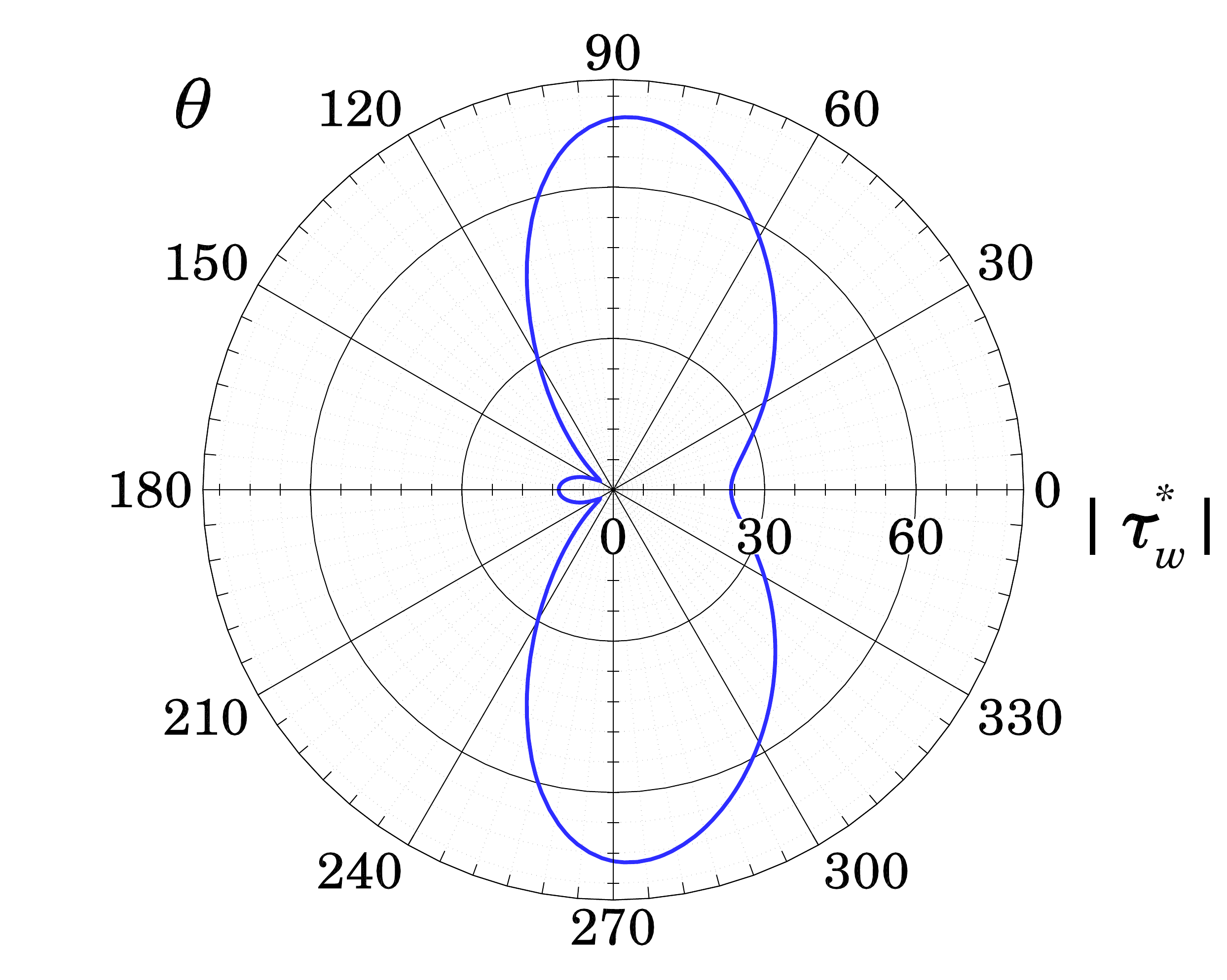}
       \label{f:wss_wec_t23_22}}
   \hspace{-0.1in}
   \subfloat[]{
       \includegraphics[height=\fsize,keepaspectratio]
       {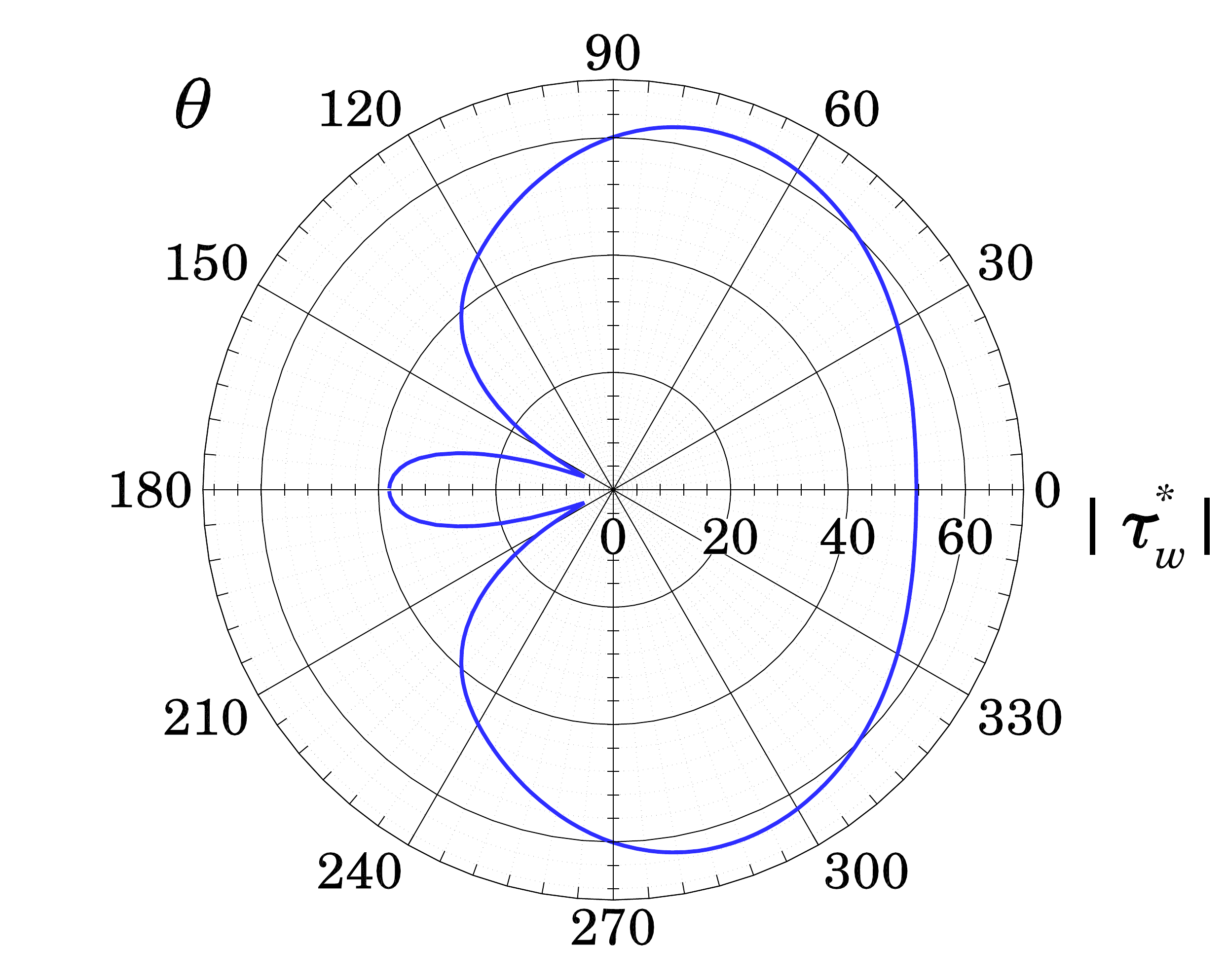}
       \label{f:wss_wec_t23_45}}
   \hspace{-0.1in}
   \subfloat[]{
       \includegraphics[height=\fsize,keepaspectratio]
       {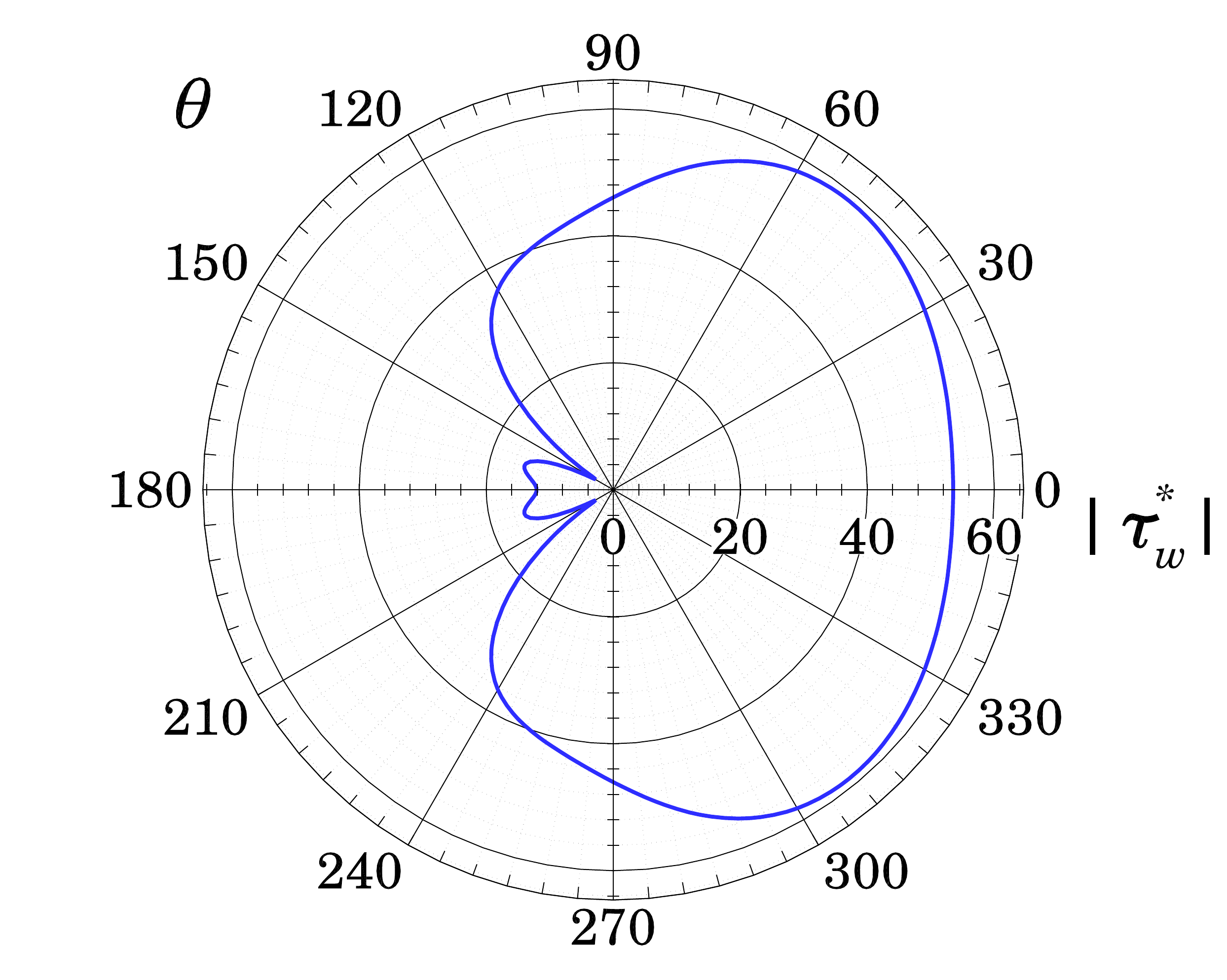}
       \label{f:wss_wec_t23_90}}
   \hspace{-0.1in}
   \subfloat[]{
       \includegraphics[height=\fsize,keepaspectratio]
       {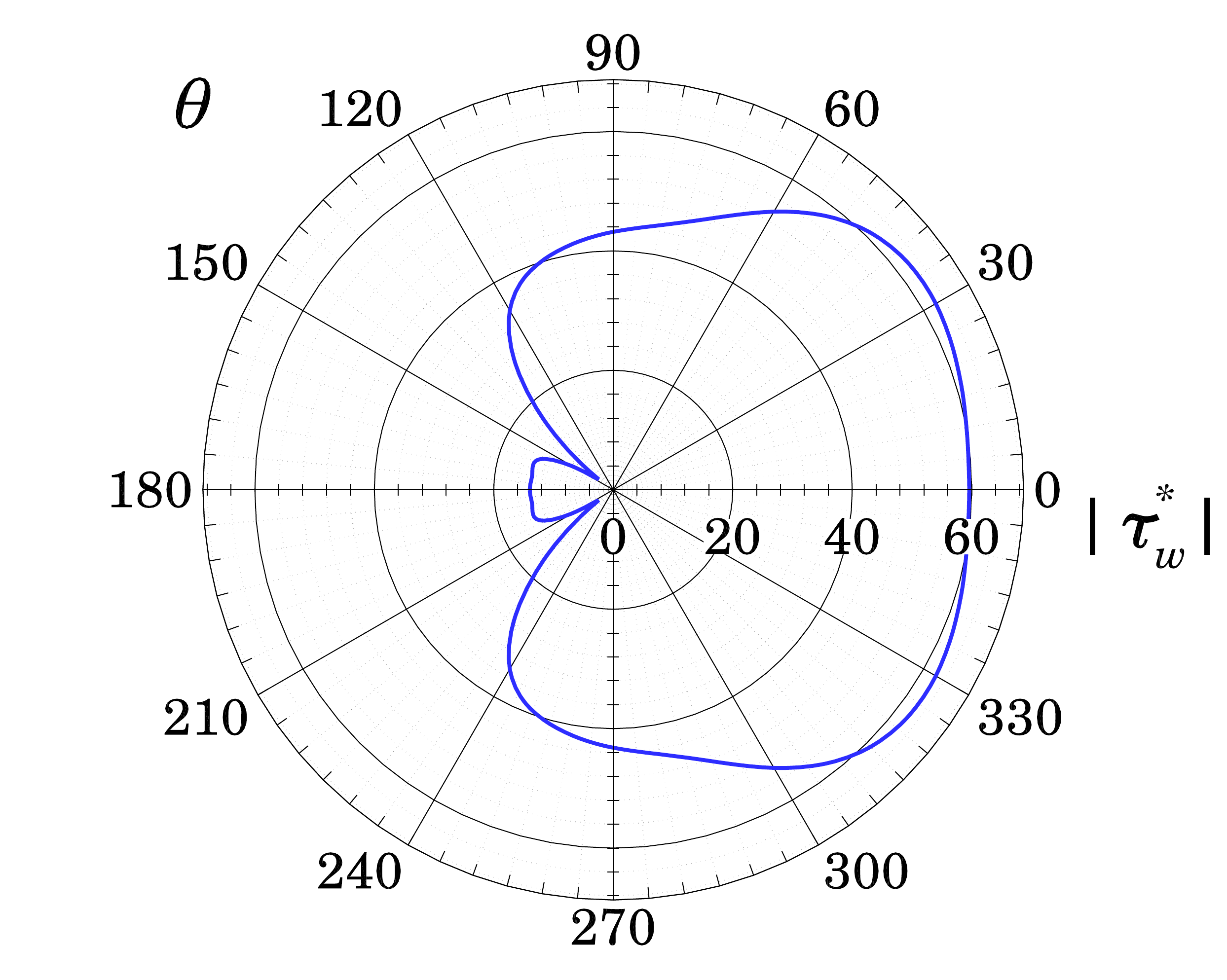}
       \label{f:wss_wec_t23_135}}
   
   \caption{WEC: cross-sectional results at $t^\star=0.23$ of (a)-(d) streamwise velocity $u^\star_s$, (e)-(h) secondary velocity magnitude $|\bmm{u}^\star_{\theta r}|$, (i)-(l) wall shear stress magnitude $|\bmm{\tau}^\star_w|$. Columns from left to right represent toroidal locations $\phi=\{22^\circ,45^\circ,90^\circ,135^\circ\}$.}
   \label{f:var_wec_t23}
\end{figure*}

\begin{figure*}
   \renewcommand{\fsize}{0.162\textwidth}
   \centering\setcounter{subfigure}{0}
   \renewcommand{\ftime}{230}
   \vspace{-0.2in}
   \subfloat[]{
       \includegraphics[height=\fsize,keepaspectratio]
       {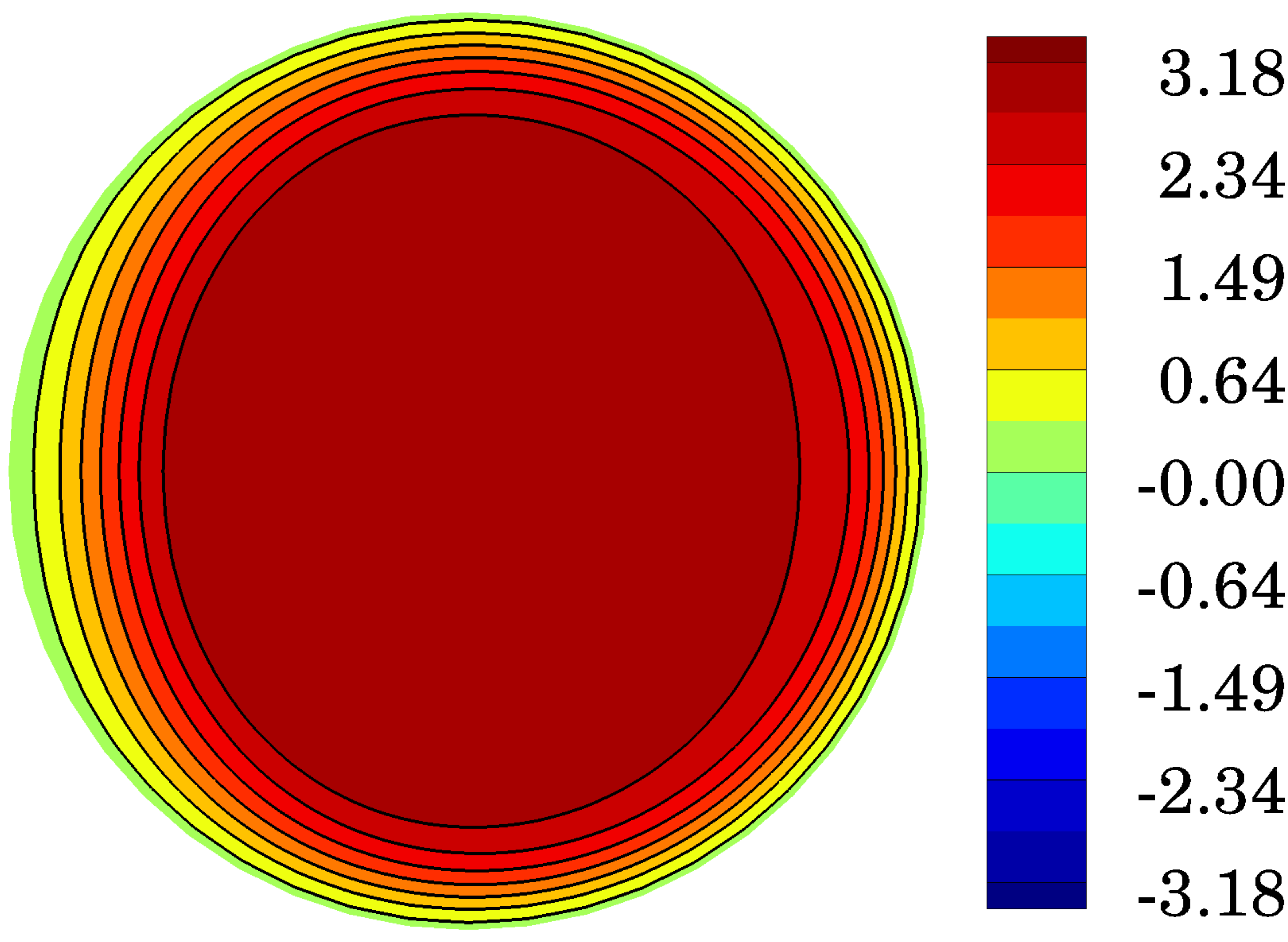}
       \label{f:us_uec_t23_22}}
   \hspace{0.025in}
   \subfloat[]{
       \includegraphics[height=\fsize,keepaspectratio]
       {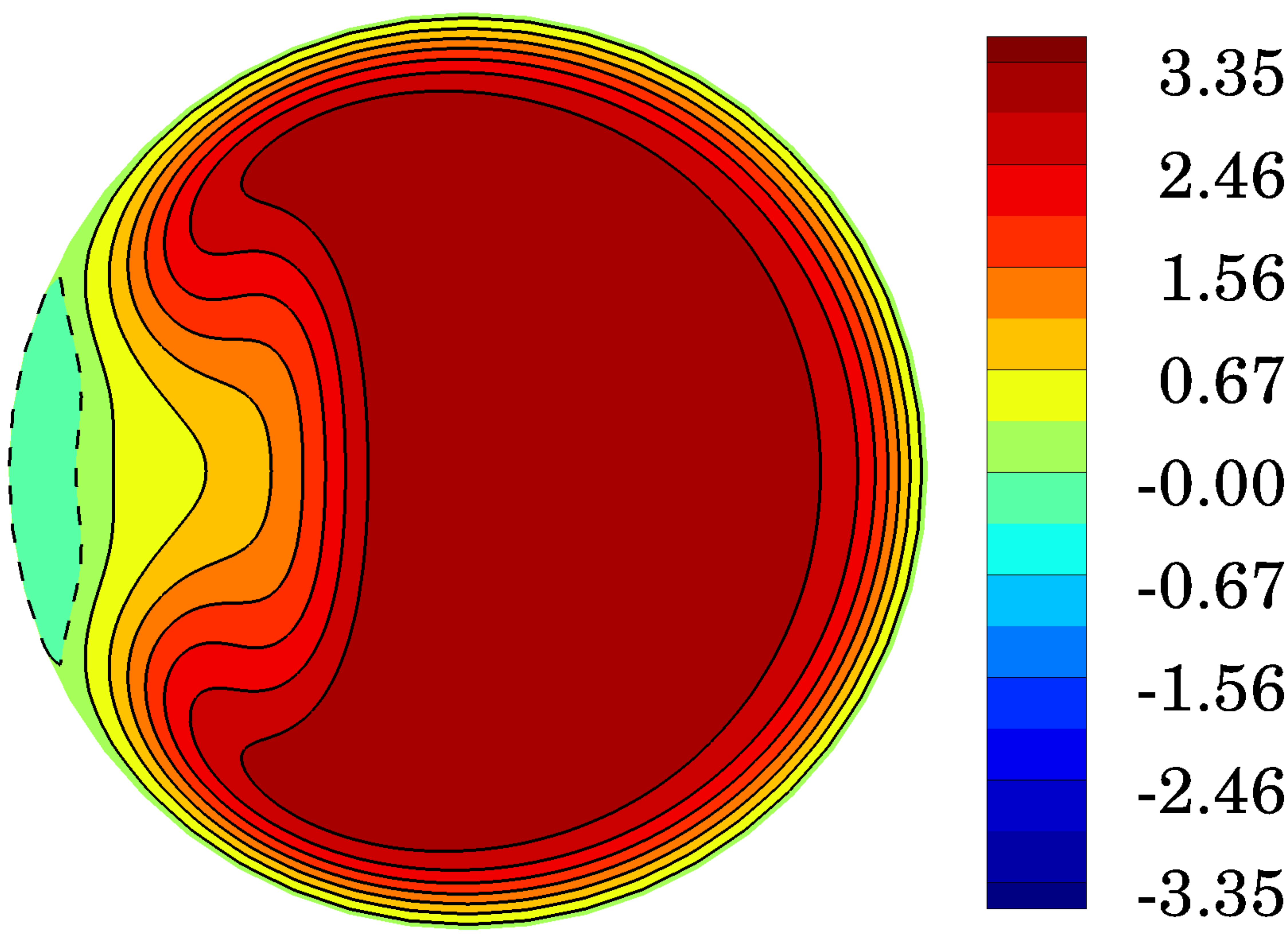}
       \label{f:us_uec_t23_45}}
   \hspace{0.025in}
   \subfloat[]{
       \includegraphics[height=\fsize,keepaspectratio]
       {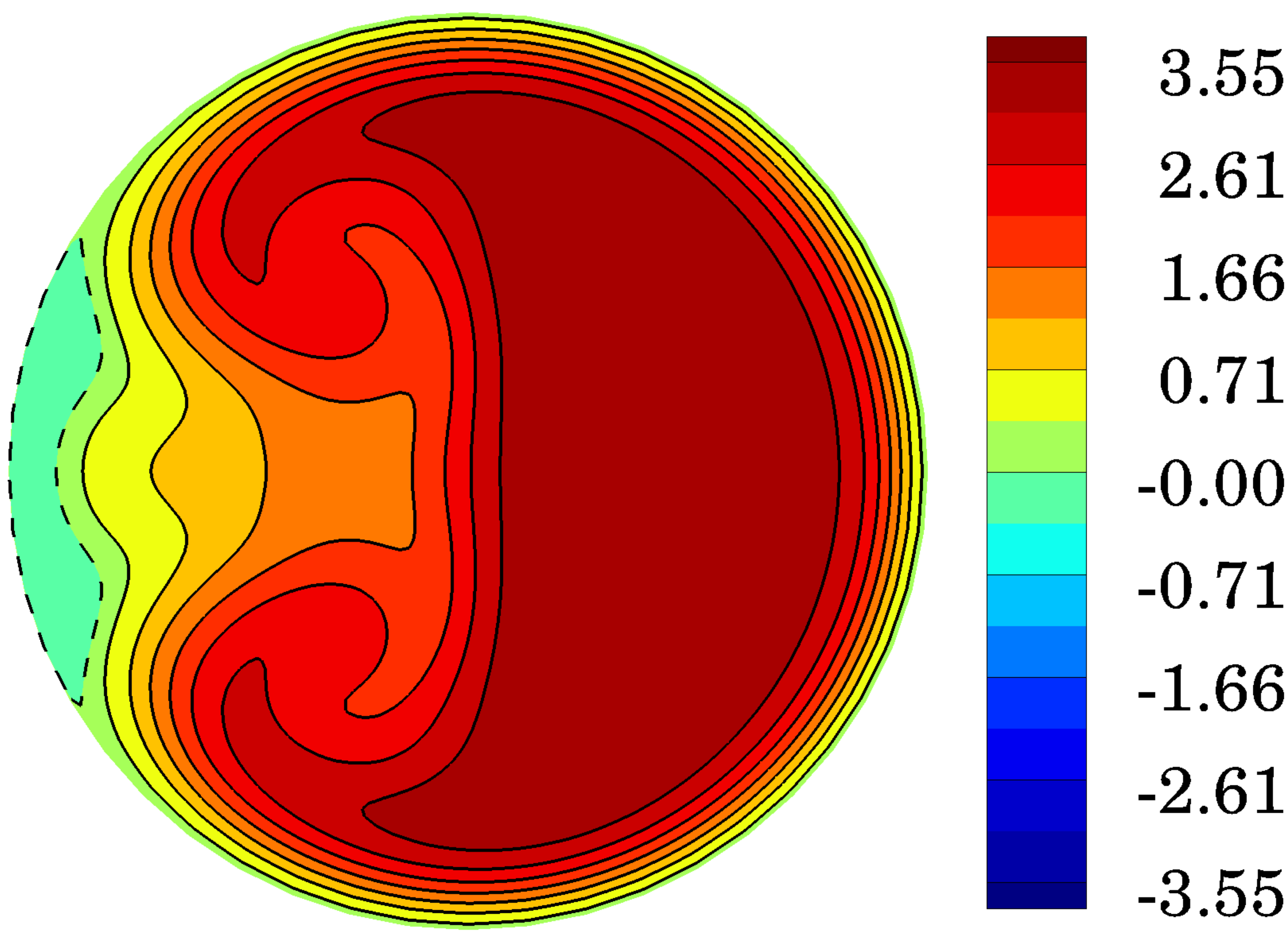}
       \label{f:us_uec_t23_90}}
   \hspace{0.025in}
   \subfloat[]{
       \includegraphics[height=\fsize,keepaspectratio]
       {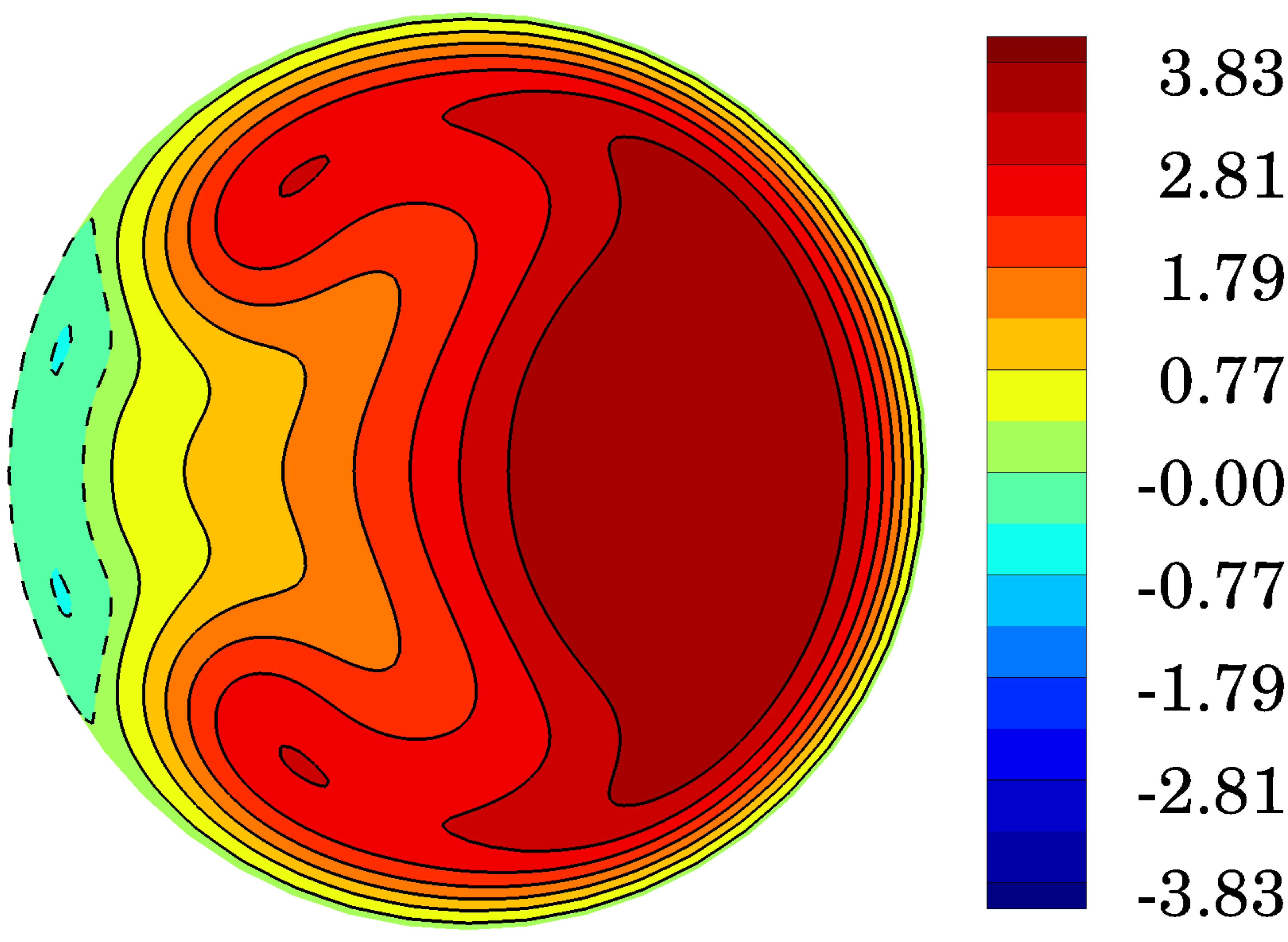}
       \label{f:us_uec_t23_135}}
   \\[-0.1in]
   \subfloat[]{
       \includegraphics[height=\fsize,keepaspectratio]
       {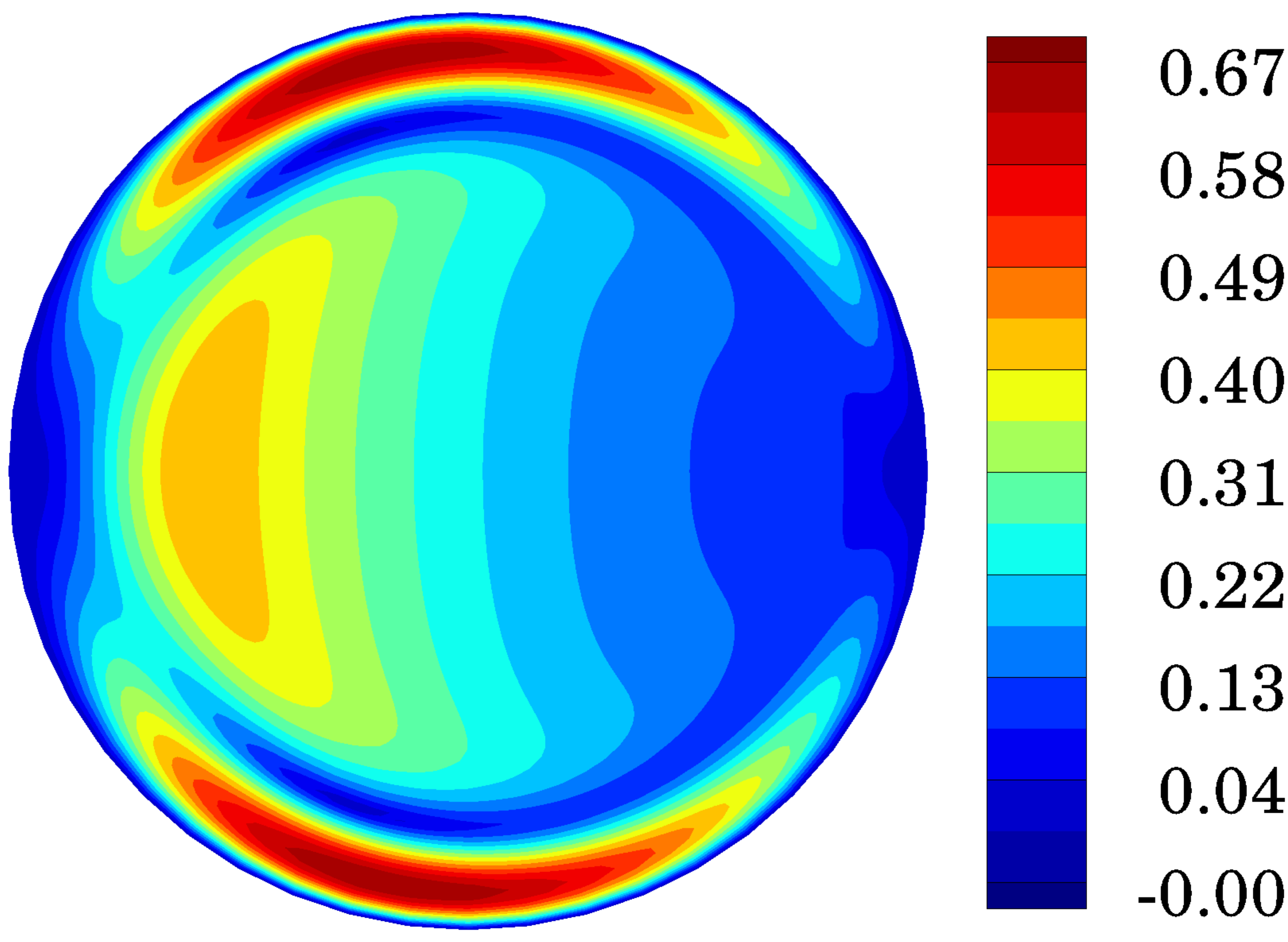}
       \label{f:uplanar_uec_t23_22}}
   \hspace{0.025in}
   \subfloat[]{
       \includegraphics[height=\fsize,keepaspectratio]
       {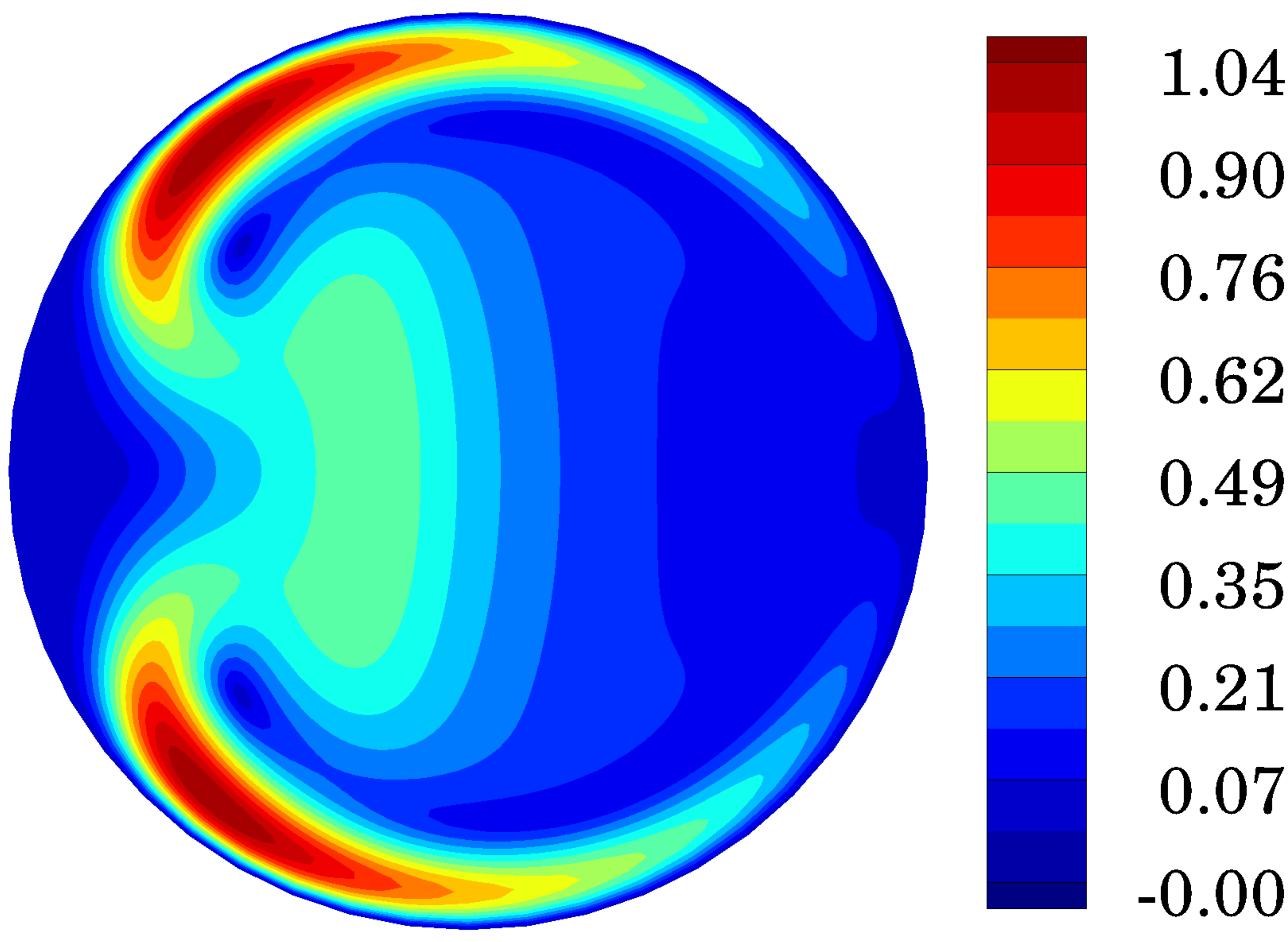}
       \label{f:uplanar_uec_t23_45}}
   \hspace{0.025in}
   \subfloat[]{
       \includegraphics[height=\fsize,keepaspectratio]
       {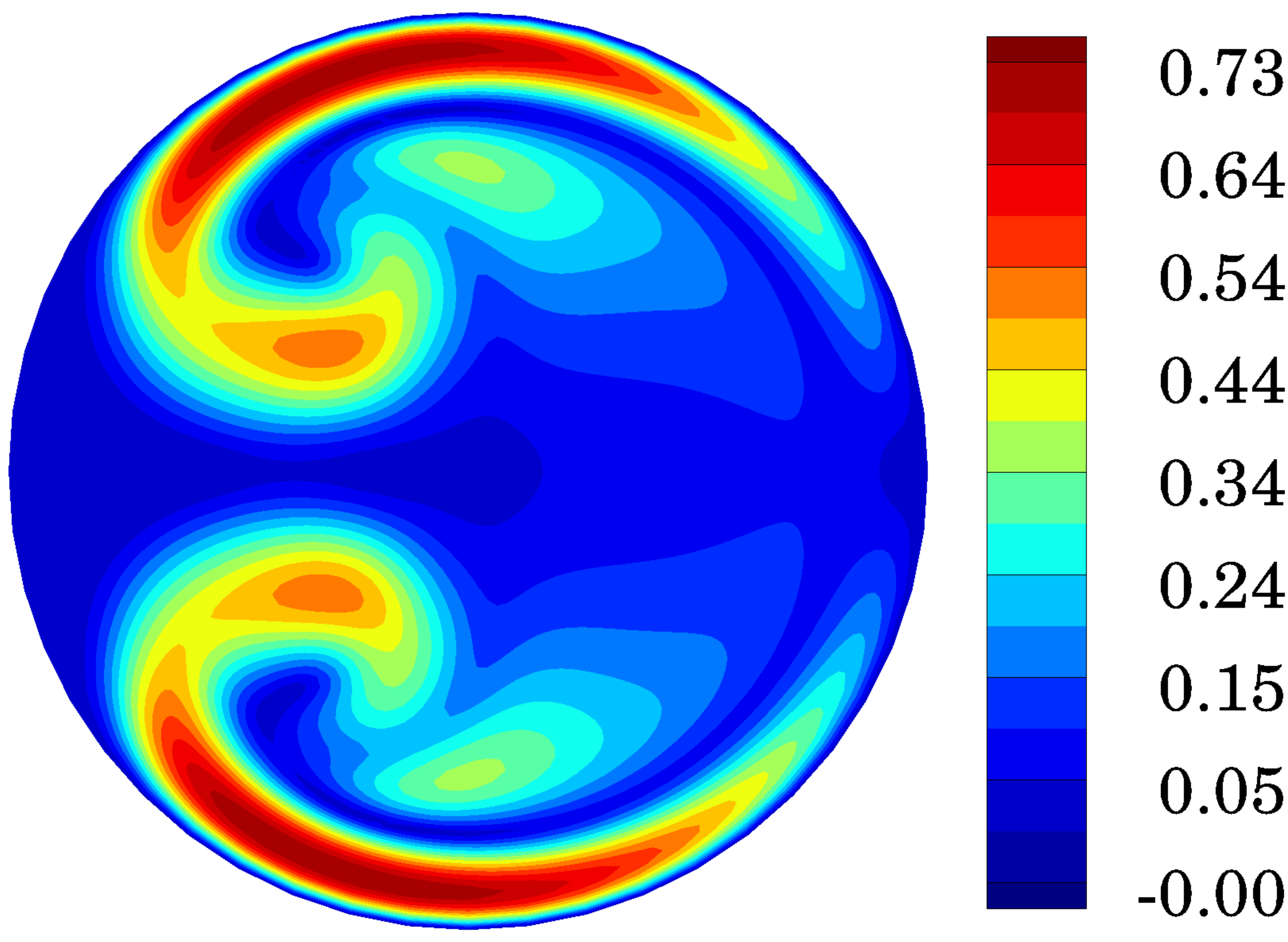}
       \label{f:uplanar_uec_t23_90}}
   \hspace{0.025in}
   \subfloat[]{
       \includegraphics[height=\fsize,keepaspectratio]
       {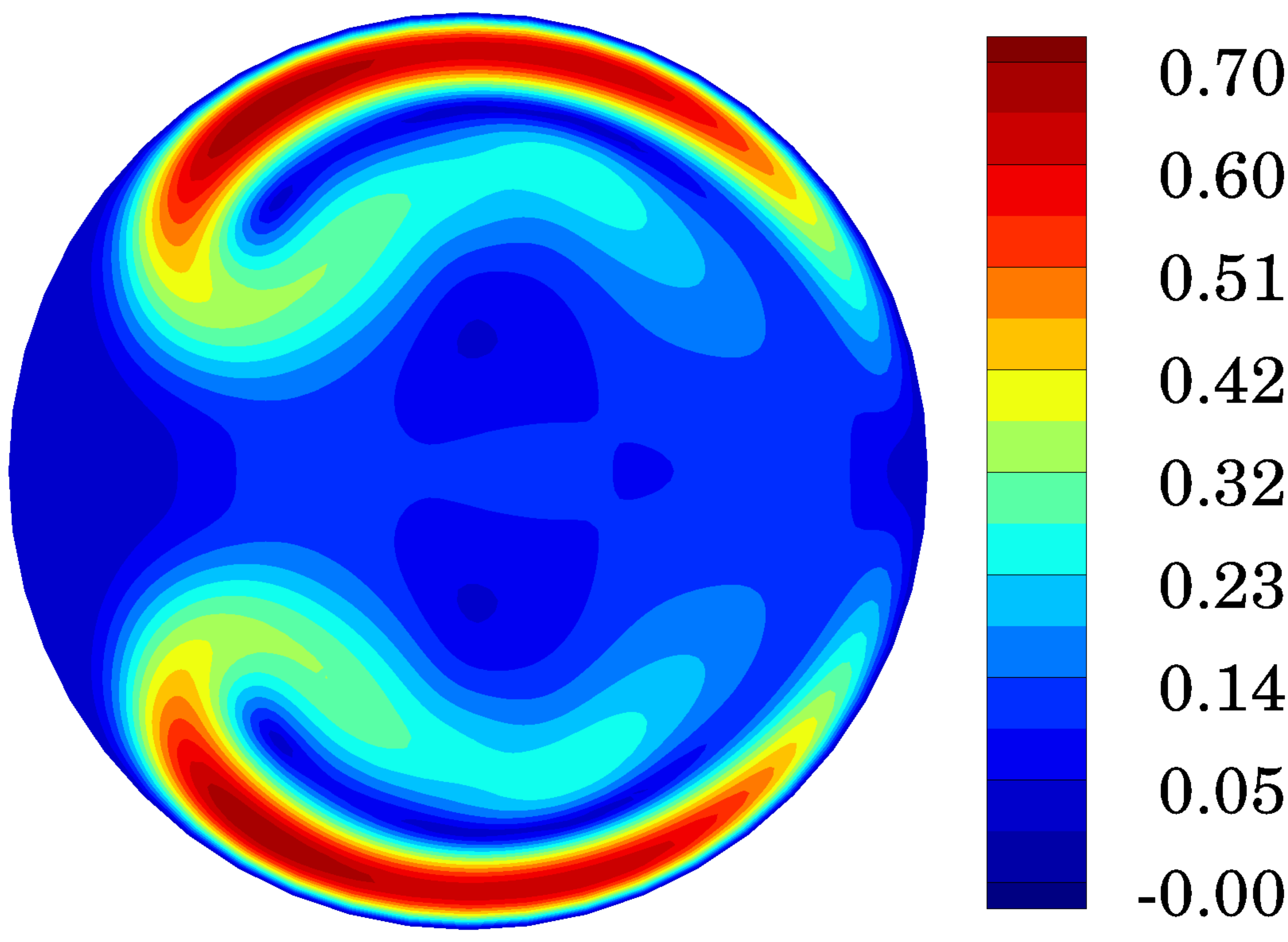}
       \label{f:uplanar_uec_t23_135}}
   \\[-0.1in]
   \renewcommand{\fsize}{0.19\textwidth}
   \hspace{-0.5in}
   \subfloat[]{
       \includegraphics[height=\fsize,keepaspectratio]
       {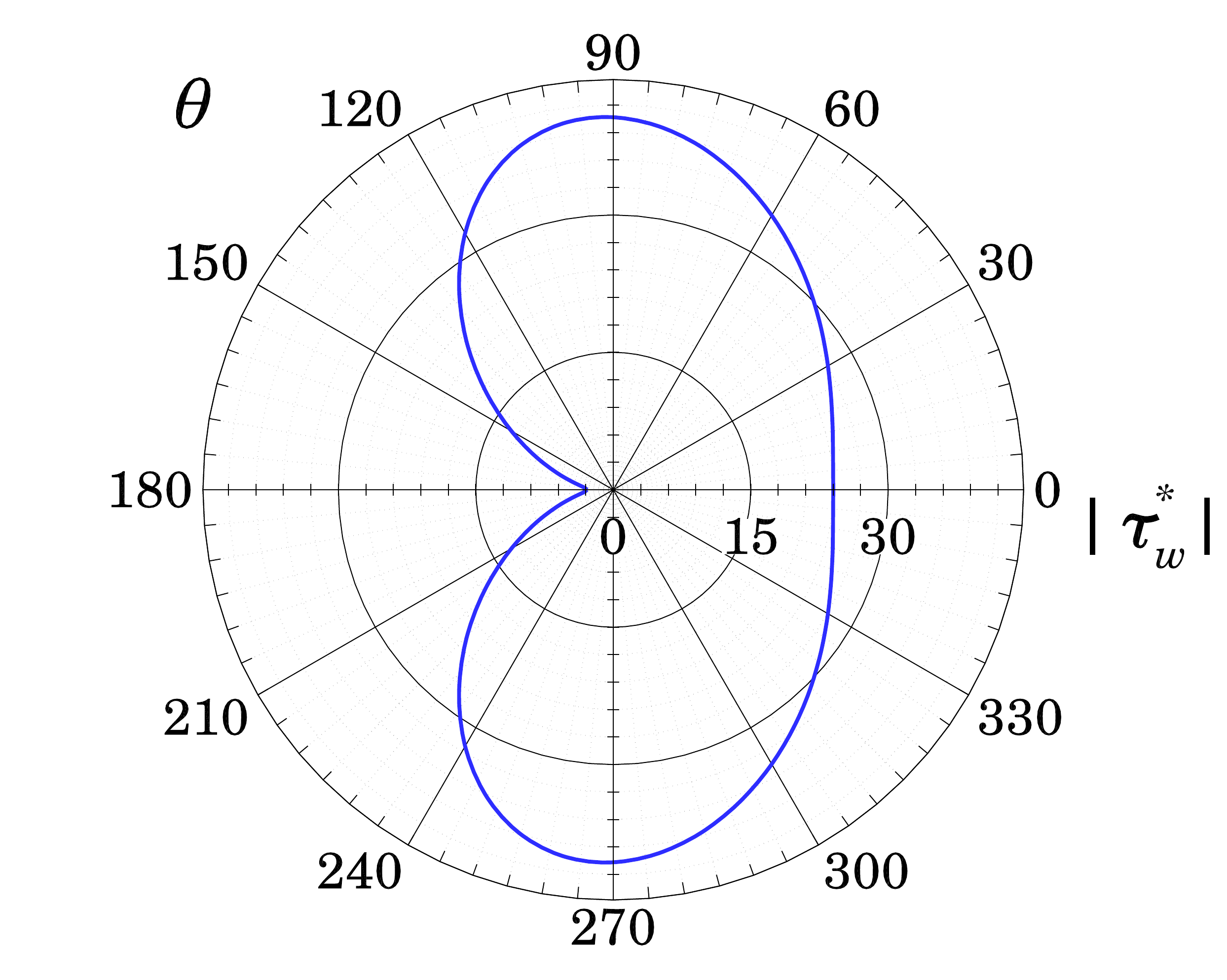}
       \label{f:wss_uec_t23_22}}
   \hspace{-0.1in}
   \subfloat[]{
       \includegraphics[height=\fsize,keepaspectratio]
       {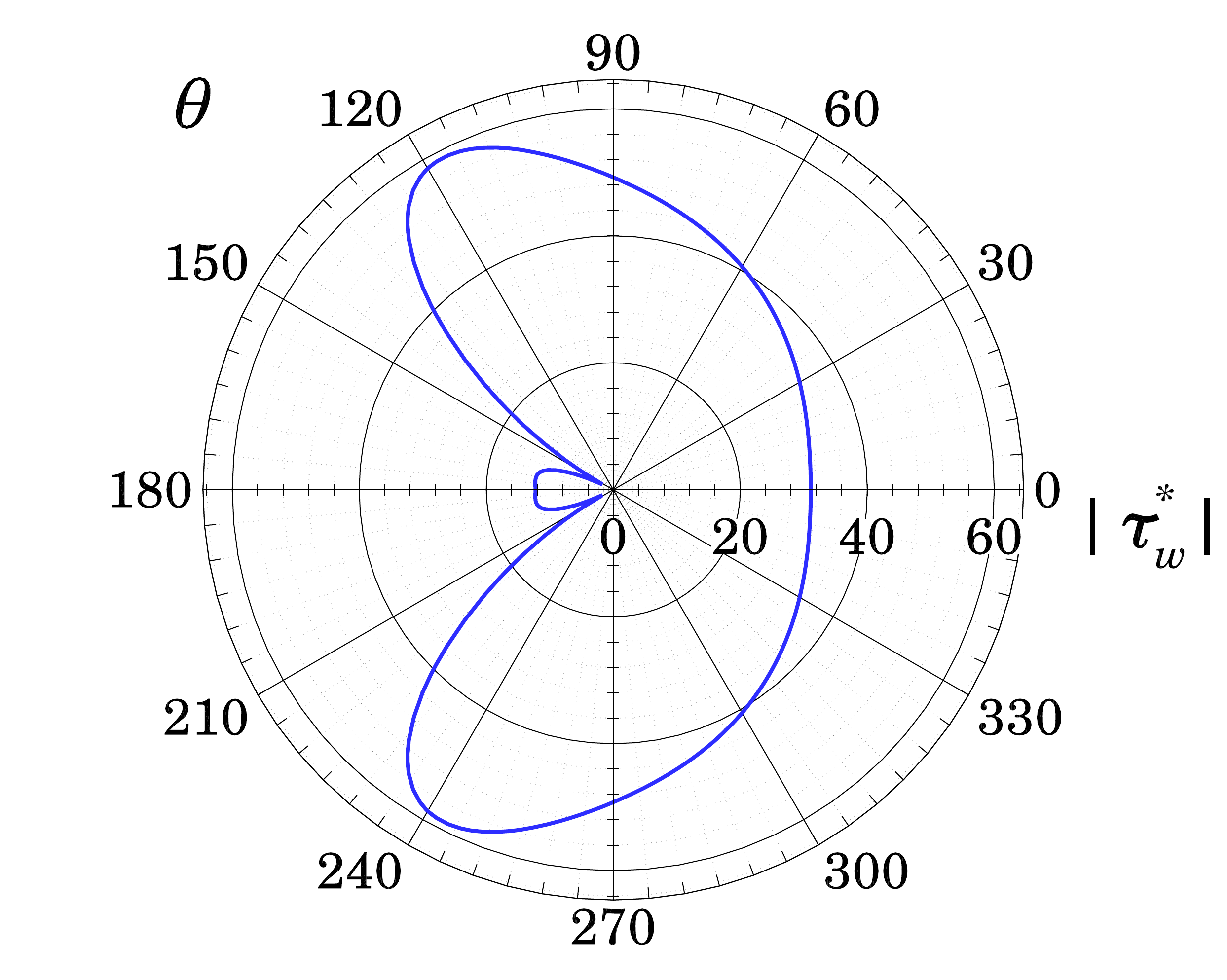}
       \label{f:wss_uec_t23_45}}
   \hspace{-0.1in}
   \subfloat[]{
       \includegraphics[height=\fsize,keepaspectratio]
       {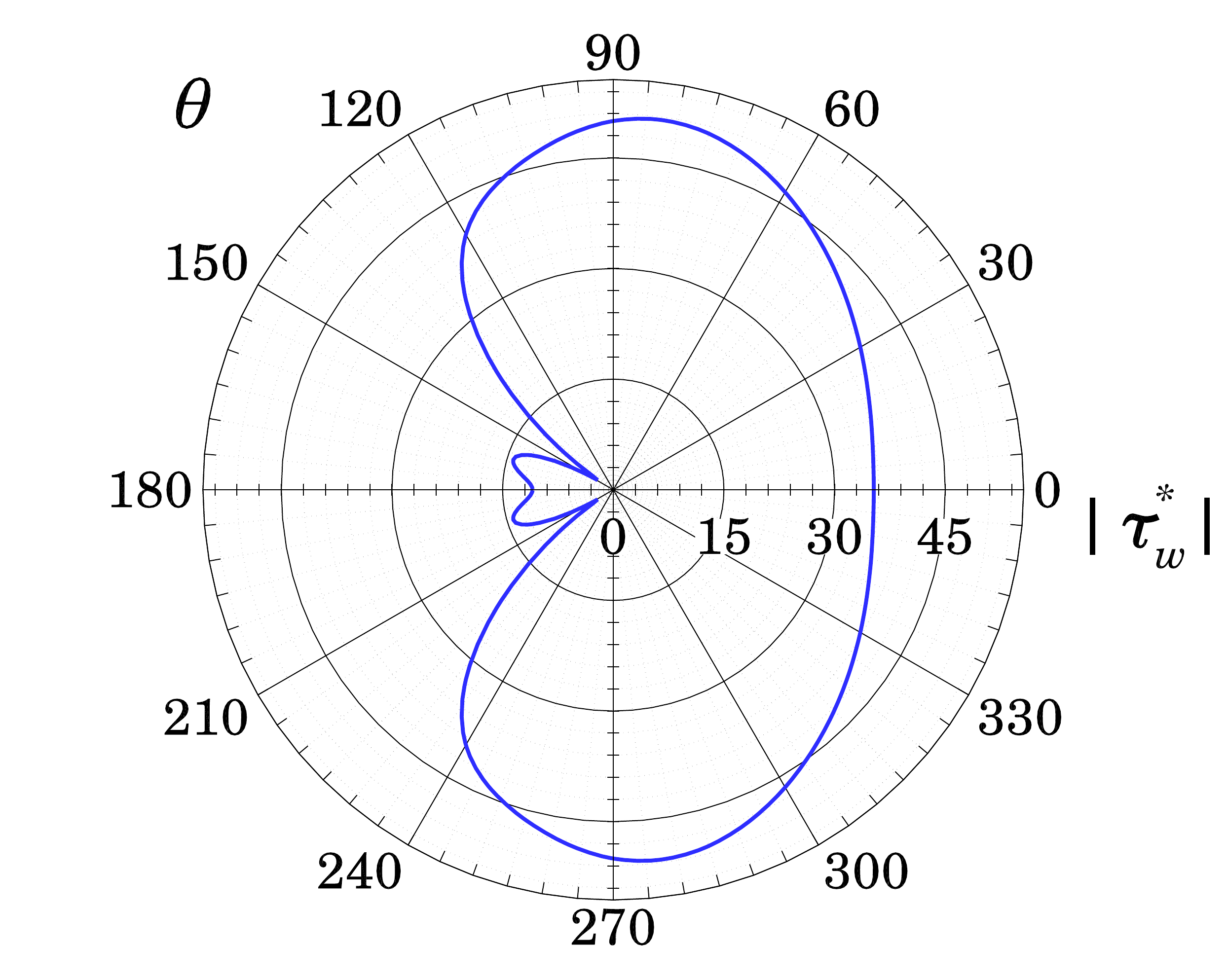}
       \label{f:wss_uec_t23_90}}
   \hspace{-0.1in}
   \subfloat[]{
       \includegraphics[height=\fsize,keepaspectratio]
       {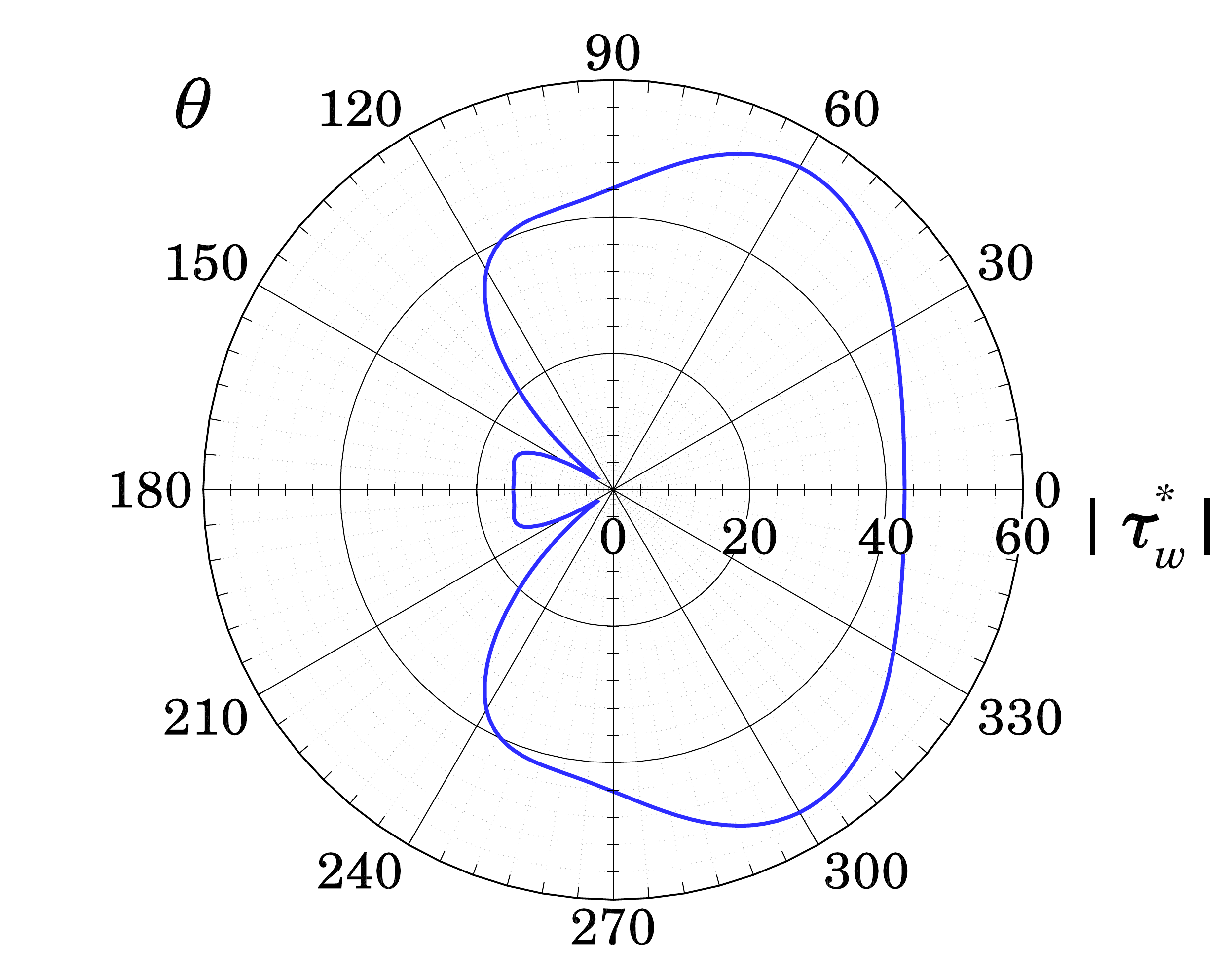}
       \label{f:wss_uec_t23_135}}
   
   \caption{UEC: cross-sectional results at $t^\star=0.23$ of (a)-(d) streamwise velocity $u^\star_s$, (e)-(h) secondary velocity magnitude $|\bmm{u}^\star_{\theta r}|$, (i)-(l) wall shear stress magnitude $|\bmm{\tau}^\star_w|$. Columns from left to right represent toroidal locations $\phi=\{22^\circ,45^\circ,90^\circ,135^\circ\}$.}
   \label{f:var_uec_t23}
\end{figure*}
%%%%%%%%%%%%%%%%%%%%%%%%%%%%%%%%%%%%%%%%%%%%%%%%%%

\subsubsection{Curvature entrance condition}
\label{s:curved_pipe_wec_uec}

The fully developed pulsatile velocity profile is obtained from experimental PIV (\cite{najjari-plesniak:2018}) and is subsequently referred to as the {\it Womersley entrance condition} (WEC). The second pulsatile entrance condition studied, motivated by the fact that flow upstream to a curved artery may not be fully developed (\cite{ku:1997}), is one where the flow is undeveloped (i.e. uniform). We refer to this inlet velocity condition as the {\it Uniform entrance condition} (UEC). Under UEC, the pulsatile flow rate is equivalent to that used under WEC. With all else being equal, this allows us to solely study the effect of flow development on the evolution of secondary flow patterns and wall shear stresses. These entrance conditions are applied at the inlet to the curved geometry.

%%%%%%%%%%%%%%%%%%%%%%%%%%%%%%%%%%%%%%%%%%%%%%%%%%
\section{Results}
\label{s:results}

Results of nondimensional streamwise velocity $u^\star_s = u_s / \overbar{u}_{mean}$, secondary velocity magnitude $|\bmm{u}^\star_{r \theta}| = |\bmm{u}_{r \theta}| / \overbar{u}_{mean}$, and magnitude of the wall shear stress vector $\bmm{\tau}^{\star}_w = \bmm{\tau}_w d / (\mu \overbar{u}_{mean})$ are presented to compare results between fully developed and uniform entrance conditions, where $\mu$ is the dynamic viscosity of the blood analog fluid.

\subsection{Velocity profiles and secondary flow}
\label{s:secondary_flow}

In Fig.~\ref{f:us_profile_z0}, we plot various profiles of streamwise velocity $u^\star_s$ under WEC and UEC during acceleration and deceleration, the latter exhibiting reverse flow along the entire length of the inner wall and heavily skewed peak velocities towards the outer wall. We emphasize the mid-deceleration phase $t^\star=0.23$, where multiple counter-rotating patterns coexist and transitional secondary flow morphologies of interest emerge.

We observe that under both entrance conditions, the maximum value of the velocity profiles is skewed towards the inner wall near the entrance ($\phi=22^\circ$). The profiles then skew towards the outer wall as the bulk flow moves downstream and the flow decelerates. Under UEC, the profiles are generally much less skewed towards the outer wall since the flow is less developed coming into the curve, causing more flattened profiles.

In Figs.~\ref{f:var_wec_t23} and \ref{f:var_uec_t23} we show that peak streamwise velocity is shifted towards the outer wall ($\theta=0^\circ$) due to centrifugal forces, which sets up a pressure gradient within the cross-section with larger pressure at the outer wall that drives strong secondary flows along the upper and lower wall ({\it middle row}). This results in larger more intense vortical structures to form under WEC (\cite{cox-plesniak:2021}).

Larger values of $|\bmm{\tau}^\star_w|$ occur along the outer and upper/lower wall due to larger velocity gradients from the outwardly shifted velocity profile; $|\bmm{\tau}^\star_w|$ decreases sharply as the fluid moves inward along the upper/lower wall, eventually increasing again at the inner wall. This sharp inflection in $|\bmm{\tau}^\star_w|$ is due to flow reversal along the inner wall, indicated by dashed lines in the streamwise velocity. At $\phi=45^\circ$, we observe that $|\bmm{\tau}^\star_w|$ at the inner wall is more than 3x greater under the fully developed entrance condition.

\subsection{Instantaneous Wall Shear Stress}
\label{s:pulsatile_wss}

%%%%%%%%%%%%%%%%%%%%%%%%%%%%%%%%%%%%%%%%%%%%%%%%%%
%\include{F-wss-contours}
%\input{F-wss-contours.tex}
%
\begin{figure*}
   \renewcommand{\fsize}{0.31\textwidth}
   \centering\setcounter{subfigure}{0}
   \subfloat[WEC: $t^\star=0.19$]{
       \includegraphics[width=\fsize,keepaspectratio]
       {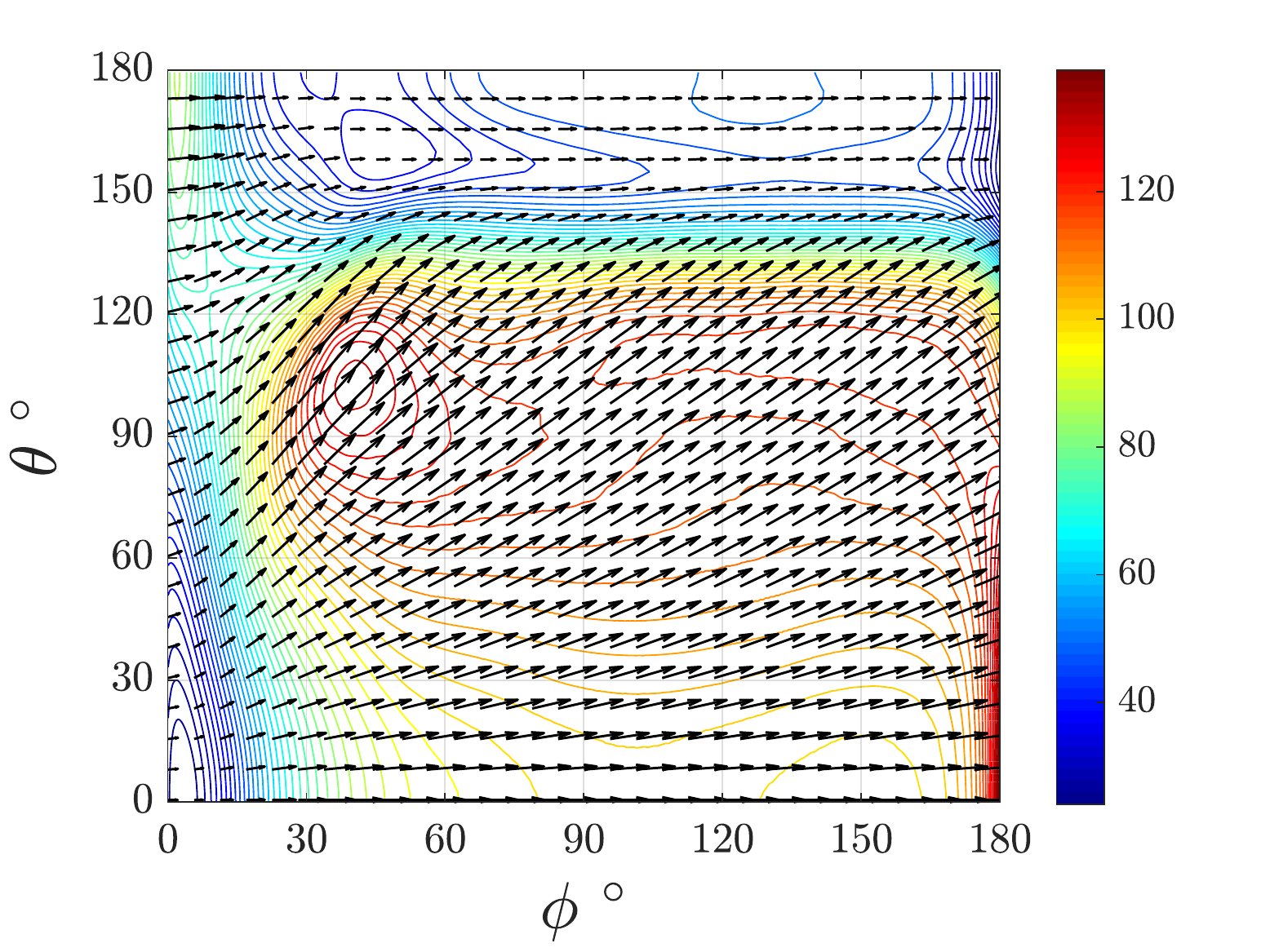}
   }
   \subfloat[WEC: $t^\star=0.21$]{
       \includegraphics[width=\fsize,keepaspectratio]
       {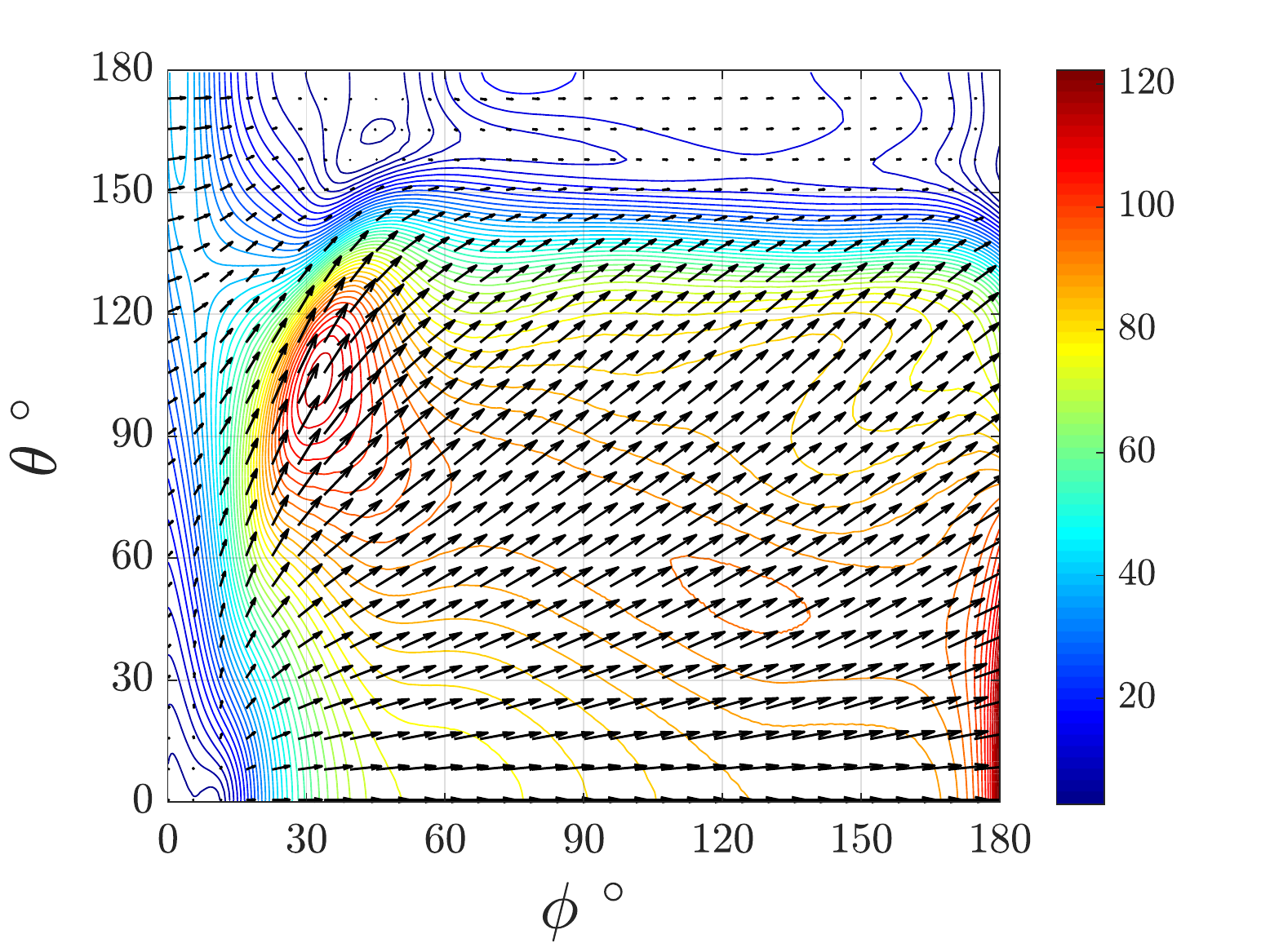}
   }
   \subfloat[WEC: $t^\star=0.23$]{
       \includegraphics[width=\fsize,keepaspectratio]
       {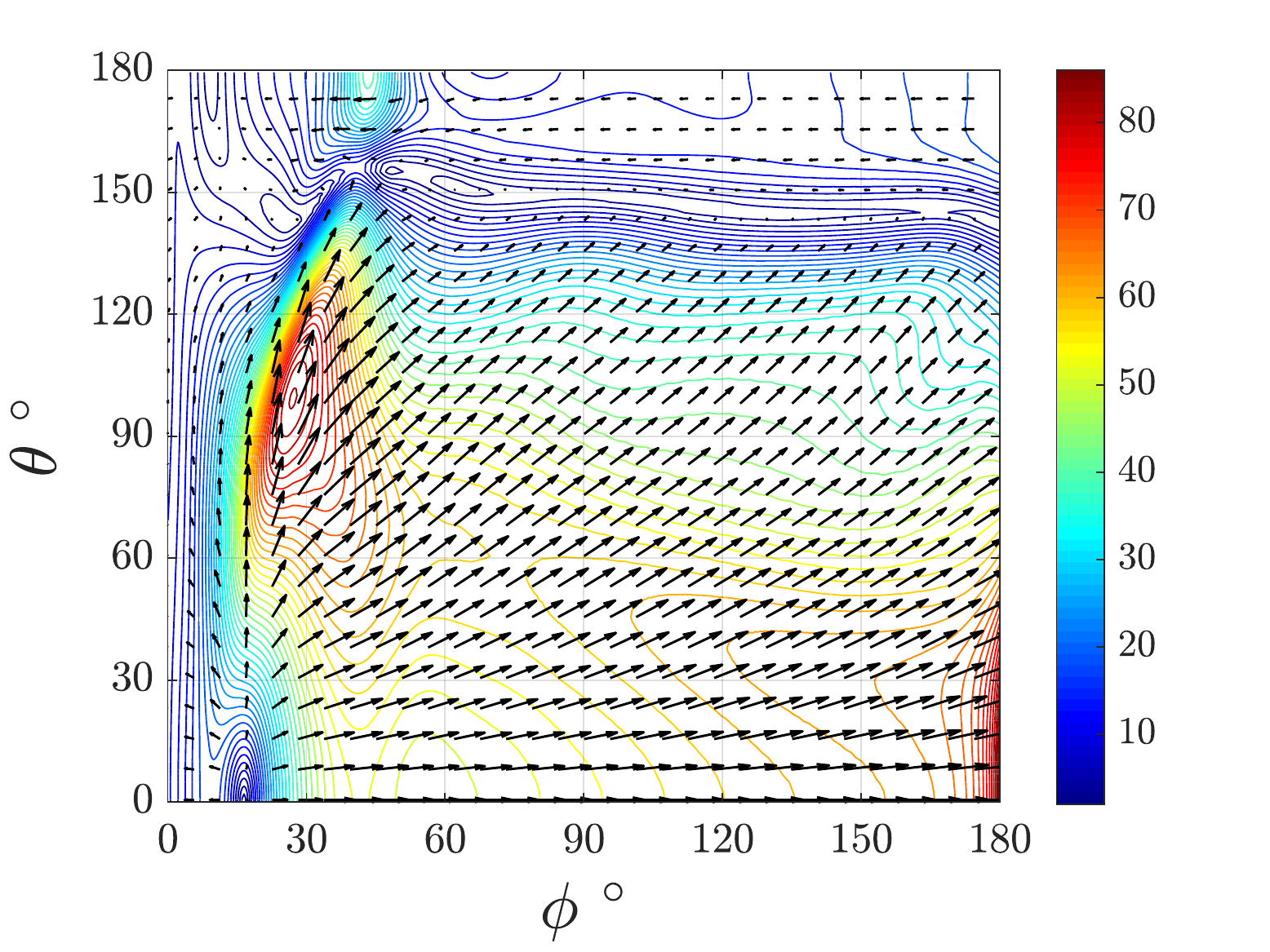}
   }
   \\[-0.15in]
   \subfloat[UEC: $t^\star=0.19$]{
       \includegraphics[width=\fsize,keepaspectratio]
       {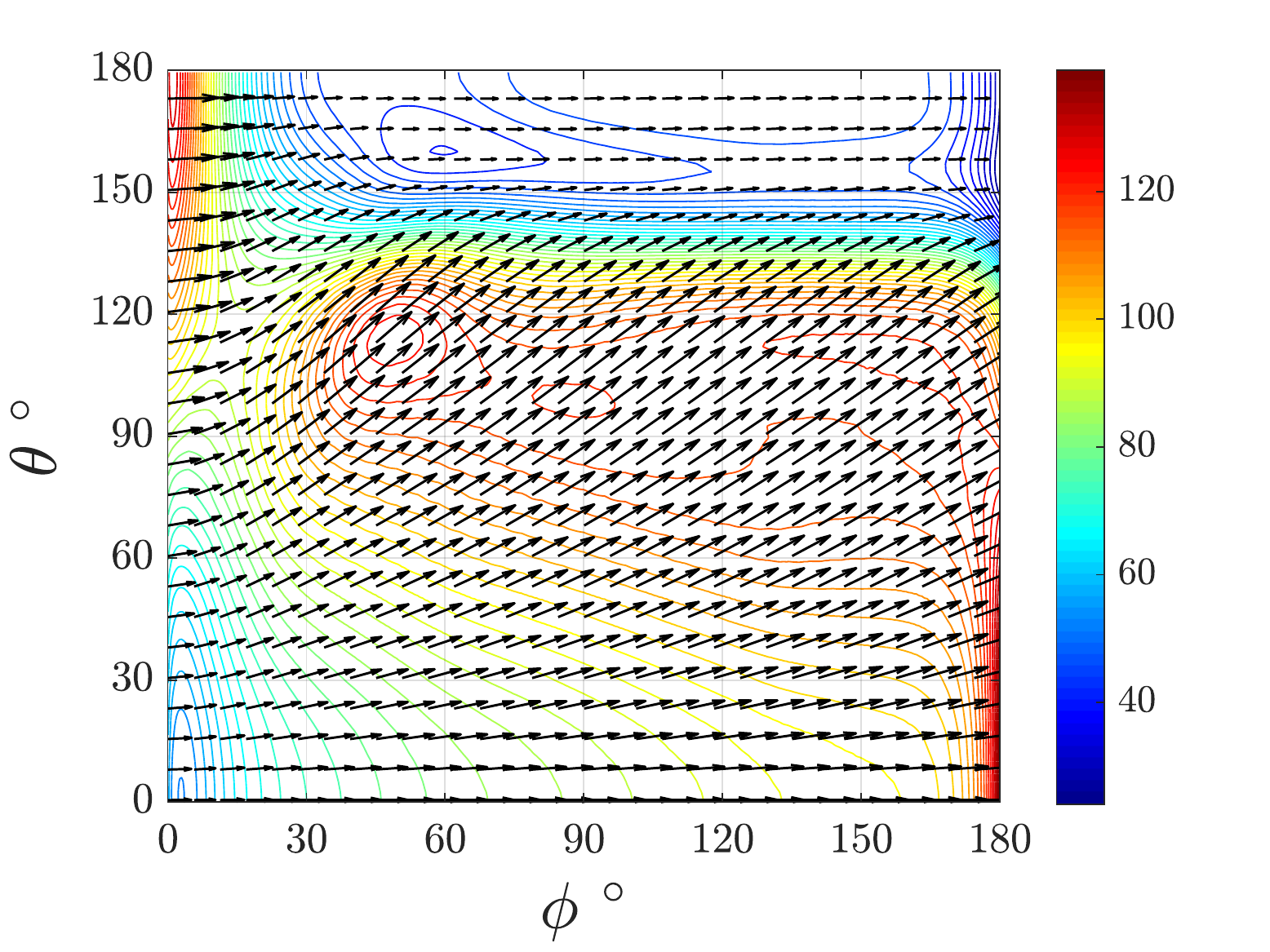}
   }
   \subfloat[UEC: $t^\star=0.21$]{
       \includegraphics[width=\fsize,keepaspectratio]
       {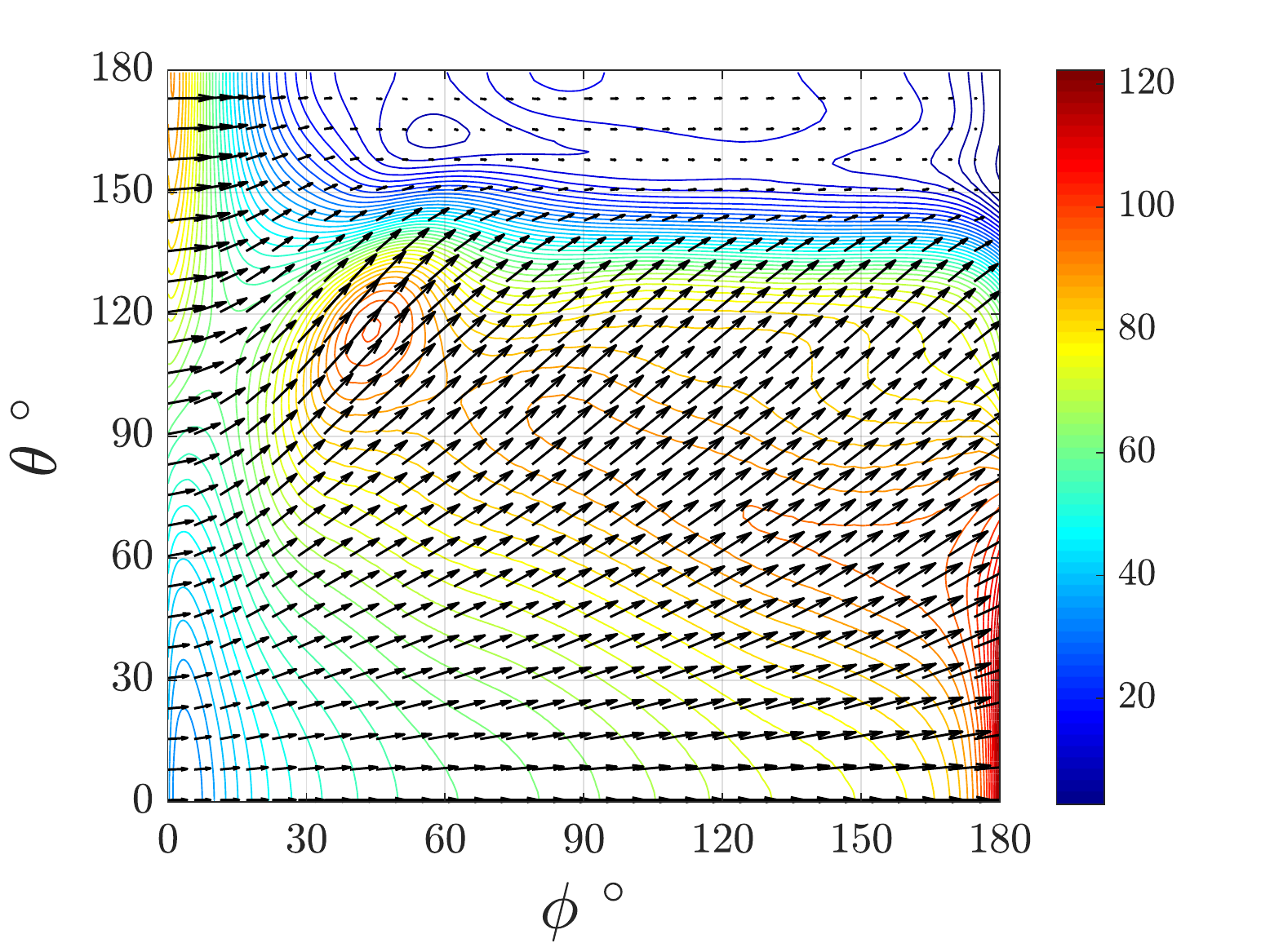}
   }
   \subfloat[UEC: $t^\star=0.23$]{
       \includegraphics[width=\fsize,keepaspectratio]
       {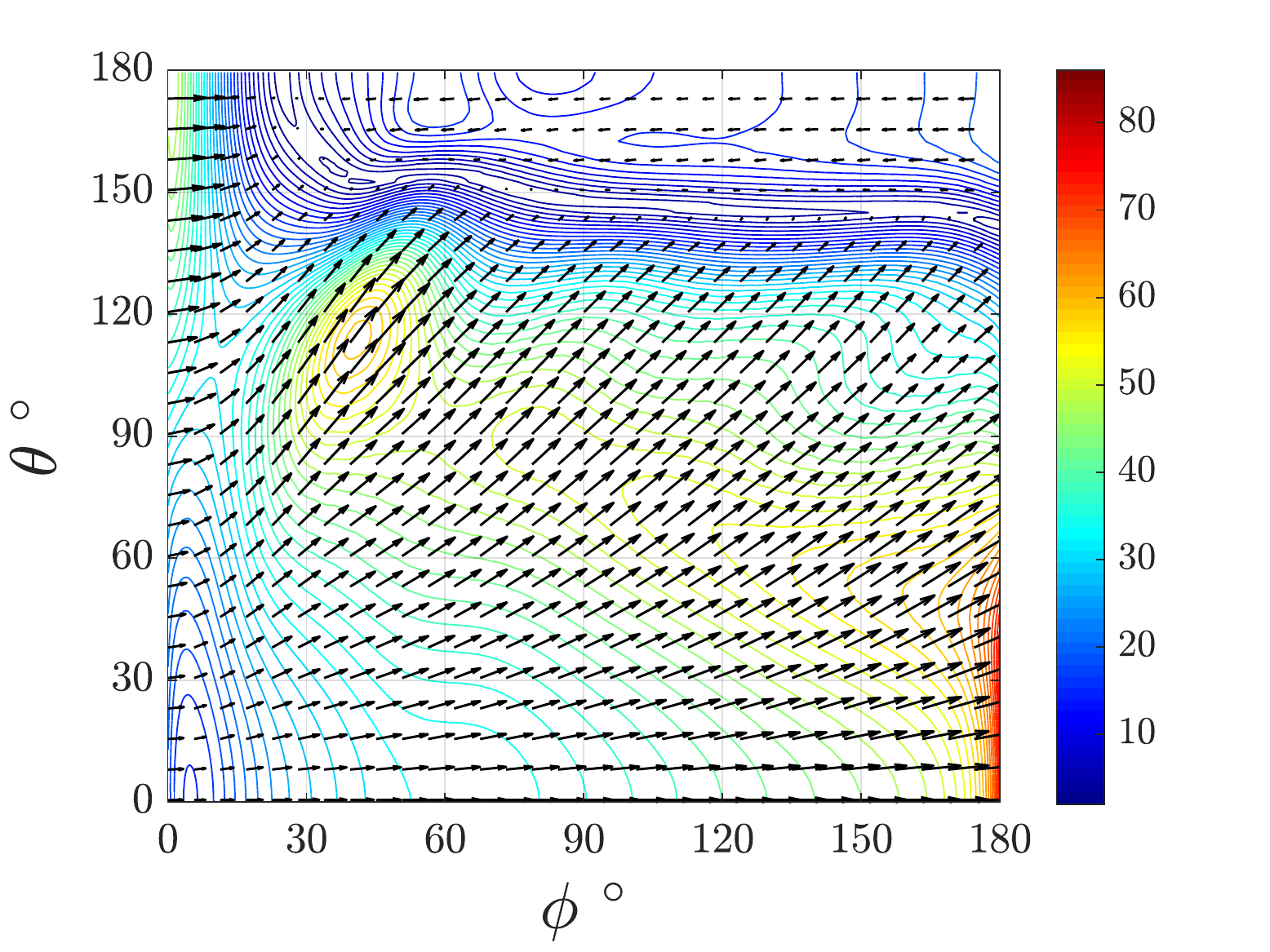}
   }
   \\[-0.15in]
   \subfloat[]{
       \includegraphics[width=\fsize,keepaspectratio]
       {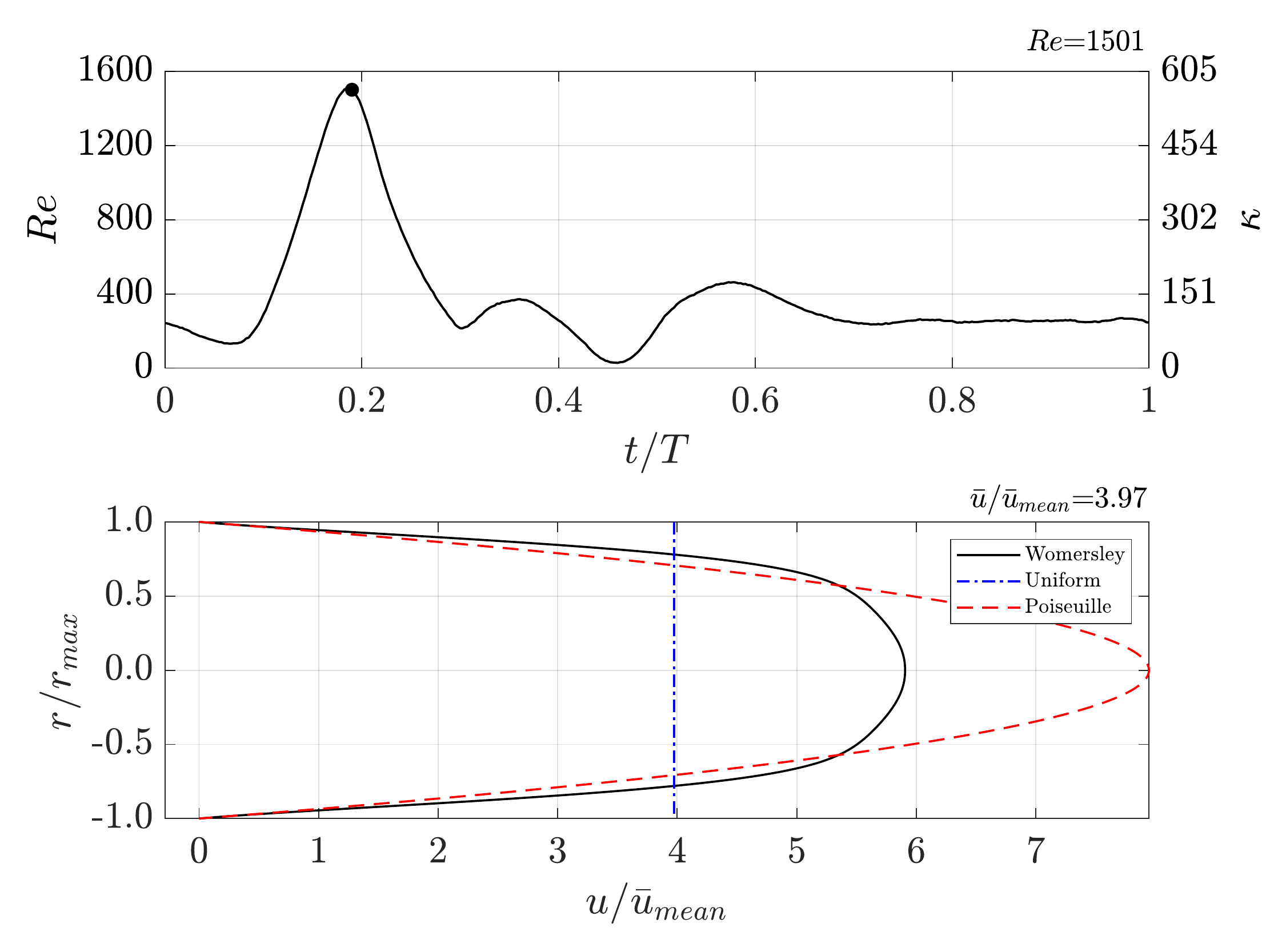}
   }
   \subfloat[]{
       \includegraphics[width=\fsize,keepaspectratio]
       {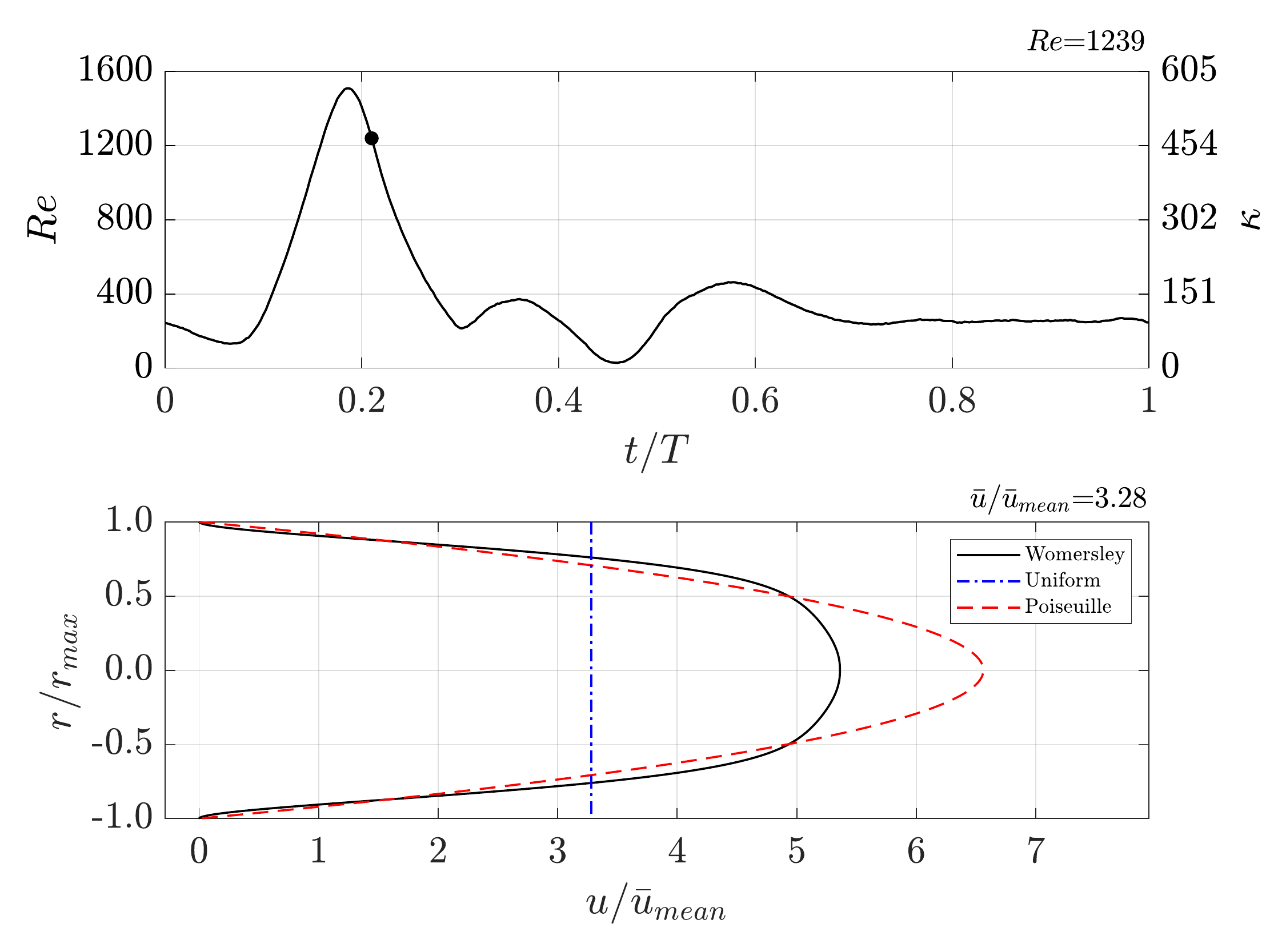}
   }
   \subfloat[]{
       \includegraphics[width=\fsize,keepaspectratio]
       {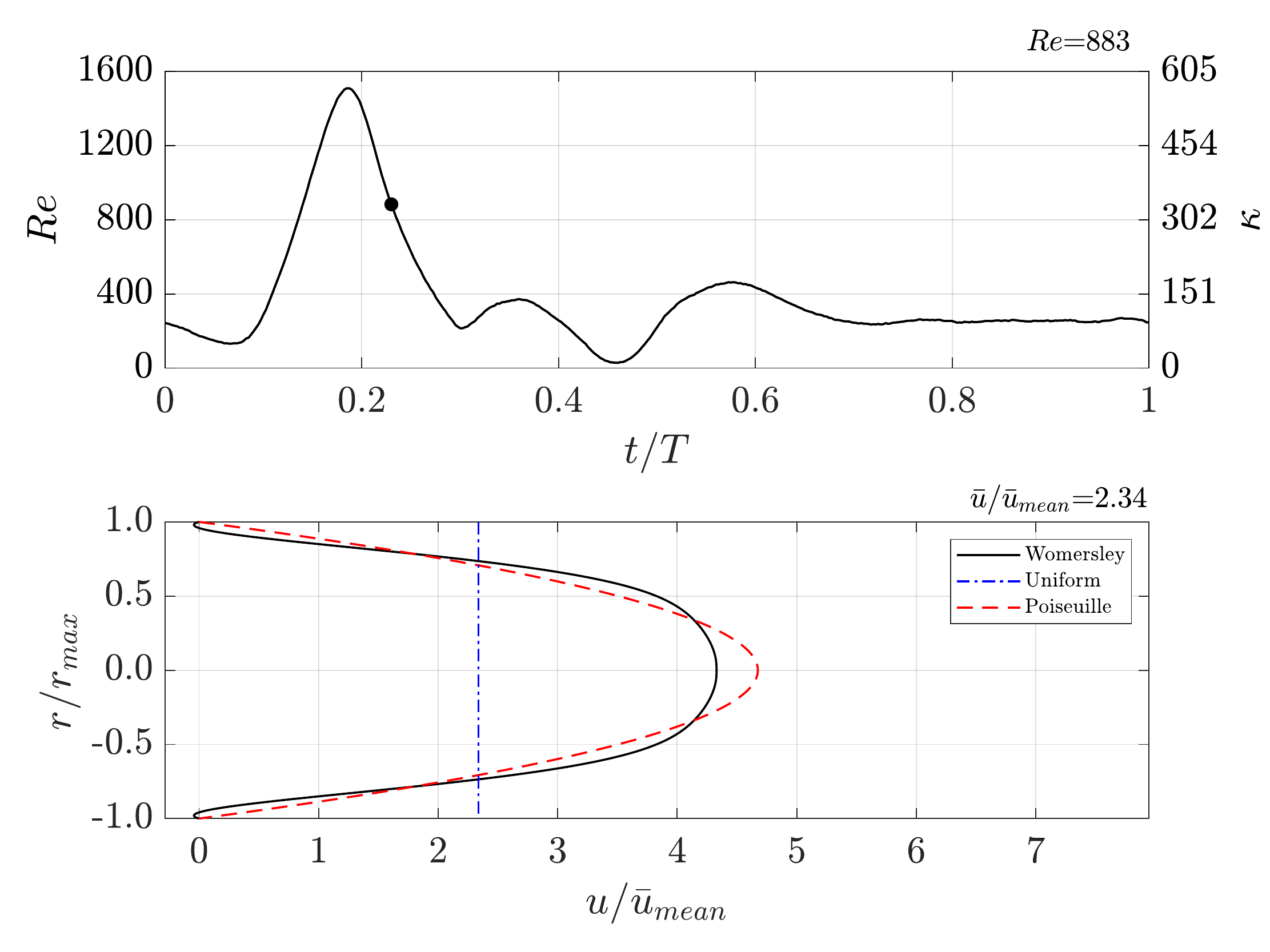}
   }
   
   \caption{Instantaneous wall shear stress $\bmm{\tau}^\star_w$ vector map of upper wall $0^\circ \leq \phi,\theta \leq 180^\circ$ during deceleration under WEC (\textit{top}) and UEC (\textit{middle}), colored by magnitude. The corresponding waveform phase and entrance velocity profiles (\textit{bottom}) are provided along with a Poiseuille profile for reference.}
   \label{f:wss_wec_uec}
\end{figure*}

\begin{figure*}
   \renewcommand{\fsize}{0.31\textwidth}
   %\ContinuedFloat
   \subfloat[WEC: $t^\star=0.25$]{
       \includegraphics[width=\fsize,keepaspectratio]
       {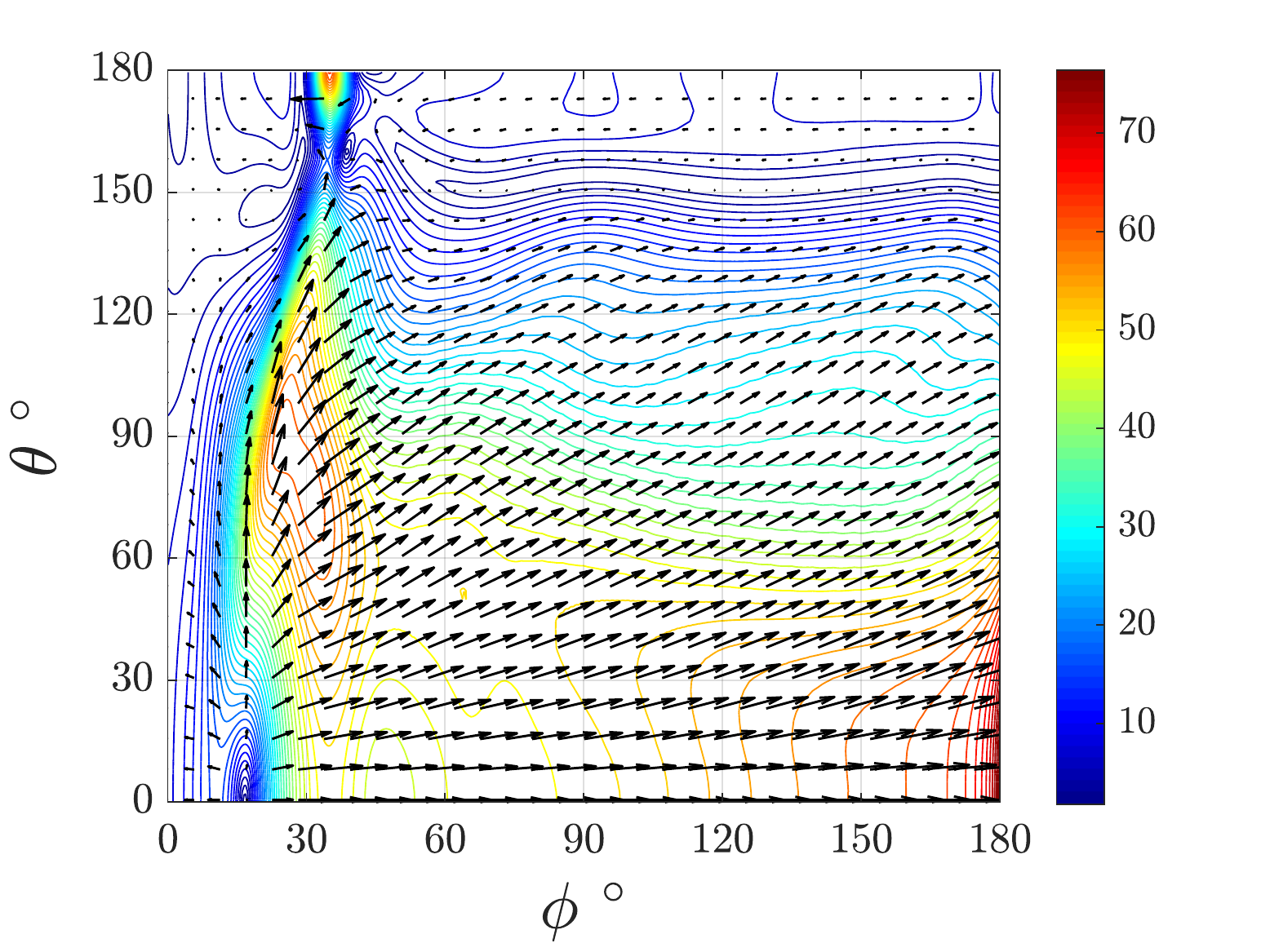}
   }
   \subfloat[WEC: $t^\star=0.27$]{
       \includegraphics[width=\fsize,keepaspectratio]
       {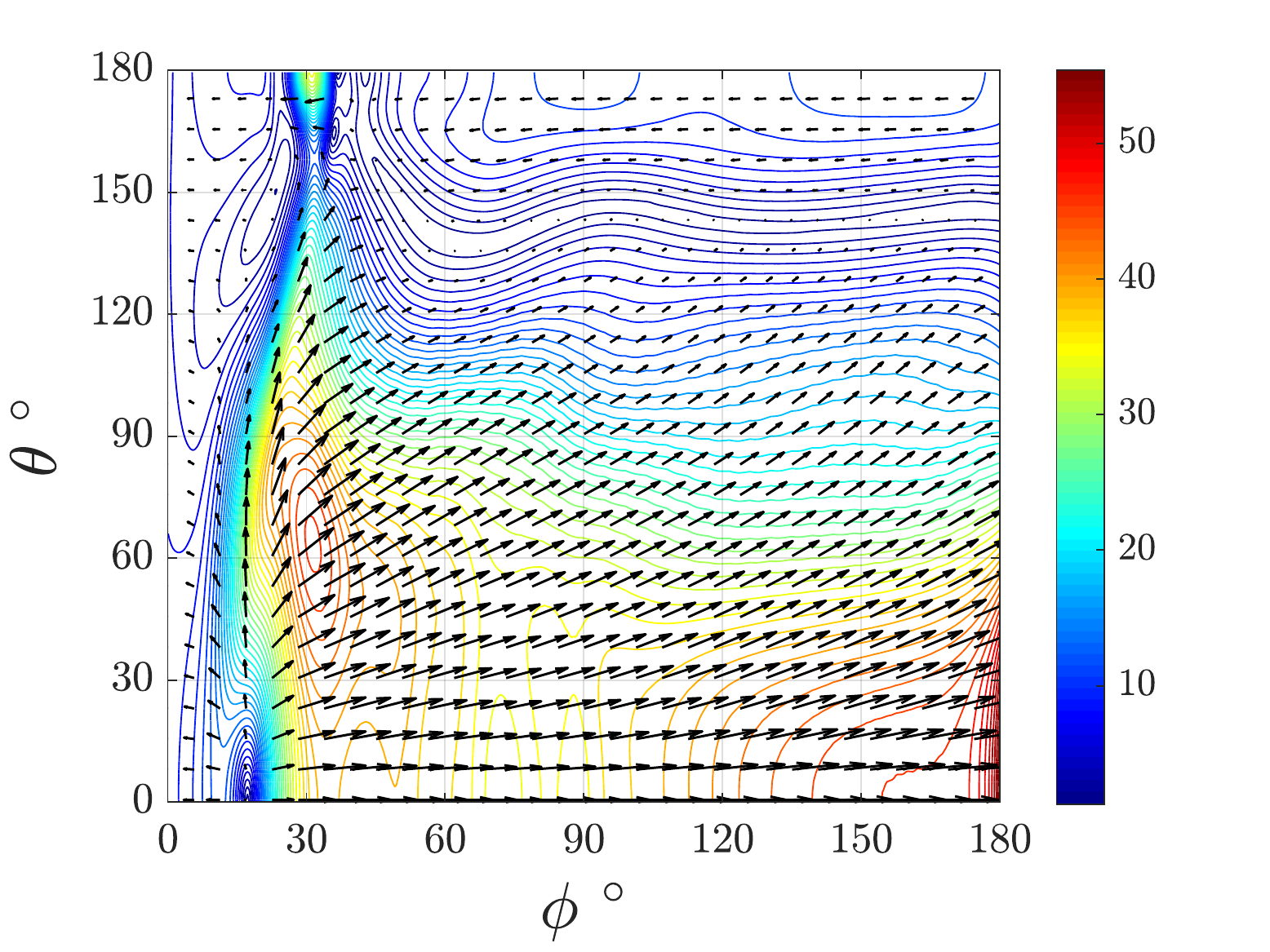}
   }
   \subfloat[WEC: $t^\star=0.29$]{
       \includegraphics[width=\fsize,keepaspectratio]
       {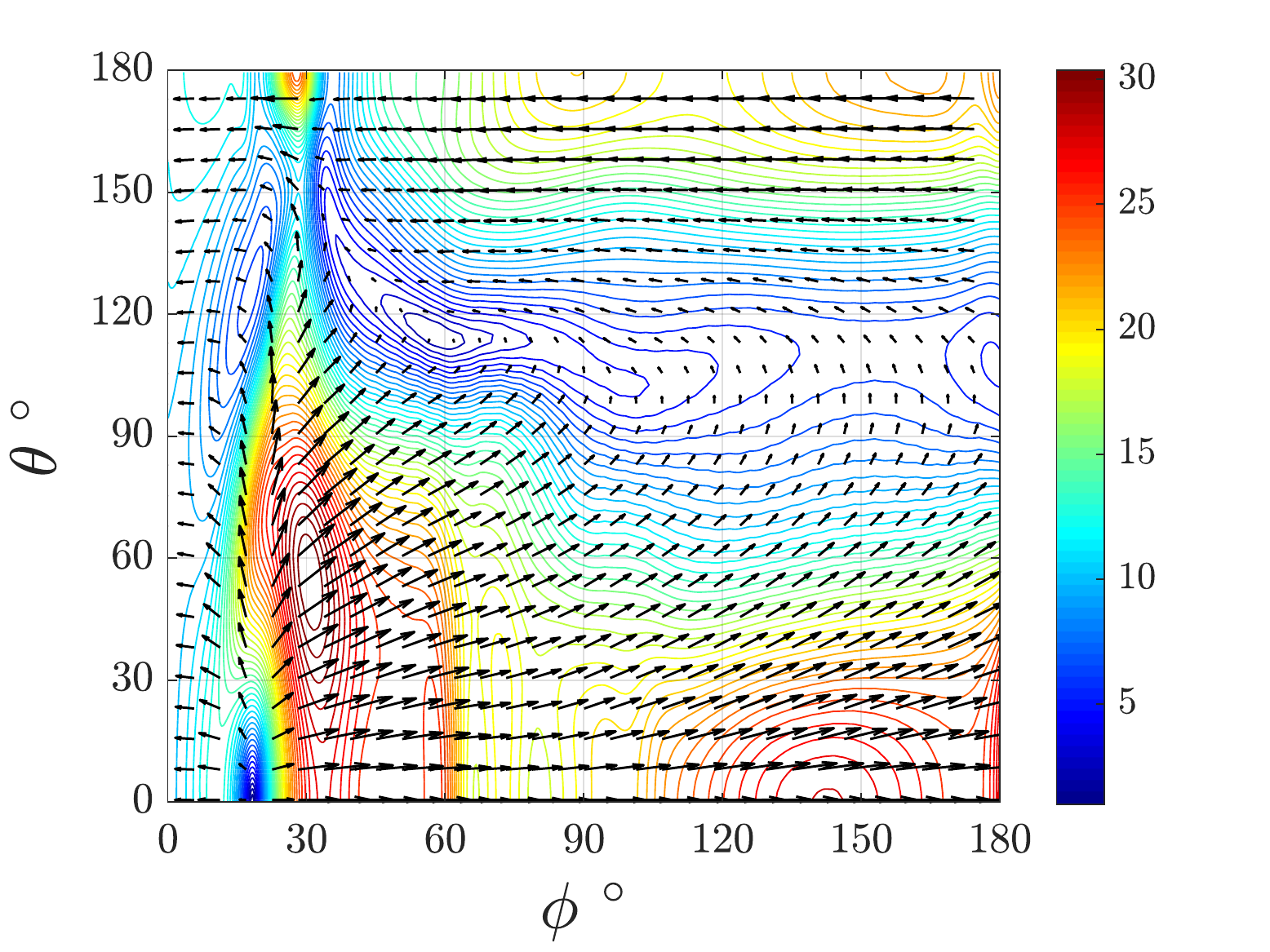}
   }
   \\[-0.15in]
   \subfloat[UEC: $t^\star=0.25$]{
       \includegraphics[width=\fsize,keepaspectratio]
       {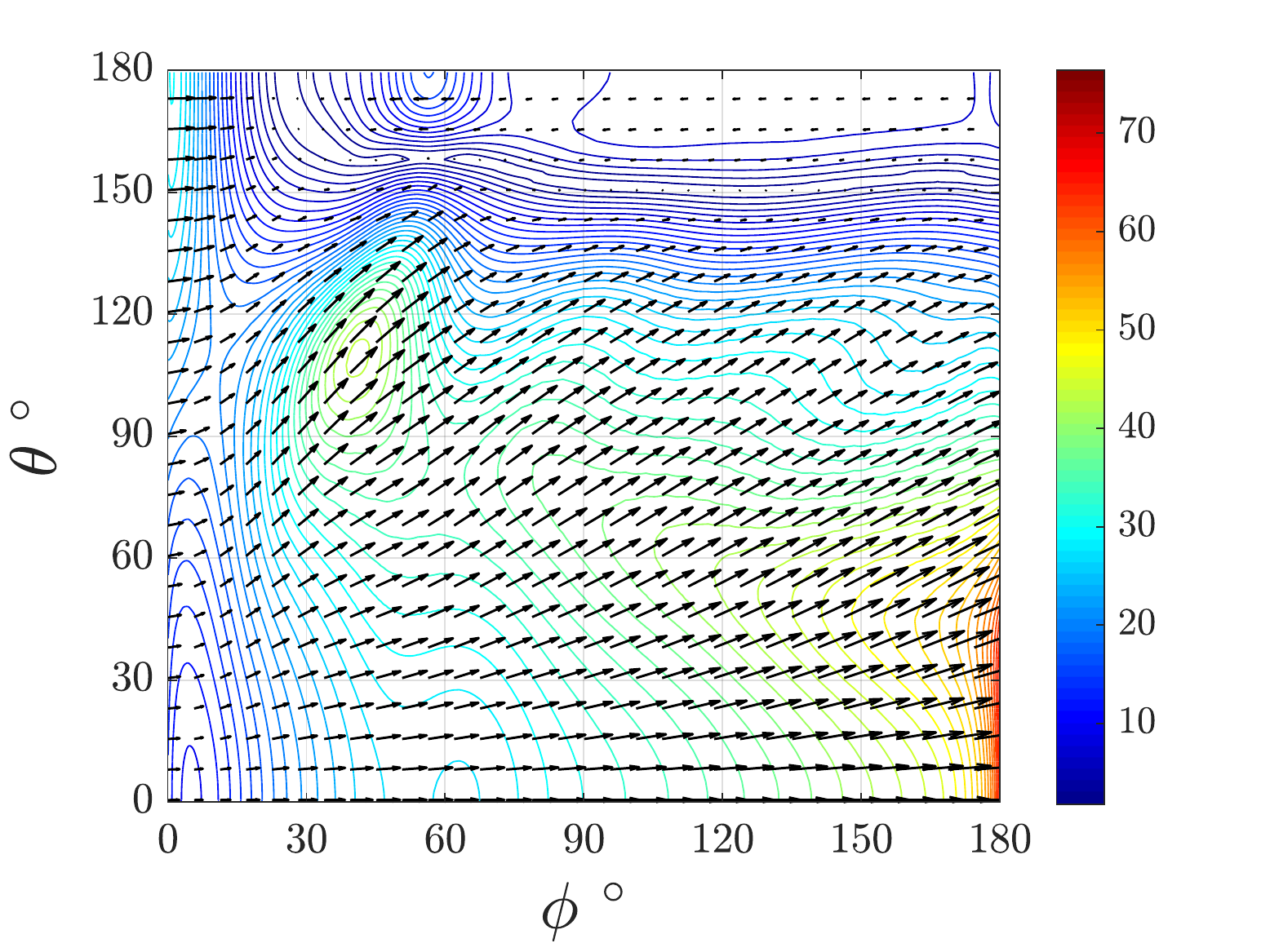}
   }
   \subfloat[UEC: $t^\star=0.27$]{
       \includegraphics[width=\fsize,keepaspectratio]
       {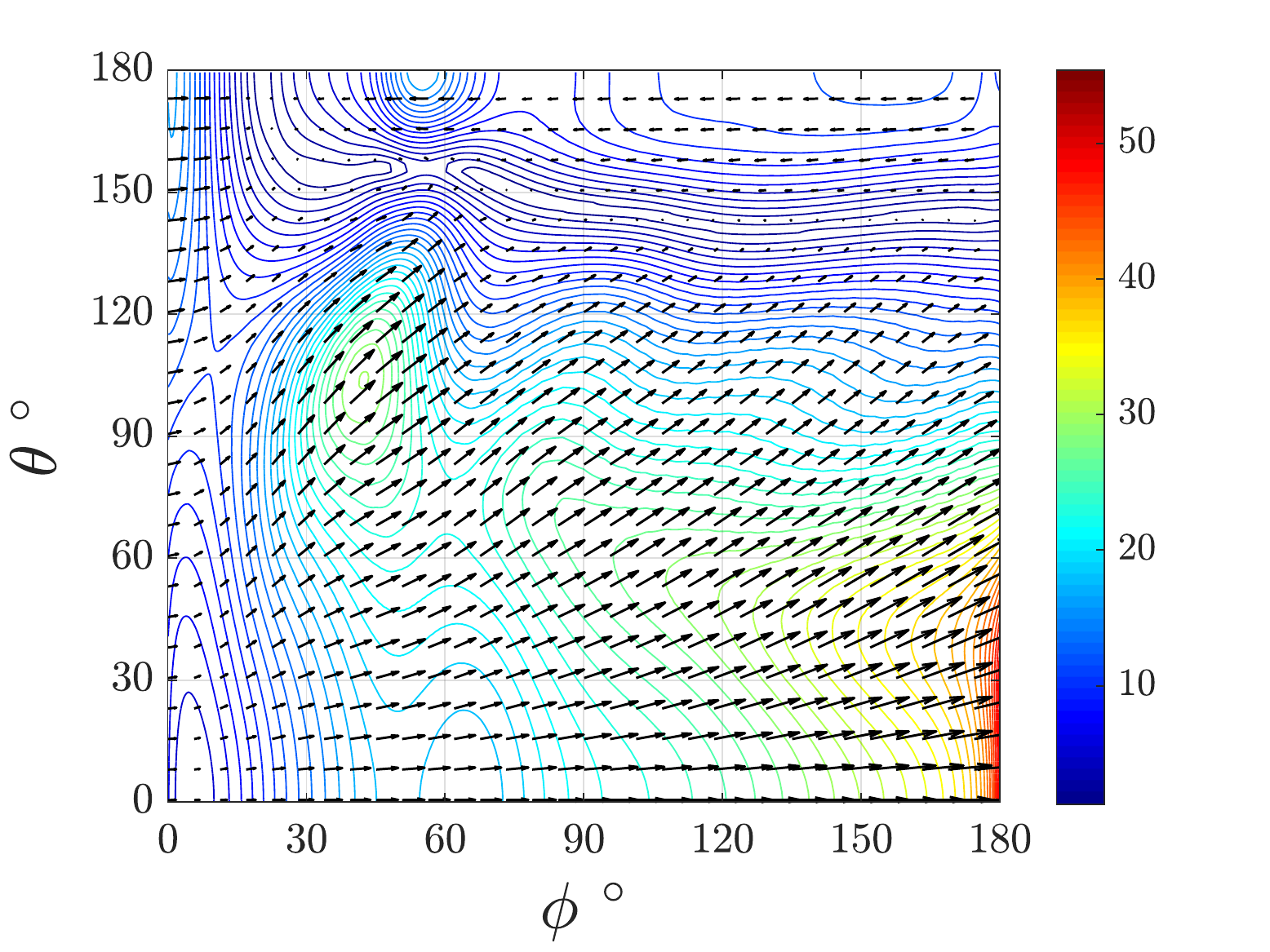}
   }
   \subfloat[UEC: $t^\star=0.29$]{
       \includegraphics[width=\fsize,keepaspectratio]
       {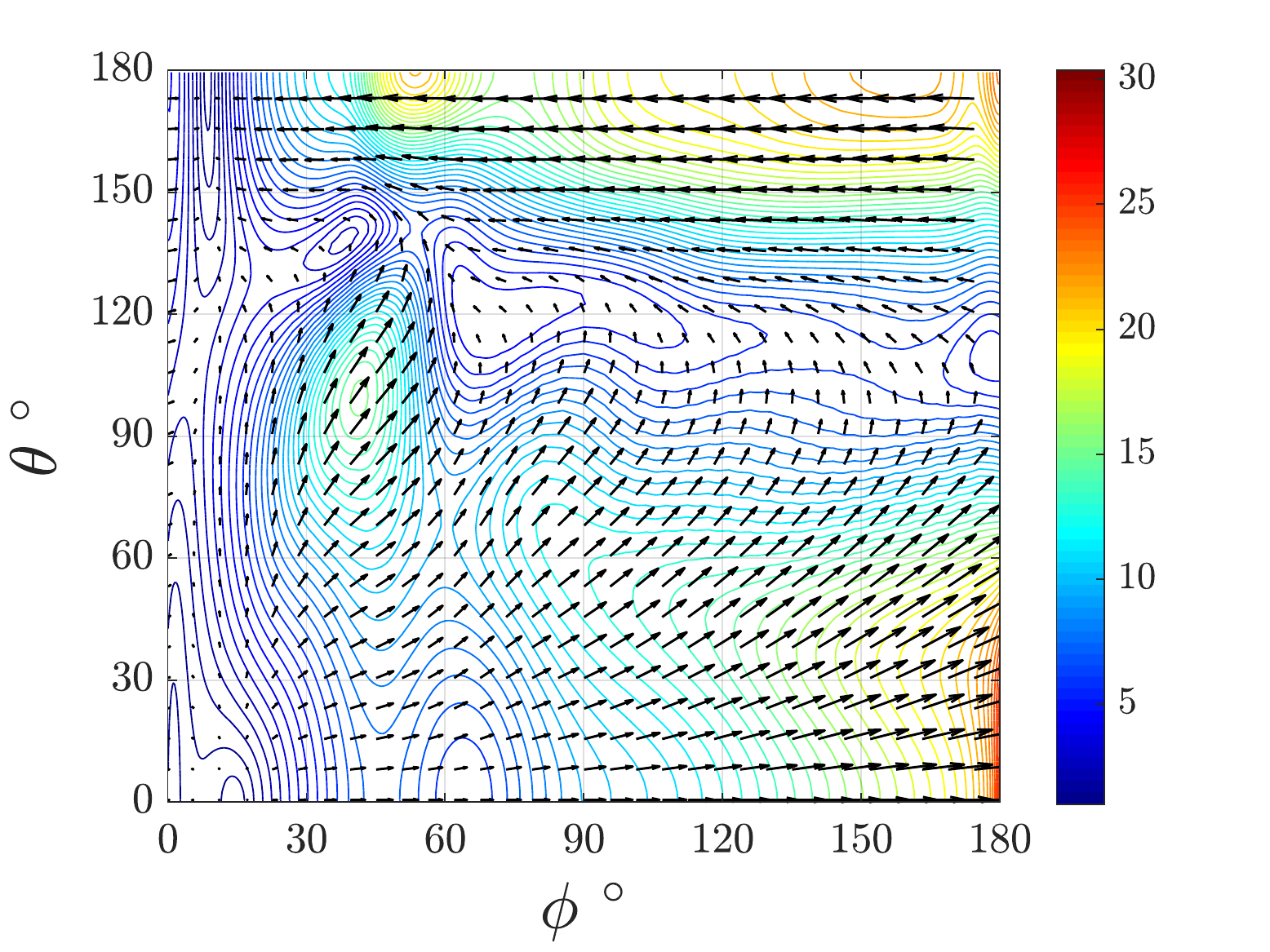}
   }
   \\[-0.15in]
   \subfloat[]{
       \includegraphics[width=\fsize,keepaspectratio]
       {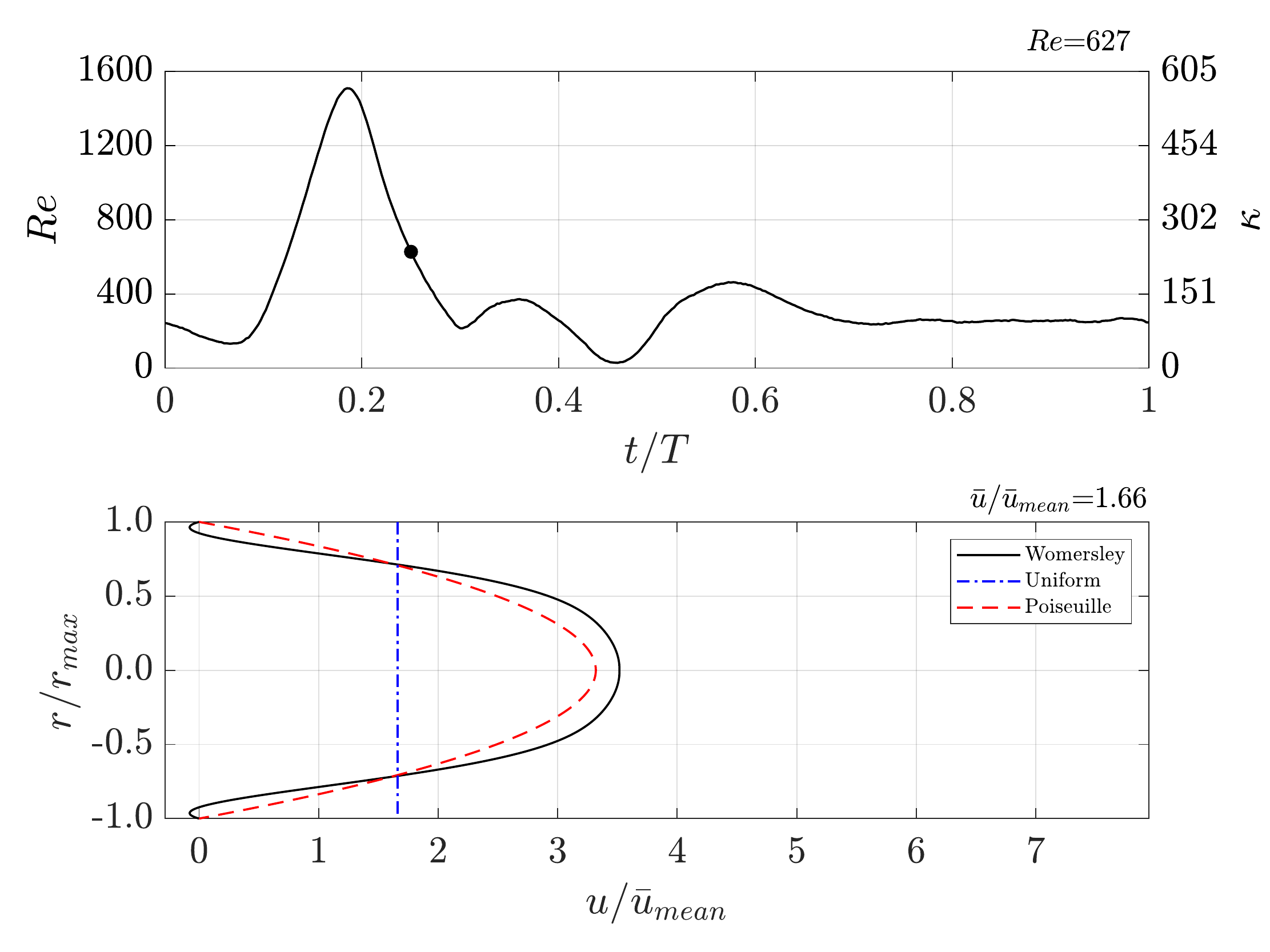}
   }
   \subfloat[]{
       \includegraphics[width=\fsize,keepaspectratio]
       {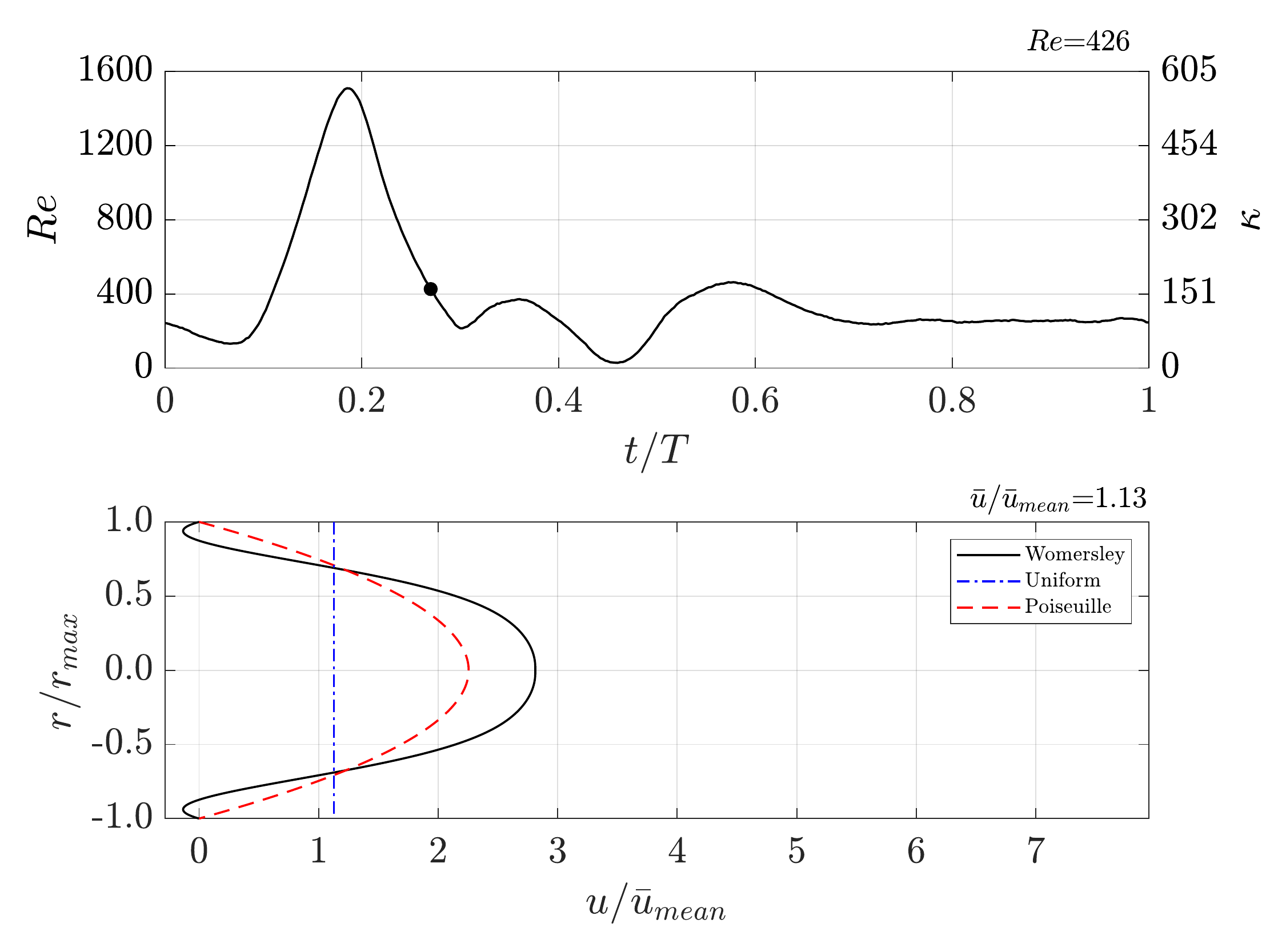}
   }
   \subfloat[]{
       \includegraphics[width=\fsize,keepaspectratio]
       {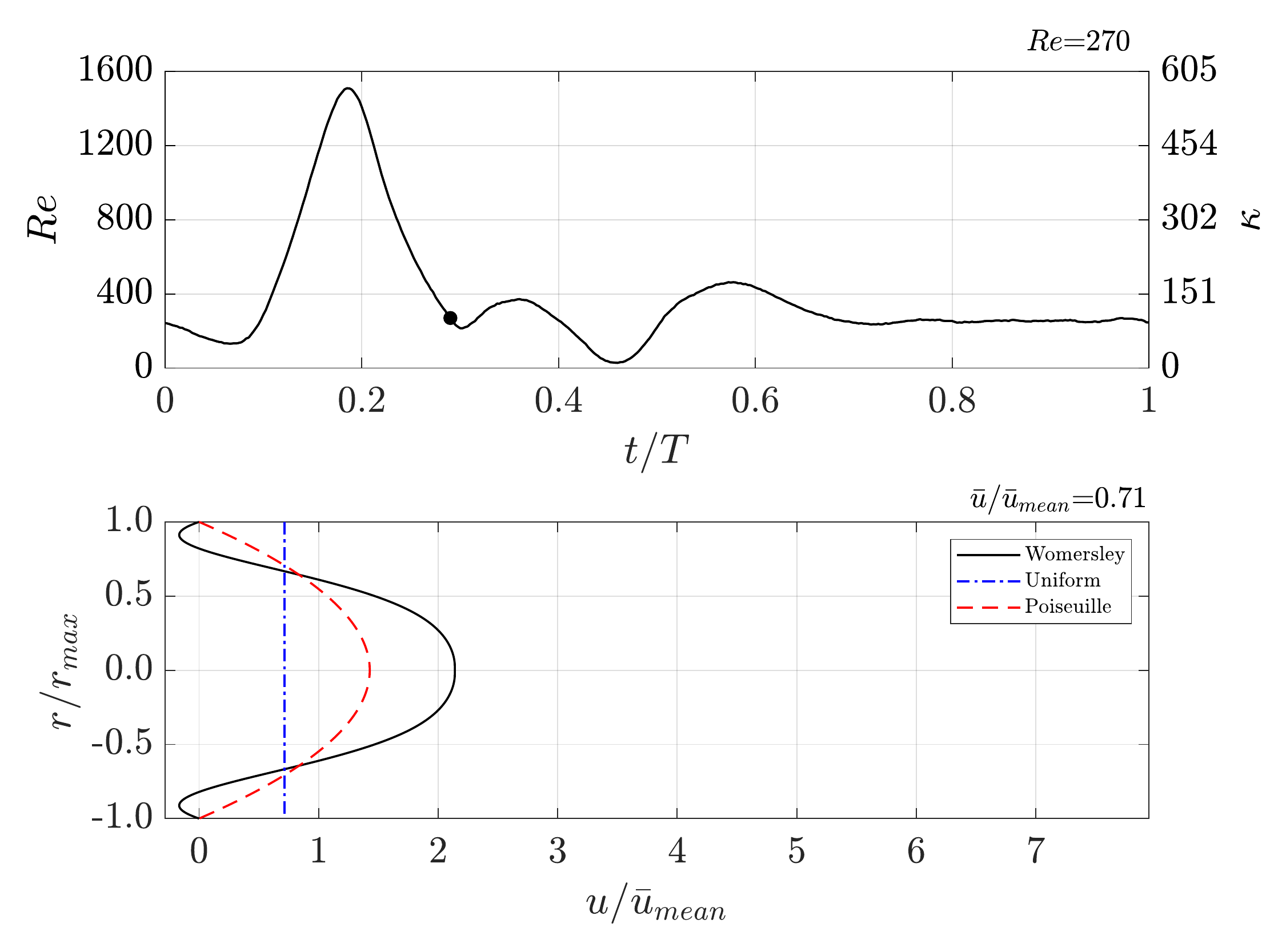}
   }
   
   \caption{Instantaneous wall shear stress $\bmm{\tau}^\star_w$ vector map of upper wall $0^\circ \leq \phi,\theta \leq 180^\circ$ during deceleration under WEC (\textit{top}) and UEC (\textit{middle}), colored by magnitude. The corresponding waveform phase and entrance velocity profiles (\textit{bottom}) are provided along with a Poiseuille profile for reference.}
   \label{f:wss_wec_uec_2}
\end{figure*}
%%%%%%%%%%%%%%%%%%%%%%%%%%%%%%%%%%%%%%%%%%%%%%%%%%

We use the following full decomposition formulation
\begin{linenomath}
\begin{align}
    \bmm{\tau}_w = \bmm{T}\bmm{n}
    - \big[ \big( \bmm{T}\bmm{n} \big) \cdot \bmm{n} \big] \bmm{n}
    \label{e:instantaneous_wss}
\end{align}
\end{linenomath}
to accurately compute the instantaneous wall shear stress vector at each solution node of the computational grid. The normal vector to the wall surface is denoted $\bmm{n}$ and the second-order stress tensor for a Newtonian, incompressible fluid is $\bmm{T}=2\mu\bmm{S}$, where the strain-rate tensor $\bmm{S}=(\nabla\bmm{u}+\nabla\bmm{u}^T)/2$ is formed from the velocity gradient tensor $\nabla \bmm{u}$ computed using the high-order spatial discretization scheme indicated in Sec.~\ref{s:numerical_scheme}. With Eq.~\ref{e:instantaneous_wss}, we can compute the direction and magnitude of the wall shear stress vector over the entire curved surface throughout the pulse cycle.

Since the flow is symmetric about the plane $z=0$, we plot results from the upper surface of the curved model only, where the orientation of the geometry is defined in Fig.~\ref{f:curved_pipe_mesh_geometry_N4}. To visualize results, we map the data onto the $\phi\theta$-axes shown in Figs.~\ref{f:wss_wec_uec} and \ref{f:wss_wec_uec_2}. IN these figures, we plot side-by-side comparisons of instantaneous wall shear stress obtained under WEC and UEC, and we display results from peak flow rate at $t^\star=0.19$ through the end of deceleration at $t^\star=0.29$. At peak flow rate, the wall shear stress vectors are mostly angled toward the inner wall due to the secondary flow, which corresponds to formation of Dean-type vortices (\cite{cox-plesniak:2021}). Along the inner and outer wall, the vectors are aligned with the streamwise direction. A local maximum in shear stress magnitude appears in both entrance conditions near $(\phi,\theta)=(45^\circ,105^\circ)$, indicating strong secondary flow in this region. Near the entrance to the curve, we also observe greater shear stress under UEC due to larger velocity gradients near the inner wall caused by higher inner wall skewness of the velocity profile (see Fig.~\ref{f:us_profile_z0}\subref{f:us_profile_22_plug_z0}). We also observe the cross-over in maximum wall shear stress from the inner wall to the outer wall as the fluid moves downstream (\cite{singh:1974}).

As systolic deceleration begins, shear stresses decrease globally while altering direction such that vectors become less aligned with the streamwise direction and more aligned with the radius of curvature, particularly near the entrance. At $t^\star=0.21$, the fluid has slowed down near the inner wall causing the shear stress vector to decrease in magnitude before reversing direction. The flow then decelerates further, increasing the amount of reverse flow along the inner wall and causing the shear stress vectors to point upstream.

Under WEC at $t^\star=0.23$, a small circular pocket of large wall shear stress appears near the inner wall at $\phi=40^\circ$. As the flow rate decelerates further, this pocket slides upstream while decreasing in magnitude. This increased inner wall shear stress does not exist under UEC because the combined effect of reverse flow and secondary flow is less intense from this entrance condition. Over the entire systolic deceleration, the spatially averaged shear stress value decreases by 86\%.

\subsection{Time-Averaged Wall Shear Stress}
\label{s:pulsatile_tawss}

\begin{figure*}
   \centering\setcounter{subfigure}{0}
   \subfloat[]{
       \includegraphics[width=0.47\textwidth,keepaspectratio]
       {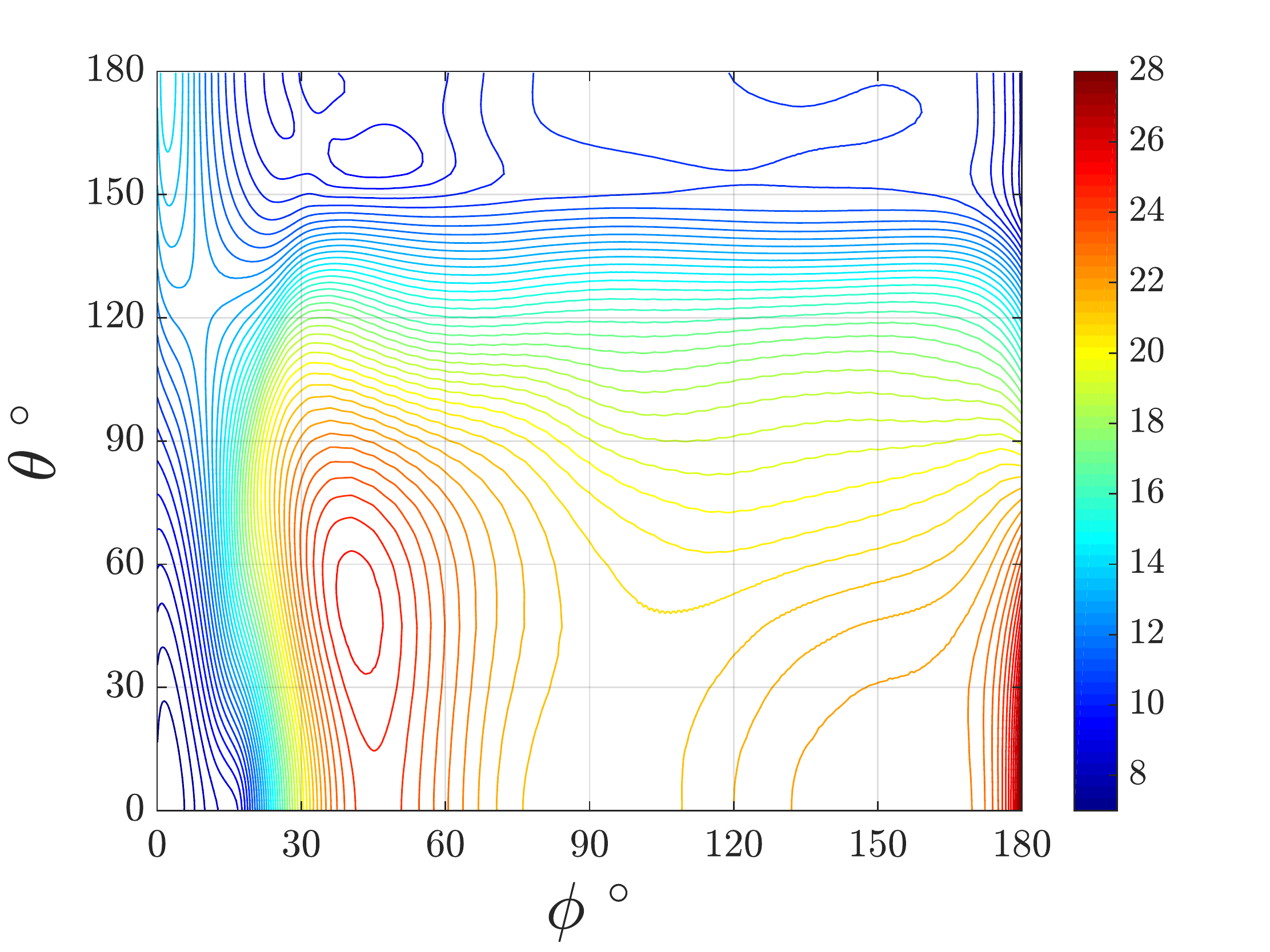}
   }\hfill
   \subfloat[]{
       \includegraphics[width=0.47\textwidth,keepaspectratio]
       {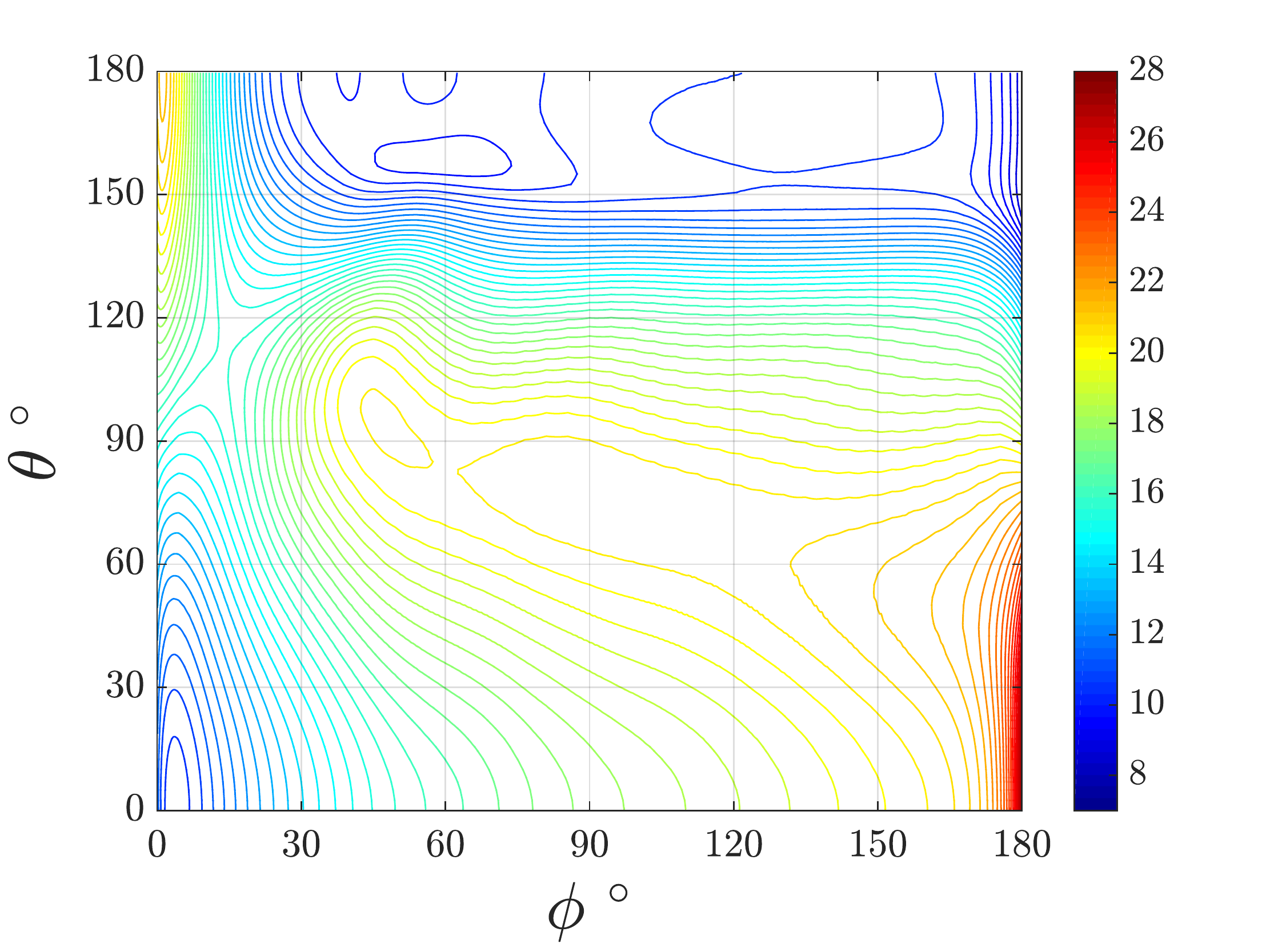}
   }
   
   \caption{Time-averaged wall shear stress, TAWSS: (a) WEC, (b) UEC. Higher values in (a) at $(\phi,\theta) \approx (45^\circ,45^\circ)$ resulting from strong secondary flow.}
   \label{f:tawss}
\end{figure*}

Values of time-averaged wall shear stress are computed from the instantaneous wall shear stress vector $\bmm{\tau}^{\star}_w$ at each node along the wall by Eq.~(\ref{e:instantaneous_wss}) and integrated over the nondimensional pulsatile waveform period $t^\star$ using
\begin{linenomath}
\begin{align}
    %%TAWSS = \frac{1}{T}\int_{0}^{T} \lvert \bmm{\tau}_w \rvert dt\\
    \mathrm{TAWSS} = \int_{0}^{1} \lvert \bmm{\tau}^{\star}_w \rvert\ \drm{t}^\star.
    \label{e:tawss}
\end{align}
\end{linenomath}
Results of TAWSS at each grid node on the wall are plotted in Fig.~\ref{f:tawss}. Under WEC, the outer wall experiences higher TAWSS than the inner wall due to the combination of a skewed velocity profile and the absence of reverse flow. Furthermore, high values of TAWSS occur near $(\phi,\theta)=(45^\circ,45^\circ)$ due to high wall shear stress resulting from strong secondary flow and no reverse flow at that location.

Under both WEC and UEC, high values of TAWSS occur at the exit for $\theta < 60^\circ$. Low values occur mostly along the inner wall for $15^\circ < \phi < 180^\circ$ and near the entrance for $\theta < 90^\circ$ and $0^\circ < \phi < 15^\circ$. We also observe higher values of TAWSS just downstream of the entrance near the inner wall under UEC caused by the inner wall skewness in the velocity field. The inner-to-out switch in maximum wall shear stress described in Sec.~\ref{s:pulsatile_wss} is also reflected in the plot of TAWSS just downstream of the entrance.
%The min/max contour levels correspond to $2.50/8.75~\mathrm{dyn/cm^2}$, assuming $\rho=1060~\mathrm{kg/m^3}$ for the density of blood plasma.

\subsection{Oscillatory Shear Index}
\label{s:pulsatile_osi}

The oscillatory shear index (\cite{he-ku:1996}) computed as
\begin{linenomath}
\begin{align}
    %%OSI = \frac{1}{2}\Big( 1 - \frac{\lvert \bmm{\tau}_{mean} \rvert}{TAWSS} \Big)\\
    \mathrm{OSI} = \frac{1}{2}\Bigg( 1 - \frac{\lvert \bmm{\tau}^{\star}_{mean} \rvert}{\mathrm{TAWSS}} \Bigg)
\end{align}
\end{linenomath}
where
\begin{linenomath}
\begin{align}
    %%\bmm{\tau}_{mean} = \frac{1}{T} \int_{0}^{T} \bmm{\tau}_w dt\\
    \bmm{\tau}^{\star}_{mean} = \int_{0}^{1} \bmm{\tau}^{\star}_w\ \drm{t}^\star
\end{align}
\end{linenomath}
is a uniaxial metric that has been used by many authors to identify the oscillatory nature of vascular flows. The theoretical value of OSI can vary from zero to 0.5, the minimum occurring when the mean wall shear stress equals the time-averaged wall shear stress and the maximum occurring when the mean stress is zero. In our results, the maximum OSI obtained is about half the maximum theoretical value. From the formulation, we can see that OSI can identify regions of flow reversal; however, it is insensitive to shear stress magnitude.

\begin{figure*}
   \centering\setcounter{subfigure}{0}
   \subfloat[]{
       \includegraphics[width=0.47\textwidth,keepaspectratio]
       {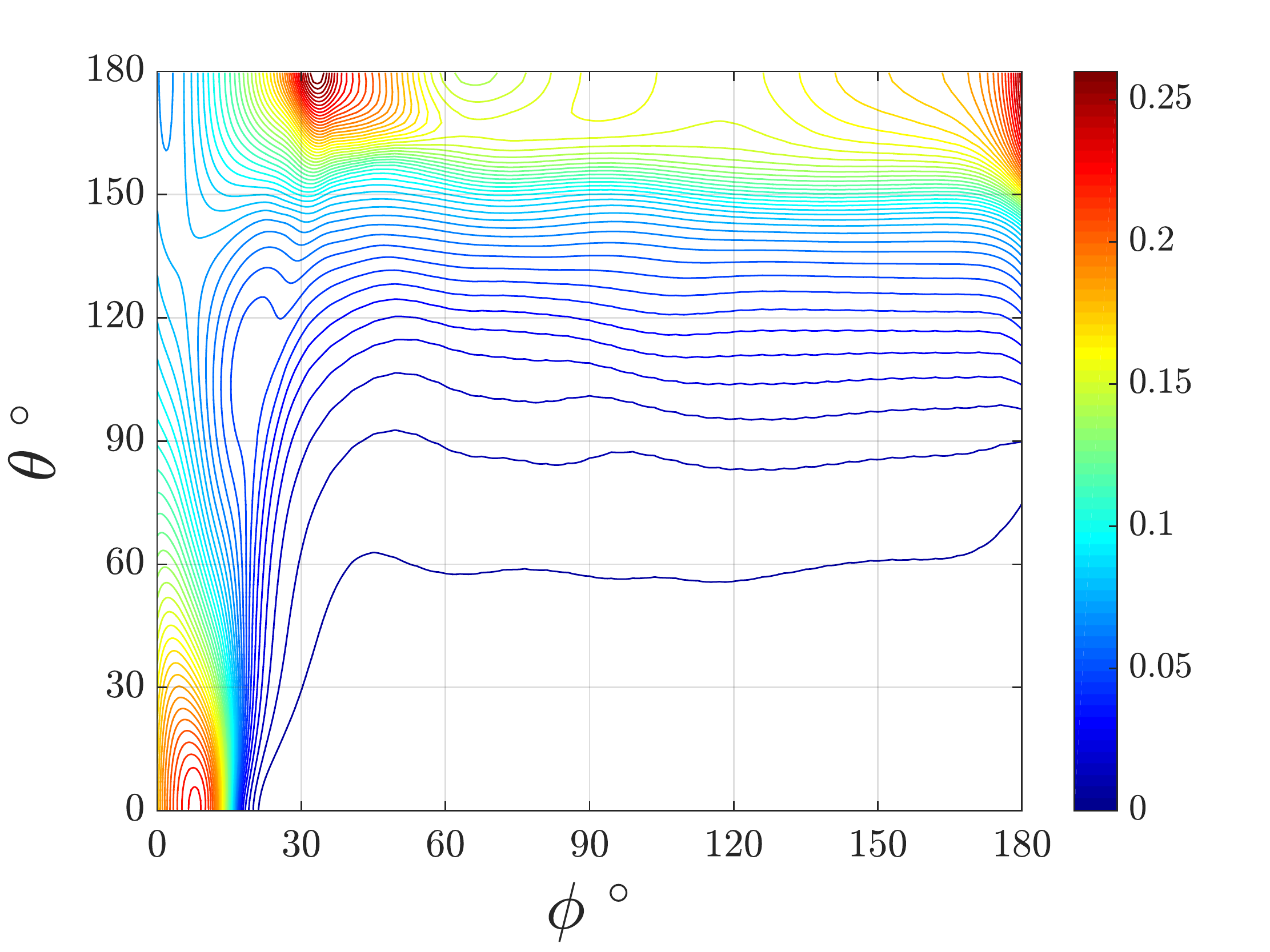}
   }\hfill
   \subfloat[]{
       \includegraphics[width=0.47\textwidth,keepaspectratio]
       {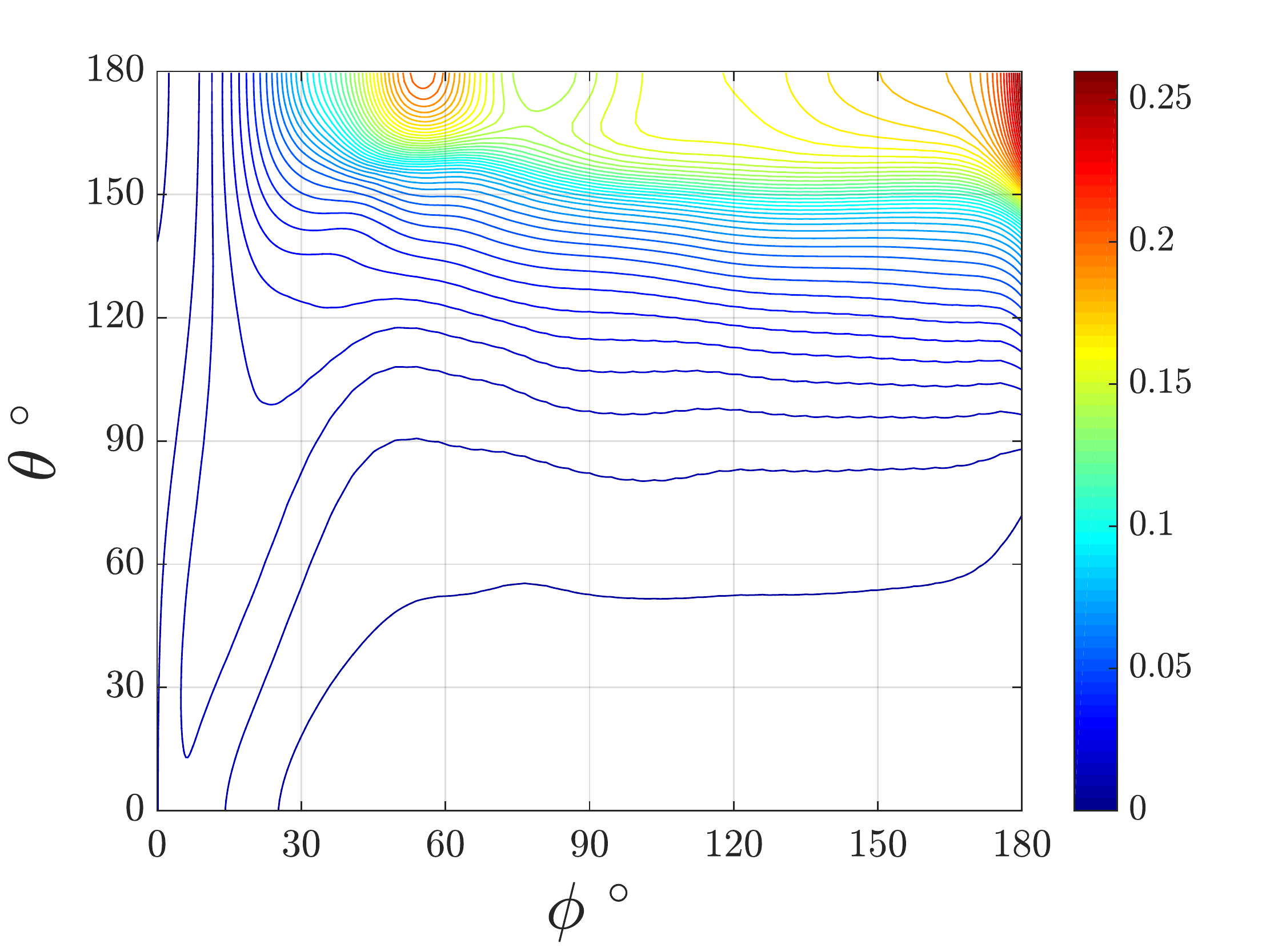}
   }
   
   \caption{Oscillatory shear index, OSI: (a) WEC, (b) UEC. Higher values in (a) occur at $(\phi,\theta) \approx (33^\circ,180^\circ)$ and $(\phi,\theta) \approx (8^\circ,0^\circ)$, reflecting larger localized flow reversal at these locations.}
   \label{f:osi}
\end{figure*}

Values of OSI are plotted in Fig.~\ref{f:osi}. Results from WEC and UEC are quite similar for $\phi > 60^\circ$. The sharp gradient indicated by the cluster of contour lines at $\theta=150^\circ$ reflects the higher values near the inner wall that are mainly due to flow reversal during the deceleration phase of the pulsatile waveform. For $\phi < 60^\circ$, however, the results from WEC and UEC are noticeably different. Under WEC, a pocket of high OSI occurs at the inner wall where $\phi \approx 33^\circ$. This value reflects the large local flow reversal at this location, indicated by the vector plot in Figs.~\ref{f:wss_wec_uec} and \ref{f:wss_wec_uec_2} between $0.23 < t^\star < 0.27$. Furthermore, large values of OSI are also observed along the outer wall near the entrance where $\phi<15^\circ$. The value in this region also indicates flow reversal, which is inherent to the fully developed entrance condition near the wall. Under UEC, we do not observe any increased OSI near the entrance since the velocity condition is uniform and always streamwise positive. A pocket of increased OSI does occur along the inner wall; however, it appears further downstream at $\phi \approx 56^\circ$ and is approximately 75\% of the value obtained under WEC.

%%%%%%%%%%%%%%%%%%%%%%%%%%%%%%%%%%%%%%%%%%%%%%%%%%
%\include{F-tawss}
%\input{F-tawss.tex}
%
%%%%%%%%%%%%%%%%%%%%%%%%%%%%%%%%%%%%%%%%%%%%%%%%%%

\subsection{Relative Residence Time}
\label{s:pulsatile_rrt}

The relative residence time metric (\cite{himburg:2004}) is a relative concept since all particles move in the flow and therefore exhibit zero residence time. The formulation of this metric is based upon the idea that an entrained particle located a small distance from the wall travels a certain distance during one cardiac cycle.
%%
%\begin{align}
%    %%RRT \sim \frac{1}{TAWSS \big( 1 - 2\ OSI \big)}\\
%    L(a) = \frac{Ta}{\mu}\ \mathrm{TAWSS}\ \big( 1 - 2\ \mathrm{OSI} \big)
%\end{align}
%%
%where $T$ is the cardiac cycle period. Since RRT is inversely proportional to $L$, we can write
It can be defined as
\begin{linenomath}
\begin{align}
    %%RRT \sim \frac{1}{TAWSS \big( 1 - 2\ OSI \big)}\\
    \mathrm{RRT} \sim \frac{1}{\mathrm{TAWSS}\ \big( 1 - 2\ \mathrm{OSI} \big)}.
\end{align}
\end{linenomath}
Similar to OSI, it is a uniaxial metric whereby OSI modifies the effect of time-averaged wall shear stress on the relative residence time. As OSI approaches zero it has little effect on RRT, which becomes inversely proportional to the time-averaged wall shear stress. On the other hand, OSI has an increasingly larger effect on RRT as it approaches the upper limit of 0.5, alluding to the local oscillatory nature of the shear.

\begin{figure*}
   \centering\setcounter{subfigure}{0}
   \subfloat[]{
       \includegraphics[width=0.47\textwidth,keepaspectratio]
       {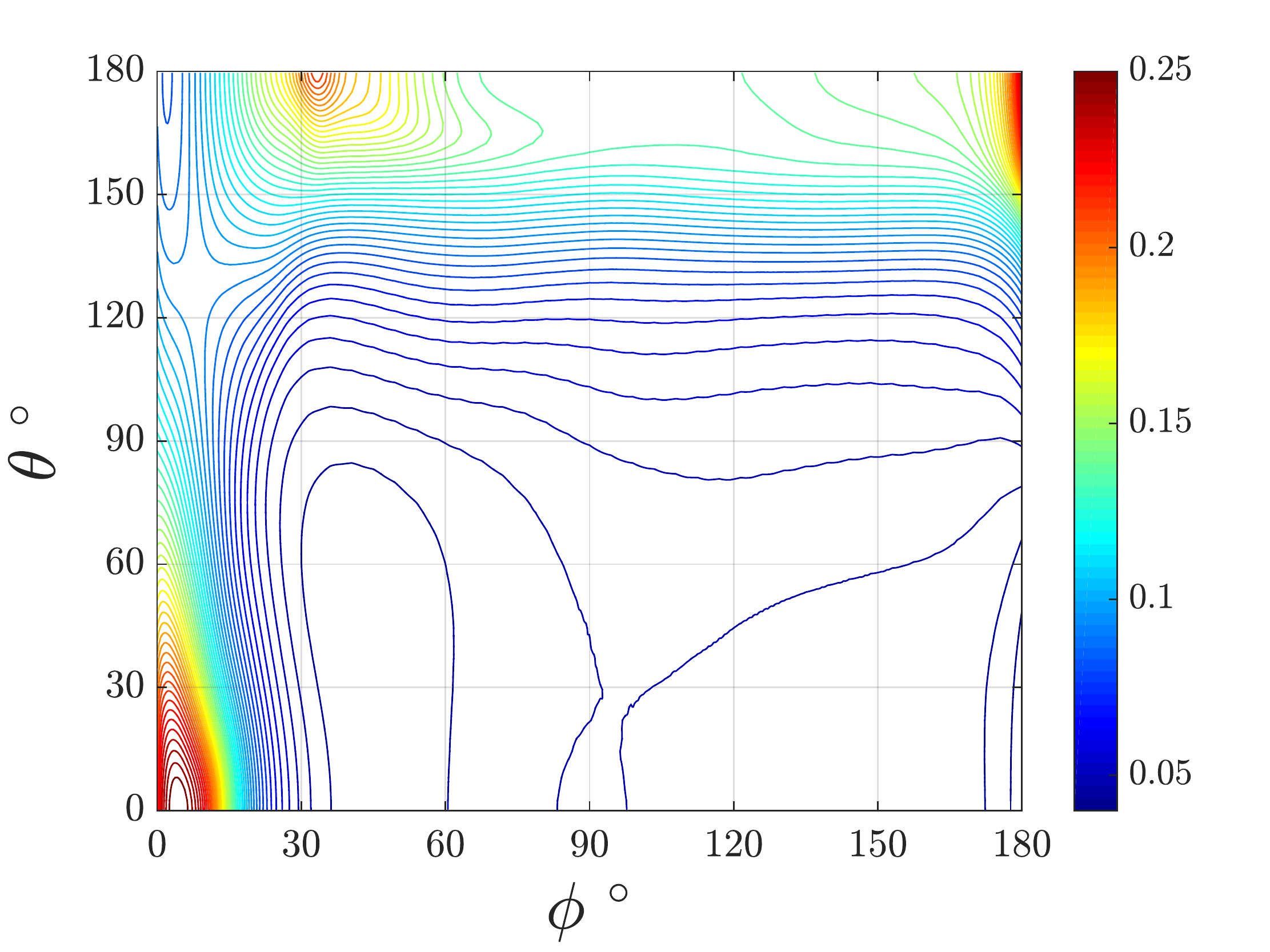}
   }\hfill
   \subfloat[]{
       \includegraphics[width=0.47\textwidth,keepaspectratio]
       {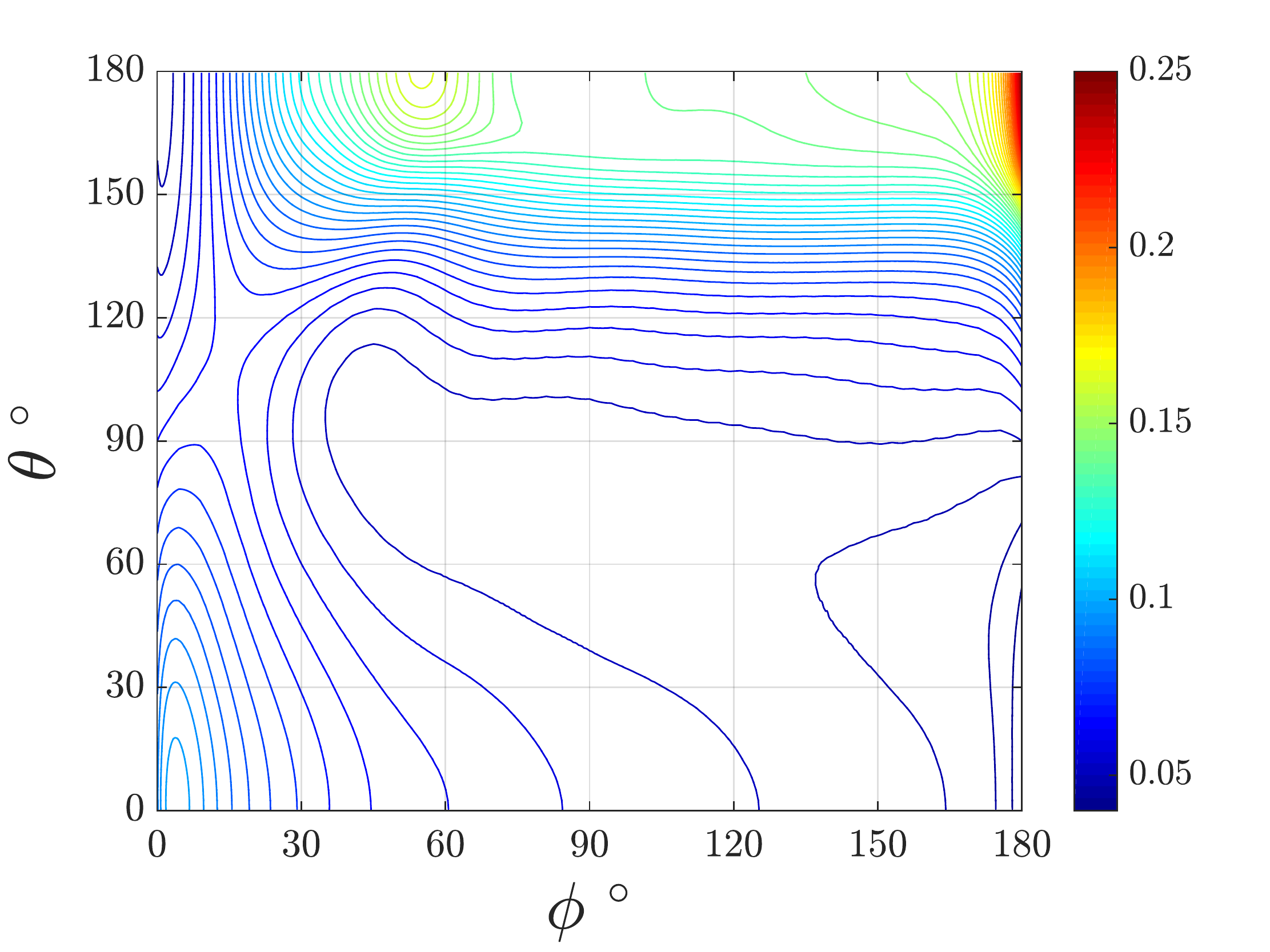}
   }
   
   \caption{Relative residence time, RRT: (a) WEC, (b) UEC. The images look similar to those in Fig.~\ref{f:osi}, supporting the fact that higher values of OSI have a larger effect on RRT.}
   \label{f:rrt}
\end{figure*}

Results of RRT are plotted in Fig.~\ref{f:rrt}. Since the maximum value of OSI is $\sim$0.25 and the minimum value of TAWSS is $\sim$8, the maximum value of RRT is also $\sim$0.25. The pattern of RRT looks quite similar to OSI, supporting the description above that higher values of OSI have a larger effect on RRT. Therefore, the combination of secondary and reverse flow causes a similar effect on RRT as it does on OSI. Further comparison of the results also reveals that WEC produces much larger RRT values near the entrance along the outer wall due to flow reversal---a pattern similar to the one computed in OSI.

%%%%%%%%%%%%%%%%%%%%%%%%%%%%%%%%%%%%%%%%%%%%%%%%%%
%\include{F-osi}
%\include{F-rrt}
%\include{F-transwss}
%\input{F-osi.tex}
%\input{F-rrt.tex}
%\input{F-transwss.tex}
%

\begin{figure*}
   \centering\setcounter{subfigure}{0}
   \subfloat[]{
       \includegraphics[width=0.47\textwidth,keepaspectratio]
       {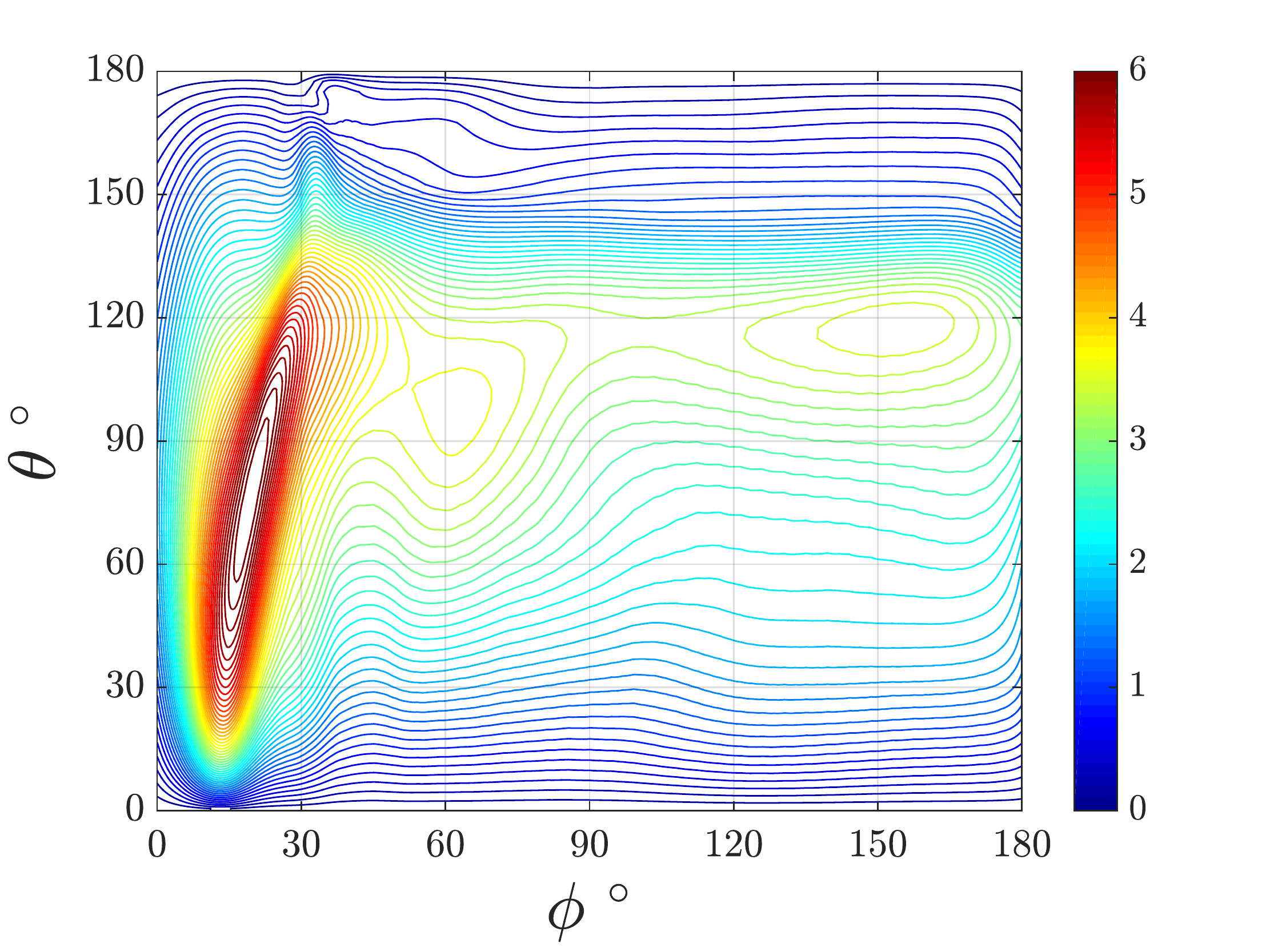}
   }\hfill
   \subfloat[]{
       \includegraphics[width=0.47\textwidth,keepaspectratio]
       {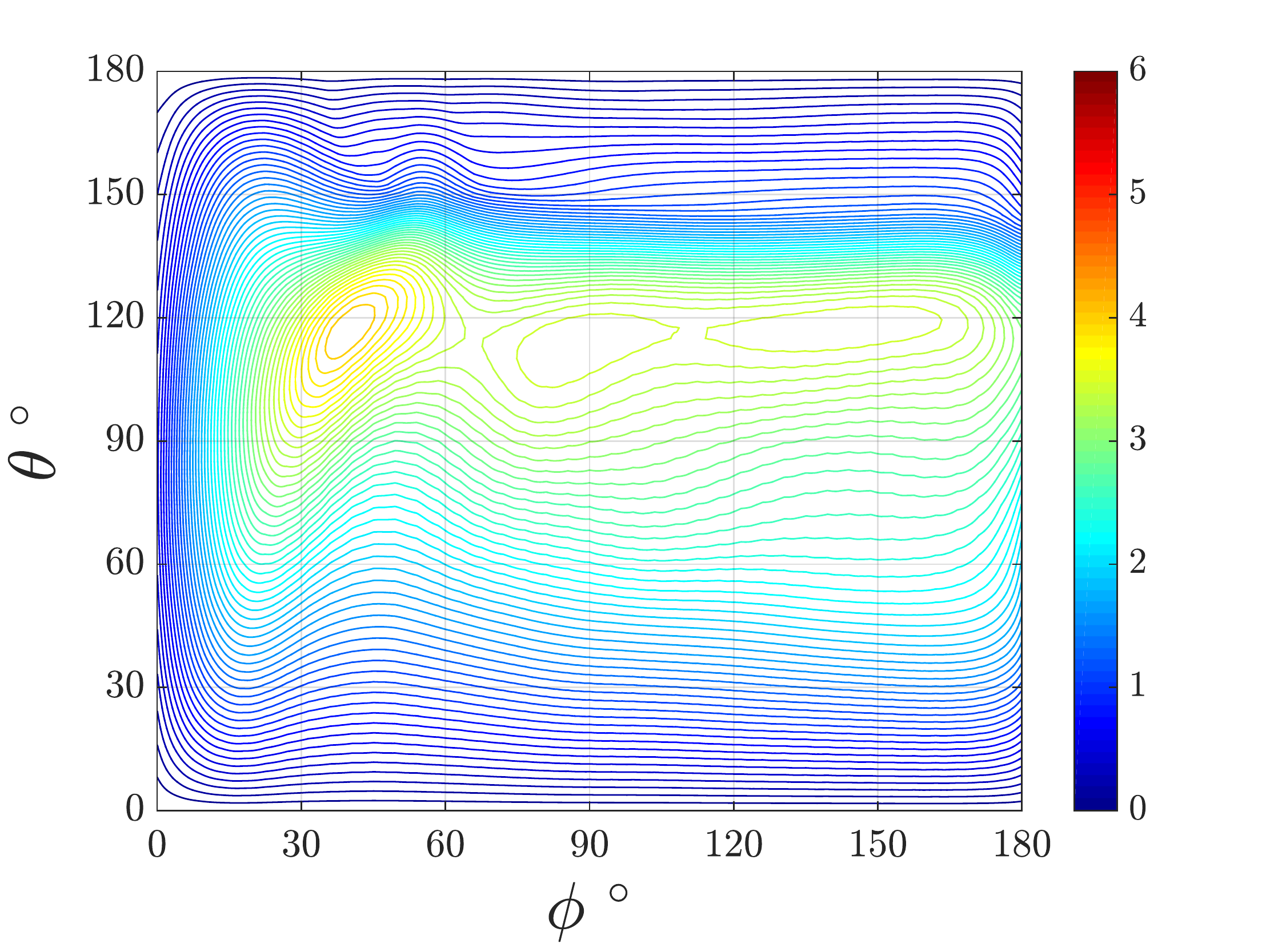}
   }
   
   \caption{Transverse wall shear stress, TransWSS: (a) WEC, (b) UEC. The more complex flow under a fully developed condition causes a higher degree of multidirectionality in the wall shear stress vector, resulting in a 50\% increase in maximum TransWSS in (a).}
   \label{f:transwss}
\end{figure*}
%%%%%%%%%%%%%%%%%%%%%%%%%%%%%%%%%%%%%%%%%%%%%%%%%%

\subsection{Transverse Wall Shear Stress}
\label{s:pulsatile_transwss}

Although widely popular, metrics such as TAWSS, OSI and RRT presented thus far are unable to distinguish between uniaxial and multidirectional flows. To account for multidirectionality of the shear stress vector, the transverse wall shear stress metric (\cite{peiffer-sherwin-weinberg:2013}) was designed. This metric is computed as the time-average of the magnitude of wall shear stress components perpendicular to the mean shear vector at each node on the wall. The formulation for this metric is
\begin{linenomath}
\begin{align}
    %TransWSS = \frac{1}{T}\int_{0}^{T}
    %		   \left| \bmm{\tau}_w \cdot
    %		   \left(
    %		   \bmm{n} \times \frac{\bmm{\tau}_{mean}}{\lvert \bmm{\tau}_{mean} \rvert}
    %		   \right)
    %   	   \right| dt\\
    \mathrm{TransWSS} = \int_{0}^{1}
    \Bigg| \ \bmm{\tau}^{\star}_w \cdot
    \Bigg(
    \bmm{n} \times \frac{\bmm{\tau}^{\star}_{mean}}{\lvert \bmm{\tau}^{\star}_{mean} \rvert}
    \Bigg) \
    \Bigg| \ \drm{t}^\star
    \label{e:transwss}
\end{align}
\end{linenomath}
where $\bmm{n}$ is the vector normal to the wall. Integrating this value over the cardiac cycle produces the TransWSS metric at each nodal position on the wall, which can theoretically range in value from zero to TAWSS. Low values of TransWSS indicate alignment of the shear vector primarily with a single direction, whereas high values signify multidirectionality.

Maps of TransWSS for our pulsatile simulations are shown in Fig.~\ref{f:transwss}. Along the outer and inner walls at $\theta=0^\circ$ and $\theta=180^\circ$, respectively, TransWSS is zero because the flow is always uniaxial close to the walls at the $z=0$ plane of symmetry. Downstream of $\phi=60^\circ$, results appear similar. Upstream of $\phi=60^\circ$, however, the images are quite different---a region of high TransWSS occurs under WEC for $15^\circ < \theta < 130^\circ$ and $10^\circ < \phi < 30^\circ$ with a maximum value of 6.1 \cmmnt{(1.88 dyn/cm$^2$)} at $(\phi,\theta) \approx (20^\circ,78^\circ)$. This multidirectionality of the WSS vector is a direct result of varying intensities of secondary flow under pulsatility. Under UEC, a maximum value of 4.0 \cmmnt{(1.25 dyn/cm$^2$)} occurs at $(\phi,\theta) \approx (40^\circ,118^\circ)$.

From these results, we observe that the fully developed entrance condition produces TransWSS values approximately 50\% greater than the uniform condition. Since Dean-type vortices occur due to secondary flow induced by the curvature and the magnitude of secondary flow varies throughout the pulsatile cycle, we can conclude that alternating strength of Dean-type vortices near the entrance correlate to high TransWSS in that region under WEC. This observation aligns with previous research (\cite{mohamied-sherwin-weinberg:2017}).

%%%%%%%%%%%%%%%%%%%%%%%%%%%%%%%%%%%%%%%%%%%%%%%%%%
\section{Conclusion}
\label{s:conclusion}

We performed numerical simulations of a curved artery model using a physiological (pulsatile) waveform to investigate the influence of entrance flow development on the appearance of abnormal wall shear stresses and the production of common WSS metrics such as TAWSS, OSI, RRT and TransWSS. This work presented results under two entrance conditions, the first of which was fully developed and the second which was undeveloped (i.e. uniform). We demonstrated that a fully developed entrance condition generated up to 33\% larger OSI values along the inner wall due to locally complex reverse flow patterns during deceleration. We also concluded that the fully developed condition generated up to 50\% larger TransWSS values near the entrance to the curve due to multidirectionality of the wall shear stress vector caused by strong fluctuations in secondary flow. This signifies that alternating strength of Dean-type vortices directly correlates to high TransWSS. These results highlight the importance of curvature entrance flow conditions in producing varying degrees of oscillatory and multidirectional flow and its effect on shear stress metrics that are commonly used to assess the progression of atherosclerosis.

%%%%%%%%%%%%%%%%%%%%%%%%%%%%%%%%%%%%%%%%%%%%%%%%%%
\section*{Acknowledgments}
\label{s:acknowledgments}

This study was conducted under the support of the Presidential Merit Fellowship and the Center for Biomimetics and Bioinspired Engineering at The George Washington University.

%%%%%%%%%%%%%%%%%%%%%%%%%%%%%%%%%%%%%%%%%%%%%%%%%%
%\appendix
%\section{Appendix}
%\label{s:appendix}


% Loading bibliography style file
% \bibliographystyle{cas-model2-names}
\bibliographystyle{spmpsci}
%\bibliographystyle{plain}

%\bibliography{cox_job_2022}
%\bibliography{job-v1.bbl}

%%%%%%%%%%%%%%%%%%%%%%%%%%%%%%%%%%%%%%%%%%%%%%%%%%
%\newpage
%\hypersetup{colorlinks=false,linktoc=none}
%\let\oldnumberline\numberline
%\renewcommand{\numberline}{Fig.~\oldnumberline}
%\setlength{\cftfigindent}{0pt}
%\pagenumbering{gobble}
%\listoffigures\pagestyle{empty}

\end{document}